\documentclass[a4paper,fleqn,usenatbib]{mnras}

\usepackage{newtxtext,newtxmath}
\usepackage{threeparttable}

\usepackage[T1]{fontenc}
\usepackage{ae,aecompl}

\usepackage{graphicx}	
\usepackage{amsmath}	

\usepackage{amssymb}	
\usepackage{subcaption}
\usepackage{lscape}
\usepackage{xcolor}

\hypersetup{citecolor=magenta}

\newcommand\refrep[1]{\textcolor{black}{#1}}
\newcommand\update[1]{\textcolor{black}{#1}}
\newcommand\dtwo[1]{\textcolor{black}{#1}}
\newcommand\HI{H\textsc{i}}
\newcommand\sigmol{$\Sigma_{\text{H}_2}$}

\graphicspath{{./}{Figures/}}

\title[A resolved, multi-wavelength study of gas-rich dwarf galaxies in the Fornax cluster]{A resolved, multi-wavelength study of gas-rich dwarf galaxies in the Fornax cluster using MUSE, MeerKAT, and ALMA}

\author[Nikki Zabel et al.]{Nikki Zabel,$^{1}$\thanks{E-mail: nikki.zabel@uct.ac.za}
Alessandro Loni,$^{2}$
Marc Sarzi,$^{3}$
Paolo Serra,$^{4}$
Arjun Chawla,$^{3}$
Timothy A. Davis,$^{5}$ \newauthor
Dane Kleiner,$^{6}$
S. Ilani Loubser,$^{7}$
Reynier Peletier$^{8}$
\\
$^{1}$Department of Astronomy, University of Cape Town, Private Bag X3, Rondebosch 7701, South Africa \\
$^{2}$INAF - Astronomical Observatory of Capodimonte, Salita Moiariello 16 80131, Naples, Italy \\
$^{3}$Armagh Observatory and Planetarium, College Hill, Armagh BT61 9DG, UK \\
$^{4}$INAF - Osservatorio Astronomico di Cagliari, Via della Scienza 5, I-09047, Selargius, CA, Italy \\
$^{5}$Cardiff Hub for Astrophysics Research \& Technology, School of Physics \& Astronomy, Cardiff University, Queens Buildings, Cardiff CF24 3AA, UK \\
$^{6}$Netherlands Institute for Radio Astronomy (ASTRON), Oude Hoogeveensedijk 4, 7991 PD, Dwingeloo, The Netherlands \\
$^{7}$ Centre for Space Research, North-West University, Potchefstroom 2520, South Africa \\
$^{8}$ Kapteyn Institute, University of Groningen, Landleven 12, NL-9747 AD, Groningen, The Netherlands \\
}

\date{Accepted 2024 November 1. Received 2024 October 24; in original form 2024 September 18}

\pubyear{2024}

\begin{document}
\label{firstpage}
\pagerange{\pageref{firstpage}--\pageref{lastpage}}
\maketitle

\begin{abstract}
We combine new and archival MUSE observations with data from the MeerKAT Fornax Survey and the ALMA Fornax Cluster Survey to study the ionised, atomic, and molecular gas in six gas-rich dwarf galaxies in the Fornax cluster in detail. We compare the distributions and velocity fields of the three gas phases with each other, with MUSE white-light images, and with the stellar velocity fields. Additionally, we derive the resolved molecular Kennicutt-Schmidt relation for each object, and compare these with existing relations for field galaxies and for the Fornax and Virgo clusters. Finally, we explore global measurements such as gas deficiencies and star formation rates to paint as complete a picture of their evolutionary state as possible. We find that all six gas-rich dwarf galaxies have very disturbed ISM, with all three gas phases being irregular both in terms of spatial distribution and velocity field. Most objects lie well below the Kennicutt-Schmidt relations from the literature. Furthermore, they are quite deficient in \HI\ (with def\textsubscript{\HI} between $\sim$1 and $\sim$2 dex), and moderately deficient in H$_2$ (with def\textsubscript{H\textsubscript{2}} between $\sim$0 and $\sim$1), suggesting that, while both cold gas phases are affected simultaneously, \HI\ is removed in significant quantities before H$_2$. We suggest that these dwarfs are on their first infall into the cluster, and are in the process of transitioning from star-forming to passive. A combination of tidal interactions, mergers/pre-processing, and ram pressure stripping is likely responsible for these transformations.
\end{abstract}




\section{Introduction}
\label{sec:intro}
It is well-known that the galaxy population in dense environments is different from that in the ``field''. Most notably, galaxy clusters harbour a relatively large fraction of early-type galaxies \citep{Oemler1974, Dressler1980}, \dtwo{which are typically gas-poor with little to no star formation}. This abundance of ellipticals \update{(and lenticulars)} suggests that cluster galaxies are affected by environmental quenching. This idea is supported by the observation that cluster galaxies generally have smaller atomic gas reservoirs than field galaxies (e.g. \citealt{Haynes1984, Cayatte1990, Solanes2001, Gavazzi2005, Chung2009, Loni2021, Molnar2022}). Broadly, there are two categories of environmental quenching that can occur: quenching through hydrodynamical interactions with the intra-cluster medium (ICM), i.e. ram pressure stripping (RPS, \citealt{Gunn1972}) and viscous stripping \citep{Nulsen1982}; and quenching through galaxy-galaxy interactions \dtwo{and interaction with the cluster potential (e.g. \citealt{Mastropietro2021})}, including \update{tidal interactions,} ``harassment'' \citep{Moore1996} and \update{even} mergers. \dtwo{Lastly, there is ``starvation'', which is the phenomenon of galaxies depleting their gas reservoirs forming stars, as they lack accretion of fresh gas, in the absence of a circumgalactic medium, for example.} Which of these environmental processes is dominant is \update{a topic of ongoing} research, and likely varies between environments. With a significant fraction of galaxies residing in groups or clusters, especially in the local Universe \citep{Zabludoff1998, Robotham2011, Tanoglidis2021}, it is important to obtain a better understanding of the relative importance of these \dtwo{environmental} quenching mechanisms.

Because of their lower gravitational potentials, the environmental mechanisms described above are expected to be more pronounced in dwarf galaxies than in their higher-mass counterparts. Dwarfs are the most common galaxies in the Universe, as well as the building blocks of more massive galaxies. This is particularly \update{relevant} in clusters, where dwarf ellipticals are the most common galaxies (e.g. \citealt{Binggeli1985, Binggeli1988, Ferguson1989, Drinkwater2001b, Drinkwater2001a, Venhola2018}). \dtwo{It has often been found that} these early-type dwarfs are metal-poor with old stellar populations  (e.g. \citealt{Koleva2009, Sybilska2017}), \dtwo{suggesting that they either formed at high redshift and have since evolved passively, or, more likely, are remnants of galaxies that were actively star-forming when they entered the cluster. However, the Fornax cluster harbours a significant population of dwarf galaxies (both early-type and late-type) that experienced brief and intense bursts of star formation before their gas was removed completely, quenching their star formation \citep{Romero-Gomez2023}. It has also been found that the metallicities of early-type galaxies are independent of their environment for the entire stellar mass range $10^{4}\ \text{M}_\odot \lesssim M_\star \lesssim 10^{12}\ \text{M}_\odot$ \citep{Romero-Gomez2024}.}

It was long thought that early-type dwarfs are completely quiescent and have a minimal to \dtwo{non-existent} interstellar medium (ISM). However, in more recent years \dtwo{a minority of} dwarf ellipticals have been discovered with significant amounts of gas and dust, and/or evidence of recent star formation, both in the field and in \update{nearby} clusters (e.g. \citealt{Young1997a, Young1997b, Johnson1997, Kormendy2001, Drinkwater2001b, Rijcke2003a, Buyle2005, Alighieri2007, DeLooze2010, Mentz2016}). While some of this material could have been re-accreted through interactions with other galaxies, it is likely that at least a fraction of them are the remnants of late-type \dtwo{dwarf} galaxies that have been transformed by the cluster environment. The presence of visible structures in \update{these objects}, such as spiral arms, bars, nuclei, and embedded stellar discs, supports this hypothesis \citep{Jerjen2000, Barazza2002, DeRijcke2003b, Lisker2006a, Lisker2006b, Chilingarian2007, Janz2012, Hamraz2019, Su2021}. \update{In Fornax, the} relatively large velocity dispersion \dtwo{of dwarf galaxies} \update{within the cluster} ($429~\pm~41~ \text{km s}^{-1}$ vs. $308~\pm~30~\text{km s}^{-1}$ for the more massive population, \citealt{Drinkwater2001a}) is further evidence for this, as is the fact that the \dtwo{optical shapes} of some dwarfs are supported by rotation (e.g. \citealt{Simien2002, Pedraz2002, DeRijcke2003b, Zee2004, Toloba2009, Toloba2011, Toloba2012, Rys2013, Barroso2019, Scott2020, Bidaran2020}). \update{Lastly,} a morphology-density relation also exists for dwarf galaxies, similar to that for their higher-mass counterparts \citep{Binggeli1988, Tanoglidis2021}. On the other hand, there is some observational evidence that contradicts the idea that dwarf ellipticals in clusters originate from late-type galaxies, including some of them having a number of globular clusters incompatible with that in their proposed precursors (e.g. \citealt{Miller1998, Peng2008, Janssen2012}), and, arguably, the presence of bright nuclei (e.g. \citealt{Binggeli1985, Ferguson1994, Ferrarese2006}).

\dtwo{The nearest galaxy clusters provide the opportunity to} study the \update{ISM of the} dwarf population in detail. The Fornax cluster is one of the two galaxy clusters closest to us \update{(at $\sim$20 Mpc, \citealt{Tonry2001})}. It is located in the Fornax filament of the cosmic web \citep{Nasonova2011, Raj2024}, and visible from the Southern Hemisphere. Fornax is a relatively \dtwo{low-mass} cluster, with a \update{virial mass of $\sim 4.5 \pm 1 \times 10^{13}\ \text{M}_\odot$ ($R_\text{vir}$ = 0.7 Mpc, \citealt{Drinkwater2001a}).}
This is only a fraction of the mass of, for example, the Virgo cluster (the other nearby galaxy cluster at a distance similar to that to Fornax; $\sim$16.5 Mpc, \citealt{Mei2007}), which has a mass closer to 10$^{15}$ M$_\odot$ \citep{Fouque2001}. The galaxy number density in Fornax, on the other hand, is relatively high, with a central density 2-3 times that in Virgo \citep{Jordan2007}. \update{While Fornax has some substructure (e.g. \citealt{Iodice2019b}), it is relatively dynamically evolved compared to the Virgo cluster (e.g. \citealt{Mei2007}).} It has one infalling sub-group, located $\sim 5^{\circ}$ from the cluster core towards the south-east, centred around brightest group galaxy (BGG) NGC1316, or Fornax A \citep{Ekers1983, Iodice2017}. 

\update{With the advent of a new generation of telescopes and instruments in the Southern Hemisphere, the Fornax cluster has been the subject of a number of recent studies and surveys in a range of wavelenghts, including: the optical Fornax Deep Survey (FDS, e.g. \citealt{Peletier2020, Iodice2016, Venhola2017, Venhola2018, Iodice2019a}), \HI\ surveys with Parkes Observatory (e.g. \citealt{Waugh2002}), the Australia Telescope Compact Array (ATCA, \citealt{Lee-Waddell2018, Loni2021}) and the MeerKAT Fornax Survey (MFS, e.g. \citealt{Serra2023, Loni2023, Kleiner2023}), integrated field unit (IFU) surveys with the Multi-unit spectroscopic explorer (MUSE, Fornax3D, e.g. \citealt{Sarzi2018, Iodice2019b, Poci2021, Lara-Lopez2022, Ding2023}) and the Sydney-AAO Multi-IFU (SAMI, e.g. \citealt{Scott2020, Eftekhari2022, Romero-Gomez2023}), far-infrared with the \textit{Herschel} Space Observatory \citep{Davies2013, Fuller2014}, Atacama Large Millimeter/submillimeter Array (ALMA) and Atacama Compact Array (ACA) CO surveys \citep{Zabel2019, Morokuma-Matsui2022}, and an X-ray survey of its centre \citep{Scharf2005, Paolillo2002, Machacek2005, Su2017, Sheardown2018}.} Dwarf galaxies in the Fornax cluster have recently been catalogued by \citet{Venhola2018} as part of the FDS, using deep imaging in the $u', g', r', \text{ and } i'$ bands, reaching depths of $\sim28$ mag arcsec$^{-2}$ depending on the band. They find a total of 564 dwarf galaxies, 470 of which ($\sim 83 \%$) are early-types. This catalogue has since been expanded with additional low surface brightness galaxies \citep{Venhola2022}.

Because low-mass galaxies are expected to rapidly lose the bulk of their ISM after they enter a cluster (e.g. \citealt{Mori2000}), gas-rich dwarfs are not expected to exist in significant numbers in these environments. \refrep{The} ALMA Fornax Cluster Survey (AlFoCS, \citealt{Zabel2019}) \refrep{has demonstrated} that eight such objects, \dtwo{in the mass range $5 \times 10^{8}\ \text{M}_\odot \lesssim M_\star \lesssim 5 \times 10^{9}\ \text{M}_\odot$ have detectable molecular gas (traced by the CO molecule)} in addition to \HI\ \citep{Loni2021}. These galaxies have disturbed and deficient cold gas reservoirs, both compared to field galaxies \update{of similar stellar mass} and higher-mass Fornax cluster members. However, from these CO and (marginally resolved) \HI\ data alone is it not possible to draw any firm conclusions about the evolutionary path of these objects and which environmental processes they may have been experiencing. In this work we present a detailed multi-wavelength analysis of \dtwo{the six of these dwarf galaxies with the most disturbed molecular gas reservoirs}, which are listed in Table \ref{tab:sample}. We present follow-up observations from VLT/MUSE, and new, high-resolution, \refrep{high-sensitivity} \HI\ observations from the MFS. These additional datasets provide information on the morphology and kinematics of the atomic and ionised gas, as well as the stellar distribution and kinematics. We aim to \refrep{utilise} the resulting multi-wavelength dataset to \update{obtain a better understanding of the evolution of these dwarfs in the environment of the Fornax cluster.}

This paper is organised as follows. The new MUSE observations and the data from ALMA and MeerKAT are described in \S \ref{sec:data}. In \S \ref{sec:methods} we describe the methods used to derive maps of the ionised gas and stellar kinematics, derive the Kennicutt-Schmidt relation, and estimate gas deficiencies. In \S \ref{sec:comp_samps} we describe the control samples with which we compare the global properties of the dwarfs in our sample. In \S \ref{sec:results} we show the results, which include resolved maps of the ionised, molecular, and atomic gas, as well as the stellar component, descriptions of each individual galaxy, and relations between various (resolved) quantities. These results are then discussed in more detail in \S \ref{sec:discussion}. Finally, we summarise our findings in \S \ref{sec:summary}. Throughout the paper we adopt the distance to the Fornax cluster (19.95 Mpc, \citealt{Tonry2001}) as a common distance to all galaxies in the sample.

\section{Data}
\label{sec:data}

\subsection{MUSE observations \& data reduction}
\label{sec:MUSE}
Optical integrated-field unit observations of \update{four dwarf galaxies, }FCC261, FCC282, FCC332, and FCC335, were carried out during the night of 23 November 2019 with the Multi Unit Spectroscopic Explorer (MUSE, \citealt{Bacon2010}) mounted to Very Large Telescope Unit Telescope 4 (VLT UT4; Yepun), in visitor mode, under programme ID 0104.A-0734 (PI: N. Zabel). Seeing conditions ranged from $\sim 1 - 1.5^{\prime\prime}$ at the start of the night to $\sim 0.5 - 1^{\prime\prime}$ during the second half of the night, while the airmass ranged from 1.088 to 1.697.

MUSE was used in Wide Field Mode (WFM), resulting in a sky coverage of \update{$\sim1.3$ arcmin$^2$}, sampled with a pixel scale of 0.2$^{\prime\prime}$. Effective resolutions range from $0.670 - 1.139^{\prime\prime}$. The observations cover a spectral range of $4750-9350$ \AA\ with resolving power $R = 3027$. This results in spectral resolutions of \update{$1.57-3.09$ \AA, with a median of 2.33 \AA\ at 7050 \AA}, sampled over 1.25 \AA\ pixel$^{-1}$. Observations were executed in blocks of four 10 minute exposures, dithered by $
\sim2-5^{\prime\prime}$, and rotated by 90\textdegree\ to mitigate the effects of systematics. Total effective observing times are between 1991 and 7478 s, with resulting 5$\sigma$ sensitivities ranging from 23.39 - 24.50 AB mag.

The data were pipeline-processed by the European Southern Observatory (ESO), using the MUSE Instrument Pipeline (version 2.8, \citealt{Weilbacher2020}) within the ESO Recipe Execution Tool (\texttt{EsoRex}, version 3.13.2, \citealt{EsoRex2015}). Standard procedures were followed including flat-field corrections (both twilight and lamp), sky, bias, and dark subtractions, and wavelength calibration. A more detailed description of the MUSE data reduction pipeline can be found in the MUSE Data Reduction Software Manual.\footnote{\url{https://data.aip.de/data/musepipeline/v2.8/muse-pipeline-manual-2.8.pdf}}

MUSE observations of FCC090 and FCC207 were taken from the ESO archive. FCC090 was observed under programme ID 296.B-5054 (PI: M. Sarzi) as part of F3D. Observations for this programme were carried out between July 2016 and December 2017, using a setup similar to the observations described above. A FoV of 1~$\times$~1~arcmin$^2$ was covered, and a wavelength range of 4650-9300 \AA, with a spectral resolution of 2.5 \AA\ at 7000 \AA. FCC207 was observed under programme IDs 098.B-0239, 094.B-0576, 097.B-0761, and 096.B-0063 (PI: Eric Emsellem) between October 2014 and January 2017. A FoV of 1.2 arcmin$^2$ was covered, and a spectral range of 4650-9351 \AA, with a spectral resolution of 2.3 \AA\ at 7000 \AA. 
For consistency with \update{the newly obtained} observations, pipeline-processed data cubes for FCC090 and FCC207 were downloaded from the ESO Archive Science Portal. They were processed following a procedure similar to that described above, but using MUSE Instrument Pipeline v2.2 for FCC090, and v1.6.4 for FCC207, both within \texttt{EsoRex} v3.12.3.

Where necessary, the astrometry of the MUSE data was corrected to ensure a reliable comparison with ancillary data. This is especially important since we aim to compare the distribution of ionised gas and that of other gas phases, and to reliably measure the resolved molecular Kennicutt-Schmidt relation (\S \ref{sub:rKS}). \update{Corrections are done by matching the MUSE white-light images with \textit{g}-band images from the FDS (which was carried out with the VST, which is known to have stable astrometry \update{thanks to its large FoV of 1 deg$^2$), and are typically of the order of $\sim$2-3 pixels, corresponding to $\sim$0.5$^{\prime\prime}$ or $\sim$50 pc at the distance of Fornax.}}

\begin{table*}
\begin{threeparttable}
\caption{Properties of the dwarf galaxies in the sample.}
\label{tab:sample}
\setlength{\tabcolsep}{1.5mm}
\begin{tabular}{lllllllllll}
\hline
Object & Alt. name(s) & RA & Dec & c$z$ & Type & ToT \\
-- & -- & (J2000) & (J2000) & (km s$^{-1}$) & -- & (s) \\
(1) & (2) & (3) & (4) & (5) & (6) & (7) \\
\hline
\hline
FCC090 & MCG 06-08-024 & $03^\mathrm{h}31^\mathrm{m}08.2^\mathrm{s}$ & $-36^\circ17{}^\prime25{}^{\prime\prime}$ & 1827 & E4 pec\dtwo{$^\dagger$} & 5400 \\
FCC207 & -- & $03^\mathrm{h}38^\mathrm{m}19.3^\mathrm{s}$ & $-35^\circ07{}^\prime45{}^{\prime\prime}$ & 1394 & dE2\dtwo{$^\ddagger$} & 12000 \\ 
FCC261 & PGC1 0013418 & $03^\mathrm{h}41^\mathrm{m}21.5^\mathrm{s}$ & $-33^\circ46{}^\prime09{}^{\prime\prime}$ & 1710 & dE3 pec & 2400 \\
FCC282 & MCG 06-09-023 & $03^\mathrm{h}42^\mathrm{m}45.3^\mathrm{s}$ & $-33^\circ55{}^\prime14{}^{\prime\prime}$ & 1266 & \dtwo{SA0} & 3600 \\
FCC332 & -- & $03^\mathrm{h}49^\mathrm{m}49.0^\mathrm{s}$ & $-35^\circ56{}^\prime44{}^{\prime\prime}$ & 1326 & E or S0 & 5400 \\
FCC335 & ESO 359-G002, MCG 06-09-034 & $03^\mathrm{h}50^\mathrm{m}36.7^\mathrm{s}$ & $-35^\circ54{}^\prime34{}^{\prime\prime}$ & 1367 & \dtwo{dE$^*$} & 10800 \\
\hline
\end{tabular}
(1) Galaxy name/FCC number. \update{(2) Common/alternative name(s).} (3) Right ascension. (4) Declination. (5) \update{Heliocentric radial velocity from \citet{Maddox2019}}. (6) Morphological type, \dtwo{as classified by \citet[][indicated with a $\dagger$]{Iodice2019a}, \citet[][indicated with a $\ddagger$]{Rijcke2003a}, \dtwo{$^*$the authors,} or the NASA/IPAC Extragalactic Database (NED)}. (7) Total MUSE exposure time.
\end{threeparttable}
\end{table*}
	
\begin{table*}
\begin{threeparttable}
\caption{Measured/derived properties of the dwarfs in the sample.}
\label{tab:props}
\setlength{\tabcolsep}{1.5mm}
\begin{tabular}{llllllllll}
\hline
Object & $M_\star$ & SFR & $M_{\text{\HI}}$ & $M_{\text{H}_2}$ & def$_\text{HI}$ & def$_{\text{H}_2}$ & 12 + log(O/H) & X\textsubscript{CO} \\
 & (log M$_\odot$) & (M$_\odot$ yr$^{-1}$) & (log M$_\odot$) & (log M$_\odot$) & (dex) & (dex) & - & $\left({\text{cm}^{-2}\ (\text{K km s}^{-1})^{-1}}\right)$ \\
(1) & (2) & (3) & (4) & (5) & (6) & (7) & (8) & (9) \\
\hline
\hline
FCC090 & 8.86 $\pm$ 0.08 & 0.042 & 7.72 $\pm$ 0.04 & 7.57 $\pm$ 0.07 & 0.99 & -0.06 & 8.21 & $1.16 \times 10^{21}$ \\
FCC207 & 8.38 $\pm$ 0.08 & 0.004 & 6.09 $\pm$ 0.08 & 6.56 $\pm$ 0.22 & 2.26 & 0.44 & 8.49 & $3.66 \times 10^{20}$ \\ 
FCC261 & 8.37 $\pm$ 0.09 & 0.004 & 7.19 $\pm$ 0.05 & 6.70 $\pm$ 0.8 & 1.16 & 0.30 & 8.18 & $1.18 \times 10^{21}$ \\
FCC282 & 9.01 $\pm$ 0.08 & 0.045 & 7.06 $\pm$ 0.05 & 7.32 $\pm$ 0.05 & 1.76 & 0.35 & 8.48 & $4.16 \times 10^{20}$ \\
FCC332 & 8.57 $\pm$ 0.08 & 0.009 & 7.18 $\pm$ 0.05 & \refrep{7.02} $\pm$ 0.06 & 1.32 & \refrep{0.19} & \refrep{8.56} & \refrep{$2.89 \times 10^{20}$} \\
FCC335 & 9.15 $\pm$ 0.08 & 0.003 & 7.31 $\pm$ 0.05 & \refrep{6.81} $\pm$ 0.05 & 1.61 & \refrep{1.00} & \refrep{8.64} & \refrep{$1.88 \times 10^{20}$} \\
\hline
\end{tabular}
(1) Galaxy name/FCC number. (2) Stellar mass (see \S \ref{sub:stellar_masses_dwarfs}). (3) Star formation rate (see \S \ref{sub:stellar_masses_dwarfs}). (4), (5), (6) (7) Atomic \& molecular gas mass and deficiency (see \S \ref{sub:deficiencies}). (8) Metallicity (see \S \ref{sub:metallicities}). (9) CO-to-H$_2$ conversion factor (see \S \ref{subsub:mh2}).
\end{threeparttable}
\end{table*}

\subsection{\HI\ data from MeerKAT}
Observations of the \HI\ in the six dwarf galaxies come from the MeerKAT Fornax Survey \citep[MFS,][]{Serra2023}. The MFS blindly maps the \HI\ content of the Fornax cluster in a 12 square degree footprint, which covers both the central region of the cluster (out to $\sim R_\text{vir}$) as well as the region around the infalling sub-group \refrep{surrounding} Fornax A to the south-west of the main cluster (see fig. 1 in 
\citealt{Serra2023}). 

To be able to fully spatially resolve the \HI\ content of the dwarfs, and in order to match the MUSE and ALMA data as closely as possible, we use the highest-resolution version of the data cubes available (i.e. the cubes that were imaged using a Briggs \textit{robust} parameter of 0.0 and a $uv$ taper of 6$^{\prime\prime}$, see table 2 in \citealt{Serra2023}). The beam of these observations is \update{11}$^{\prime\prime}$, with a corresponding \HI\ column density sensitivity of $5.0 \times 10^{19}$ cm$^{-2}$ at 3$\sigma$ over 25 km s$^{-1}$. The \HI\ velocity resolution is 1.4 km s$^{-1}$. \update{While all six dwarfs \refrep{were} observed in \HI\ with ATCA, only FCC090 was \refrep{previously} detected \citep{Loni2021}.}

\subsection{Molecular gas data from ALMA}
Observations of the molecular gas in the dwarf galaxies are from the ALMA Fornax Cluster Survey (AlFoCS, \citealt{Zabel2019}). AlFoCS targeted the CO(1--0) line in all Fornax galaxies with stellar mass ${M_\star \geq 3 \times 10^8\ \text{M}_\odot}$ that \refrep{were} detected in \HI\ with ATCA \citep{Loni2021} or in the far-infrared (FIR) with the \textit{Herschel} Space Observatory \citep{Davies2013}. AlFoCS includes galaxies in the Fornax cluster core only, as neither the ATCA nor the \textit{Herschel} observations cover the group around Fornax A. The beam of the ALMA observations is $\sim 3^{\prime\prime}$, \dtwo{and the root mean square (rms) values in the cubes are 2.3 - 3.1 mJy beam$^{-1}$} for the dwarf galaxies (see table 3 in \citealt{Zabel2019}). The velocity resolution is 10 km s$^{-1}$, except for FCC207 and FCC261, which were imaged with a 2 km s$^{-1}$ resolution to account for their narrow linewidths.

\section{Methods}
\label{sec:methods}

\subsection{Stellar kinematics}
In this work, we are primarily interested in studying the stellar kinematics, the morphology and kinematics of the ionised gas, and deriving resolved star formation rates (SFRs). To ensure consistency with past work, the analysis of the MUSE cubes is performed similarly to that of F3D, as described below.

To extract stellar kinematics, we use the Penalized PiXel-Fitting method (\texttt{pPXF}, \citealt{Cappellari2004, Cappellari2017}), which uses a maximum penalised likelihood approach (see \citealt{Cappellari2004} for more details). \dtwo{The full MILES stellar library was used to provide templates \citep{Vazdekis2012, Vazdekis2015}.} The library has an age range of 30 Myr to 14 Gyr, and a metallicity range of -2.26 $\leq$ [$M$/H] $\leq$ 0.4 dex. The \dtwo{\texttt{pPXF} fit was limited to a wavelength range of $4750 - 5850$ \AA, as this range contains most of the prominent stellar absorption features.} It has been shown that using the full wavelength range does not significantly improve the fit, whereas using a narrower wavelength range significantly cuts down computation time and prevents mis-matches in resolution between the templates and the data \citep{Sarzi2018, Iodice2019b}. To ensure high enough signal-to-noise ($S/N$) to obtain reliable measurements of the stellar kinematics, the spaxels were Voronoi-binned \citep{Cappellari2003} to a fixed $S/N$ of 50, which has been shown to provide a good balance between $S/N$ and spatial resolution \citep{Sarzi2018, Iodice2019b, Pinna2019a, Pinna2019b}. Additionally, spaxels below the isophote level with $\langle S/N \rangle = 3$ were masked. To account for remaining small mis-matches between the best fitting template and the stellar spectrum, an additional additive polynomial correction of order 10 was used. \dtwo{Given that the intrinsic FWHM of the stellar library is large, 2.51\AA, it will not be possible to measure velocity dispersions lower than $\sim$20 km s$^{-1}$ \citep{Bidaran2020}.}

\subsection{Stellar masses \& star formation rates}
\label{sub:stellar_masses_dwarfs}
Stellar masses of the dwarf galaxies in the sample were derived from Wide-field Infrared Survey Explorer (WISE) band 1 (3.4 $\mu$m) images (T. Jarrett, priv. comm.), \update{using the calibration from \citet{Jarrett2023}, which uses the IMF from \citet{Chabrier2003}}. 
Global SFRs are obtained by summing the pixels of the maps described in \S \ref{subsub:sfr_sd}, which are derived from dereddened H$\alpha$ fluxes. The resulting \refrep{values} are listed in Table \ref{tab:props}.

\subsection{Ionised gas \dtwo{measurements}}
\label{subsub:ionised_gas_meas}
Both Balmer lines (H$\alpha$ and H$\beta$) and the forbidden lines within the chosen wavelength range ([S\textsc{ii}]$_{\lambda6716}$, [S\textsc{ii}]$_{\lambda6731}$, [O\textsc{iii}]$_{\lambda5007}$, [O\textsc{i}]$_{\lambda6300}$, and [N\textsc{ii}]$_{6583}$) were fitted on a spaxel-by-spaxel basis using Gas AND Absorption Line Fitting (\texttt{GANDALF}, \citealt{Sarzi2017}). \refrep{For each spaxel the stellar velocity of the corresponding Voronoi bin was adopted.} The observed Balmer decrement was used to estimate the reddening, as well as a second \update{de-reddening} component to adjust the shape of the continuum. \dtwo{We followed the approach of \citet{Sarzi2018} to determine the detection threshold used to display our emission-line results. Specifically, we used maps for the returned A/N values for all fitted major lines to identify regions devoid of emission and used such regions to derive the A/N value distributions for each line corresponding to false-positive detections. We then restricted our maps to regions where for each line the A/N values exceeded their corresponding 99\% false-positive threshold. This approach allows \refrep{us} to retain regions that \refrep{might have been masked had we adopted} a more standard approach of excluding regions where all lines exceed the same A/N threshold (e.g. A/N $>$ 3).}
 
\subsection{\refrep{Baldwin, Phillips \& Terlevich diagrams}}
\label{sub:bpt}
To verify that the H$\alpha$ emission from which we measure $\Sigma$\textsubscript{SFR} is indeed the result of star formation rather than alternative sources of ionisation, such as the presence of an active galactic nucleus (AGN), we analyse the Baldwin, Phillips \& Terlevich (BPT, \citealt{Baldwin1981}) diagnostic for each pixel in the H$\alpha$ maps of the galaxies in the sample, using the [O\textsc{iii}]$\lambda$5007/H$\beta$ and [N\textsc{ii}]$\lambda$6583/H$\alpha$ line ratios. We use the boundary line from \citet{Kauffmann2003} as a limit for a pixel to be considered star-forming, and the line from \citet{Kewley2001} as a limit below which a pixel is considered in a transition region, where the ionisation is due to a combination of star formation and the presence of an AGN or other sources of ionisation ( ``composite''). \refrep{The boundary line from \citet{Schawinski2007} is used to distinguish between Seyfert-like and LINER-like ionisation.} This analysis shows that in FCC090, FCC207, FCC261, and FCC282, the gas is indeed ionised as a result of star formation. Maps of the BPT diagnostic in each pixel are shown in Figure \ref{fig:BPT_maps_app}, \refrep{and the corresponding BPT diagrams are presented in Appendix \ref{app_sub:bpt_diagrams}}. 

\refrep{However, the use of BPT diagrams only to determine the source of ionisation has been questioned by many studies (see \citealt{Sanchez2024} and references therein). Therefore, we use the recently developed method by \citet{Sanchez2024} as an additional test to ensure the emission is indeed star forming. This method utilises the equivalent width and velocity dispersion of the H$\alpha$ line to assess the ionising source (the WHaD diagram, see figure 1 in \citealt{Sanchez2024}). The boundary line for AGN-like emission occurs at a velocity dispersion of $\sigma_{\text{H}\alpha}\sim 55\ \text{km s}^{-1}$. The velocities of the dwarf galaxies in the sample are scattered around 40 km s$^{-1}$, and are thus below those of the AGN-like measurements in the WHaD diagram. Therefore, the WHaD diagram confirms that the H$\alpha$ emission in these four galaxies is indeed the result of star-formation.}

In FCC332 and FCC335, on the other hand, significant fractions of the measured H$\alpha$ are classified as \refrep{``composite'', ``Seyfert\dtwo{-like}'', or``LINER\dtwo{-like}''}. This means that sources other than star formation are responsible for ionising a significant amount of the ionised gas. As such, the $\Sigma$\textsubscript{SFR} and total SFR measured for these galaxies should be considered upper limits, and their depletion times should be considered lower limits. This will be discussed further in \S \ref{sub:res_ks}. Maps of the BPT classification of FCC332 and FCC335 are shown and discussed in \S \ref{sub:indiv_gals}.

\subsection{\dtwo{Gas-phase metallicities}}
\label{sub:metallicities}
Following F3D, we use oxygen abundance as a proxy for metallicity. \refrep{We} derive this abundance using the calibration from \citet{Dopita2016}:
\begin{equation}
12 + \text{log} \left( \text{O/H} \right)  = \text{log} \left(\frac{\left[ \text{N} \textsc{II} \right]}{\left[\text{S} \textsc{II} \right]} \right) + 0.264\ \text{log} \left( \frac{\left[ \text{N} \textsc{II} \right]} {\text{H} \alpha} \right) + 8.77.
\label{eq:dop16}
\end{equation}
This calibration is independent of reddening, as it uses only the H$\alpha$, [N\textsc{II}], and [S\textsc{II}] emission lines, which are close together in wavelength. Global metallicities were defined as the \update{H$\alpha$-luminosity weighted} average of the resulting metallicity maps, i.e. the spaxels were weighted by the de-reddened H$\alpha$ flux and averaged. 

\refrep{The calibration from \citet{Dopita2016} is applicable to emission from H\textsc{ii} regions. In \S \ref{sub:bpt} we assessed the source of ionisation of the H$\alpha$ emission in the galaxies in our sample, and find that sources other than star formation contribute significantly to the ionisation of the gas in FCC332 and FCC335. Therefore, for FCC332 we have estimated the oxygen abundance based on spaxels with detected star formation emission only. FCC335 has relatively weak (albeit well-detected) emission lines, with equivalent width values of $\sim 2\AA$. This is consistent with ionisation from sources other than O-stars, as indeed shown by its BPT classification (Figures \ref{fig:bpt} and \ref{subfig:FCC335_bpt_diag}). Because in this galaxy there are no clear SF regions, we refrain from estimating its oxygen abundance using the calibration from \citet{Dopita2016}. Instead, we rely on the mass-metallicity relation from \citet{Sanchez2017}, using the calibration from \citet{Pettini2004}, to estimate its global metallicity. The resulting values are listed in Table \ref{tab:props}.}

\subsection{Cold gas masses \& deficiencies}
\label{sub:deficiencies}
\subsubsection{\HI\ masses}
\refrep{We adopt \HI\ masses from the MFS}, which were derived using the standard equation:
\begin{equation}
M_\text{\HI} = 2.356 \times 10^5 \int S_\nu\: d \nu\: D^2,
\end{equation}
where \refrep{$\int S_\nu\: d \nu$} is the integrated \HI\ flux in Jy km s$^{-1}$ and $D$ the distance to the object in Mpc (here assumed to be equal to the the distance to the Fornax cluster, \dtwo{$\sim$20 Mpc, see \S \ref{sec:intro}}). We adopt the \HI\ masses derived using the cubes imaged at 41$^{\prime \prime}$ resolution, as these were found to recover \refrep{the most} \HI\ flux \citep{Serra2023}. The resulting \HI\ masses are listed in Table \ref{tab:props}.

\subsubsection{H$_2$ masses}
\label{subsub:mh2}
Molecular gas masses were derived using the following equation:
\begin{equation}
M_{\text{H}_2} = 2\: m_\text{H}\: D^2\: X_\text{CO}\: \frac{\lambda^2}{2\: k_\text{B}} \int S_\nu\: d \nu,
\label{eqn:mh2}
\end{equation}
where $m_\text{H}$ is the mass of a hydrogen atom, $D$ the distance to the object, $X_\text{CO}$ the CO-to-H$_2$ conversion factor, $\lambda$ the rest wavelength of the CO(1--0) line, $k_\text{B}$ the Boltzmann constant, and $\int S_\nu\: d \nu$ the integrated line flux. Following past work, we use the metallicity-dependent conversion factor from \citet[eqn. 25]{Accurso2017}:
\begin{equation}
\begin{split}
\text{log}\ \alpha_\text{CO} \left( \pm 0.165 \text{dex} \right) =\ &
14.752\ -\ 1.623 \left[ 12 + \text{log} \left( \text{O/H} \right) \right]\ \\
&
 +\ 0.062\ \text{log}\ \Delta \left( \text{MS} \right),
\end{split}.
\end{equation}
12 + log(O/H) is the global metallicity, derived as described in \S \ref{sub:metallicities}, and $\Delta (\text{MS})$ is the distance from the star formation main sequence (SFMS), for which we adopt the main sequence from \citet{Saintonge2017} as a reference. The positions of the dwarf galaxies in our sample with respect to the SFMS are shown in Figure \ref{fig:SFMS}. The resulting $\alpha_\text{CO}$ is then multiplied by $2.14 \times 10^{20}$ to obtain $X_\text{CO}$ in cm$^{-2}$ (K km s$^{-1}$)$^{-1}$. The resulting H$_2$ masses are listed in Table \ref{tab:props}.

\subsubsection{\HI\ deficiency}
\label{subsub:hi_deficiency}
\HI\ deficiency was originally defined by \citet{Haynes1984} as  
\begin{equation}
\text{def}_{\text{HI}} = \left\langle \text{log } \bar{\Sigma}_{\text{HI}} \right\rangle - \text{log } \bar{\Sigma}_{\text{HI}},
\label{eqn:hi_def}
\end{equation}
where $\bar{\Sigma}_{\text{HI}} \equiv S_{\text{HI}}/D^2_{\text{opt}}$, where $S_{\text{HI}}$ is the \HI\ flux in Jy~km~s~$^{-1}$ and $D_{\text{opt}}$ is the diameter of the optical disc in arcminutes. \dtwo{$\left\langle \text{log } \bar{\Sigma}_{\text{HI}} \right\rangle = 0.37$, corresponding to the type-independent \HI\ deficiency parameter, which compares all morphological types to this mean \HI\ surface density. Since early-type galaxies are expected to contain less gas, they are thus expected to have higher \HI-deficiency values by definition.}

Since such a well-defined mass-size relationship is not available for H$_2$, \citet{Zabel2022} used the following alternative definition of \HI\ deficiency to allow for a fair comparison of deficiency in both gas phases:
\begin{equation}
\text{def}_{\text{\HI}} = \text{log } M_{\text{\HI, exp}} - \text{log } M_{\text{\HI, meas}},
\end{equation}
where the expected \HI\ mass, $M_{\text{\HI, exp}}$, is the median \HI\ mass of a control sample at the stellar mass of the galaxy of interest, and $M_{\text{\HI, meas}}$ is the measured \HI\ mass. Results using this definition of \HI\ deficiency were shown to be very similar to the results using the definition in Equation \ref{eqn:hi_def} \citep{Zabel2022}. Here we use xGASS as a reference sample for the expected \HI\ mass. \update{Following \citet{Zabel2019}, we use the rolling median of the \HI\ fraction as a function of stellar mass to calculate $M_{\text{\HI, exp}}$. We divide the sample into 10 stellar mass bins and use a shift of half a bin.} Resulting \HI\ deficiencies are listed in Table \ref{tab:props}.

\subsubsection{H$_2$ deficiency}
\dtwo{Molecular gas deficiencies can be defined in the same way as \refrep{the \HI-deficiencies described} in section \ref{subsub:hi_deficiency}:}
\begin{equation}
\text{def}_{\text{H}_2} = \text{log } M_{\text{H}_2, \text{exp}} - \text{log } M_{\text{H}_2, \text{meas}},
\label{eq:def_h2}
\end{equation}
where $M_{\text{H}_2, \text{exp}}$ corresponds to the expected molecular gas mass of a galaxy, defined as the median molecular gas mass of a control sample at the corresponding stellar mass, and $M_{\text{H}_2, \text{meas}}$ is its measured global molecular gas mass. Here we use the xCOLD GASS as a reference sample for the expected H$_2$ mass. The resulting H$_2$ deficiencies are listed in Table \ref{tab:props}.

\subsection{The resolved Kennicutt-Schmidt relation}
\label{sub:rKS}
The original Kennicutt-Schmidt relation \citep{Schmidt1959, Kennicutt1998} links gas surface density ($\Sigma_{\text{H}\textsc{i} + \text{H}_2}$) and $\Sigma$\textsubscript{SFR} as follows:
\begin{equation}
\Sigma_\text{SFR} \propto \left( \Sigma_{\text{H}\textsc{i} + \text{H}_2} \right) ^{1.4 \pm 0.15}.
\end{equation}
This relation was derived using the combined molecular and atomic gas surface density, and integrated measurements of spiral (including starburst) galaxies only. Since its publication, a number of related studies have been \refrep{carried out}, some of which use $\Sigma_{\text{H}_2}$ rather than $\Sigma_{\text{H}\textsc{i} + \text{H}_2}$, show the relation on sub-galactic scales, and/or using more diverse samples of galaxies (e.g. \citealt{Wong2002, Leroy2008, Bigiel2008, Schruba2011, Lada2012, Ford2013, Utomo2017, Reyes2019, Kreckel2019, Donaire2023}). Most notably, it has been shown that the KS relation depends significantly more strongly on H$_2$ than \HI\ (e.g. \citealt{Reyes2019}). Therefore, here we will derive the resolved, molecular KS relation (hereafter rKS) for the dwarf galaxies in our sample, and compare them to relations from the literature, including existing rKS relations for the Fornax and Virgo clusters.

\subsubsection{Molecular gas surface density}
\label{subsub:mol_gas_sd}
In order to calculate the rKS relation, we convert the CO intensity maps to \sigmol maps using the following equation (which is similar to Equation \ref{eqn:mh2}):
\begin{equation}
N_{\text{H}_2} \left(\text{M}_\odot\: \text{pc}^{-2} \right) = 9.521 \times 10^{36}\: * 2 m_\text{H}\: X_{\text{CO}} \frac{\lambda^2}{2k_{\text{B}}} \int S_\nu\: d \nu,
\end{equation}
\update{where, contrary to Equation \ref{eqn:mh2}, $\int S_\nu\: d \nu$ is the integrated line flux of a single voxel.}
The remaining parameters are identical to those described in \S \ref{subsub:mh2}. 


\subsubsection{Star formation rate surface density}
\label{subsub:sfr_sd}
In order to convert our dereddened H$\alpha$ fluxes to SFRs, we use the following equation from \citet{Calzetti2012}:
\begin{equation}
\label{eqn:SFR}
\text{SFR} \left( \text{M}_\odot\ \text{yr}^{-1} \right) = 5.5 \times 10^{-42}\ L_{\text{H}\alpha} \left( \text{erg s}^{-1} \right),
\end{equation}
where $L_{\text{H}\alpha}$ is the H$\alpha$ line luminosity:
\begin{equation}
L_{\text{H}\alpha} = 4\pi D\ F_{\text{H}\alpha}.
\end{equation}
\refrep{Here} $F_{\text{H}\alpha}$ is the dereddened H$\alpha$ flux in ${10^{20}\ \text{erg\ s}^{-1} \text{cm}^{-2}}$ and $D$ the distance in cm, assumed to be the distance to the Fornax cluster (20 Mpc). \update{Following F3D, we use a two-component reddening correction, adjusting for the observed Balmer decrement as well as the stellar continuum shape}.  $\Sigma$\textsubscript{SFR} is then calculated by dividing the resulting SFRs by the spaxel area in kpc$^2$.

Prior to performing the above calculations, we convolve and regrid the dereddened H$\alpha$ maps to match the resolution of the CO maps (i.e. the beam of the ALMA observations). The resulting 0.5$^{\prime\prime}$ pixels are smaller than the ALMA synthesised beam \dtwo{(i.e. resolution)}, and are therefore not independent. To account for this, we adopt the approach of \citet{Zabel2020}, \refrep{which relates} each pixel in the $\Sigma_{\text{H}_2}$ image to the corresponding pixel in the $\Sigma_{\text{SFR}}$ as well as the neighbouring pixels within an ALMA beam size, and adopt the median as a point in the $\Sigma_{\text{H}_2} - \Sigma_{\text{SFR}}$ plane \update{(for more details \refrep{we} refer to \S 3.2 in \citealt{Zabel2020})}.


\subsubsection{Depletion time maps}
\label{subsub:tdep}
Depletion time maps are derived by dividing \refrep{the molecular gas surface density maps by the star formation rate surface density maps}:
\begin{equation}
t_\text{dep} \left( \text{Gyr} \right) = \frac{\Sigma_{\text{H}_2} \left( \text{M}_\odot\ \text{pc}^{-2} \right)}{\Sigma_\text{SFR} \left( \text{M}_\odot\ \text{yr}^{-1}\ \text{kpc}^{-2} \right)} \times 10^{-15}.
\end{equation}

We use \refrep{the resulting depletion time} maps to study spatial variations in the Kennicut-Schmidt relation (i.e. $t_\text{dep}$) within each galaxy. \update{In most dwarfs in the sample, there are substantial areas where H$\alpha$ is detected but CO is not. If we assume that H$\alpha$ and CO are co-spatial everywhere in the galaxies, but CO remains undetected because the column density is below the detection limit, we can assign an upper limit on the depletion time in those areas. Here we have done so by setting the molecular gas surface density in these regions equal to the lowest detected value in the galaxy. Similarly, FCC090 and FCC332 have (relatively small) areas where CO is detected but H$\alpha$ is not. We set the SFR surface density in these areas equal to the lowest value measured.} The resulting maps are shown in Figure \ref{fig:DT_plots}, along with the corresponding individual Kennicutt-Schmidt relations.

\section{Comparison samples}
\label{sec:comp_samps}
We compare the properties of our dwarf galaxies with a variety of relevant samples\refrep{:} the extended Galaxy Evolution Explorer Arecibo Sloan Digital Sky Survey Survey (xGASS, \citealt{Catinella2018}) and the extended CO Legacy Database for the GASS survey (xCOLD GASS, \citealt{Saintonge2017}), The \HI\ Nearby Galaxy Survey (THINGS, \citealt{Walter2008}) and the Local Irregulars That Trace Luminosity Extremes THINGS survey (LITTLE THINGS, \citealt{Hunter2012}), and the HERA CO-Line Extragalactic Survey (HERACLES, \citealt{Leroy2009}), the Virgo Environment Traced in CO survey (VERTICO, \citealt{Brown2021}) and the VLA Imaging of Virgo in Atomic gas survey (VIVA, \citealt{Chung2009}), and the remaining galaxies in the Fornax cluster from the Fornax Deep Survey (FDS, \citealt{Peletier2020}), the ALMA Fornax Cluster Survey (AlFoCS, \citealt{Zabel2019}\refrep{)}, the Fornax3D survey (F3D, \citealt{Sarzi2018}), and the MeerKAT Fornax Survey (MFS, \citealt{Serra2023}). \update{The relevant parameters from each sample are described below. To ensure fair comparisons, these measurements were corrected to match the measurements described in \S \ref{sec:methods}, where necessary and possible \refrep{(e.g. by scaling $M_{\text{H}_2}$ to match the $X_\text{CO}$ recipe used, etc.)}.}

\subsection{H$_2$ masses \& deficiencies}
H$_2$ masses from each comparison sample were corrected to use the metallicty-dependent $X_\text{CO}$ from \citet{Accurso2017}. Direct measurements of the metallicity were only used for a minority of Fornax galaxies for which MUSE observations are available from F3D. For the remaining galaxies we use the mass-metallicity relation from \citet{Sanchez2017} to estimate the metallicity.

\subsection{Stellar masses \& star formation rates}
For the \refrep{Fornax comparison sample}, if stellar masses from WISE imaging are not available (see \S \ref{sub:stellar_masses_dwarfs}), we adopt them from the FDS \update{\citep{Iodice2019a}}, whose method assumes a Chabrier IMF \citep{Chabrier2003}. If those are not available either, we use stellar masses from the z=0 Multiwavelength Galaxy Synthesis (z0MGS, \citealt{Leroy2019}), who adopt the IMF from \citet{Kroupa2003}. For the HERACLES/(LITTLE) THINGS samples we also use stellar masses from the z0MGS. Stellar masses for the VERTICO/VIVA sample also assume a Kroupa IMF \citep{Kroupa2001}. For the x(COLD )GASS samples we adopt stellar masses from the GALEX–SDSS–WISE LEGACY CATALOG (GSWLC, \citealt{Salim2016}), for which a Chabrier IMF was assumed. \update{While, in principle, these samples could be corrected to the same IMF, in practise the difference between Kroupa/Chabrier is small compared to the uncertanties from other sources, \refrep{making such corrections negligible. Therefore, we adopt the stellar masses as they are.}}

SFRs for the \refrep{Fornax comparison sample} are adopted from F3D where possible, which were derived from the dereddened H$\alpha$ flux in the star-forming spaxels in the MUSE cubes as described in \S \ref{subsub:sfr_sd} and section 4.2 in \citet{Iodice2019b}. For galaxies for which no MUSE data is available, SFRs were derived using \dtwo{mid-infrared fluxes from WISE, with a UV correction}. If those are not available either, SFRs are adopted from the z0MGS. SFRs for the z0MGS were derived using a combination of WISE band 4 (22 $\mu$m) and UV observations as described in section 3.1 in \citet{Leroy2019}. SFRs for the VERTICO sample are also from the z0MGS. SFRs in x(COLD )GASS are also derived from WISE + GALEX data, or from spectral energy distribution (SED) fitting where these data are not available, as described in section 3 of \citet{Janowiecki2017}.

\section{Results}
\label{sec:results}
Figure \ref{fig:intensity_maps} shows the distributions and velocity fields of the three gas phases in each of the dwarf galaxies. The top-left panel corresponds to the MUSE white-light image or stellar velocity field, respectively. The top-right panel shows the \HI\ distribution and velocity field. \dtwo{In these \HI\ maps outer pixels with low S/N are omitted by masking pixels with values below 5.32 Jy beam$^{-1}$ m s$^{-1}$.} The bottom-left panels correspond to the molecular gas, and the bottom-right panels to the ionised gas. The two top panels are shown on the same scale, as are the two bottom panels, \update{which are zoomed in compared to the top panels to better show the more compact molecular and ionised gas phases}. Scalebars are shown in the top-left and bottom-right panels. \update{The arrow in the top-right corner of the top-right panel indicates the projected direction towards the cluster centre.} Contours of the white-light image are overlaid in grey in each panel to aid \refrep{visual} comparison \refrep{between} the top and bottom panels. \update{Additionally, the sizes of the bottom panels are indicated in the top-left panels with dashed black rectangles.} The 11$^{\prime\prime}$ MeerKAT images are shown here, as this resolution is closest to that of the other data. However, in some cases larger-scale \HI\ features are resolved out at this resolution. Therefore, we also show the outer contour of the 22$^{\prime\prime}$ images in the \HI\ intensity map, in \dtwo{light-red}. The respective beams are shown in their corresponding colours. Similarly, the beam of the ALMA observations is shown in the bottom-left panel. The CO and \HI\ velocity fields were converted from \update{the radio definition to the optical definition (both barycentric)} to match the definitions of the stellar and ionised gas velocity fields. 

\begin{figure*}
    \begin{subfigure}{1\textwidth}
    \centering
        \includegraphics[width=0.8\textwidth]{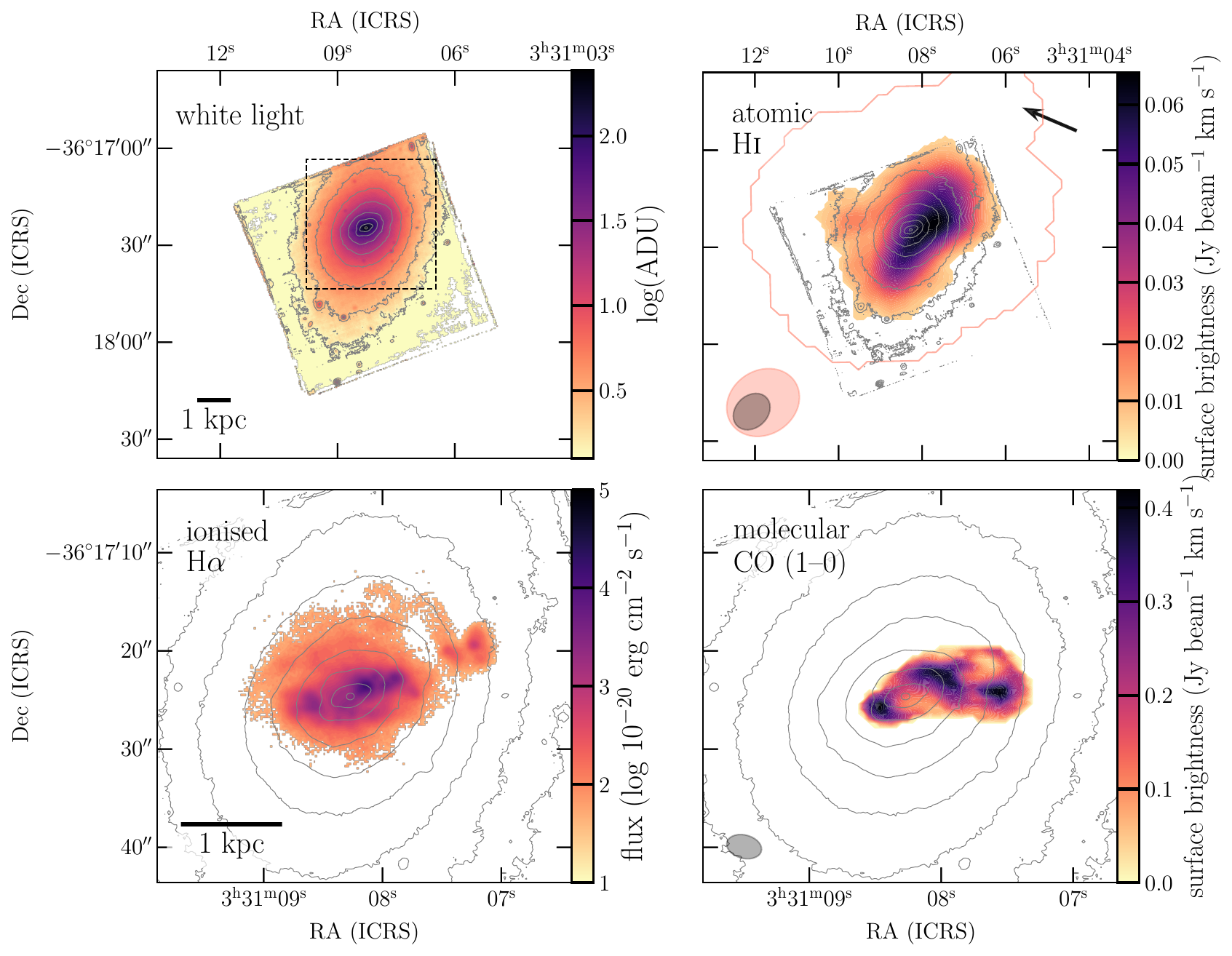}
        \caption{Intensity maps for FCC090.}
        \label{subfig:FCC090_intensity}
    \end{subfigure}
    \begin{subfigure}{1\textwidth}
        \centering
        \includegraphics[width=0.8\textwidth]{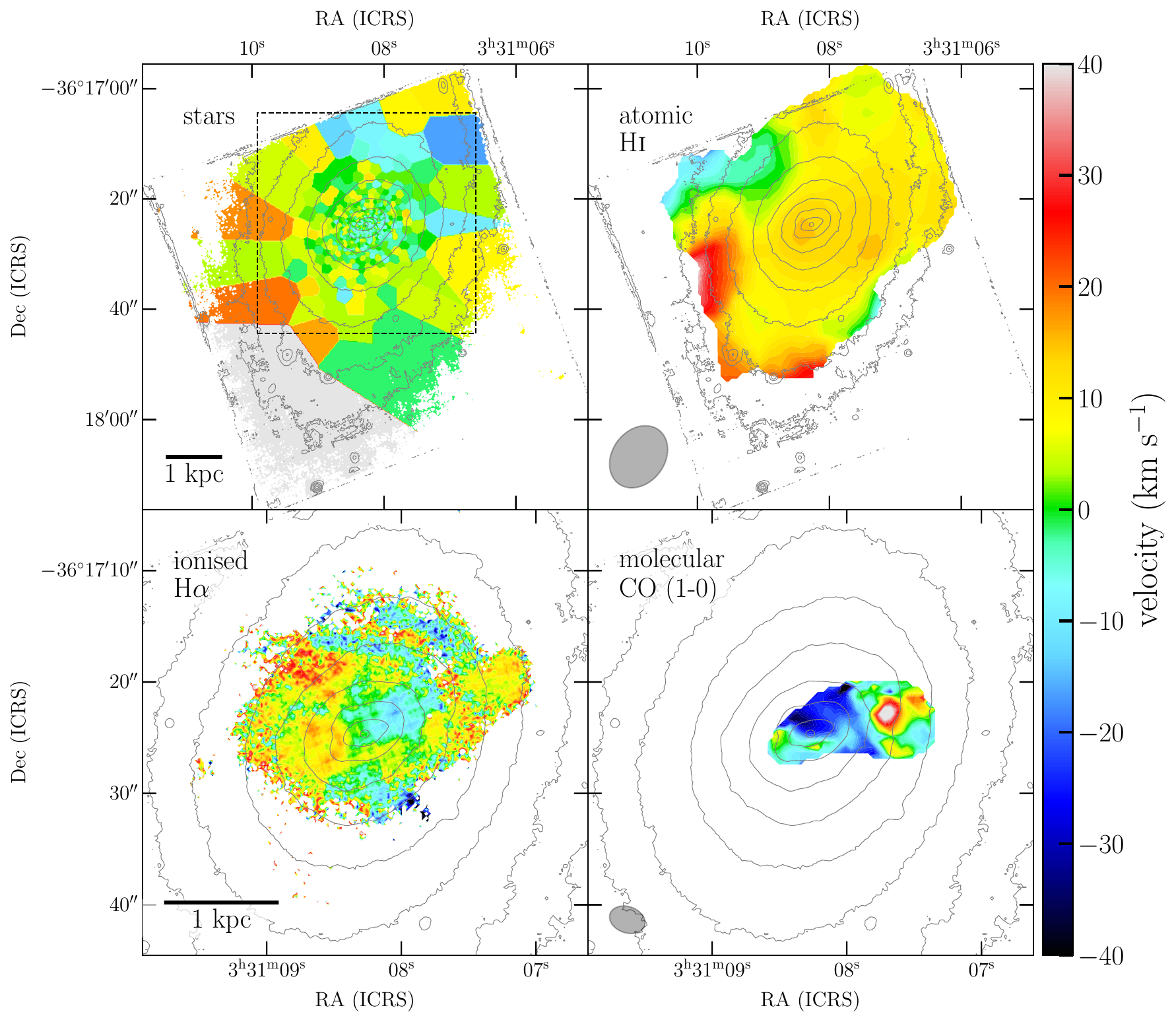}
        \caption{Velocity maps for FCC090.}
        \label{subfig:FCC090_velocity}
    \end{subfigure}
\end{figure*}

\begin{figure*}\ContinuedFloat 
    \begin{subfigure}{1\textwidth}
    \centering
        \includegraphics[width=0.8\textwidth]{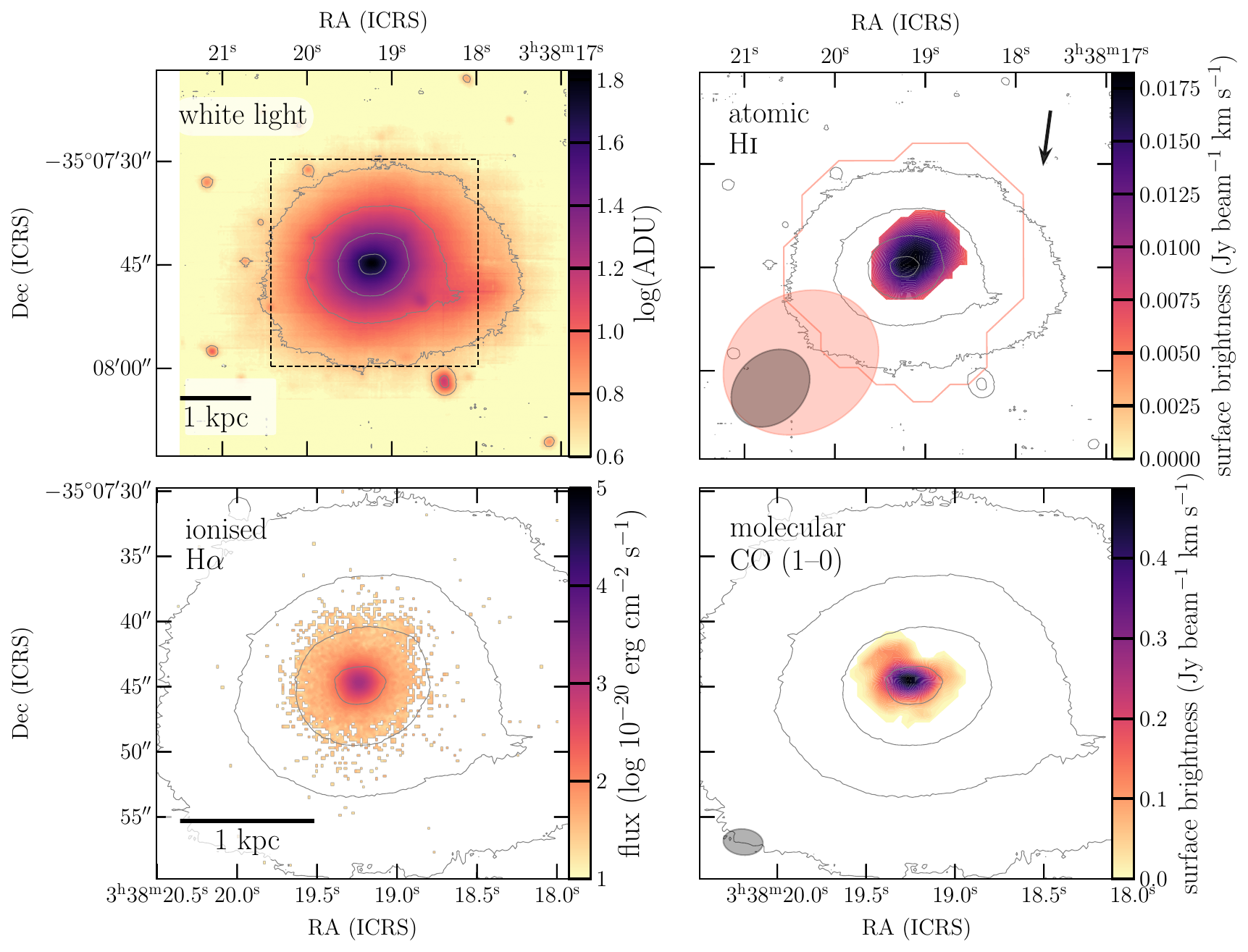}
        \caption{Intensity maps for FCC207.}
        \label{subfig:FCC207_intensity}
    \end{subfigure}
    \begin{subfigure}{1\textwidth}
        \centering
        \includegraphics[width=0.8\textwidth]{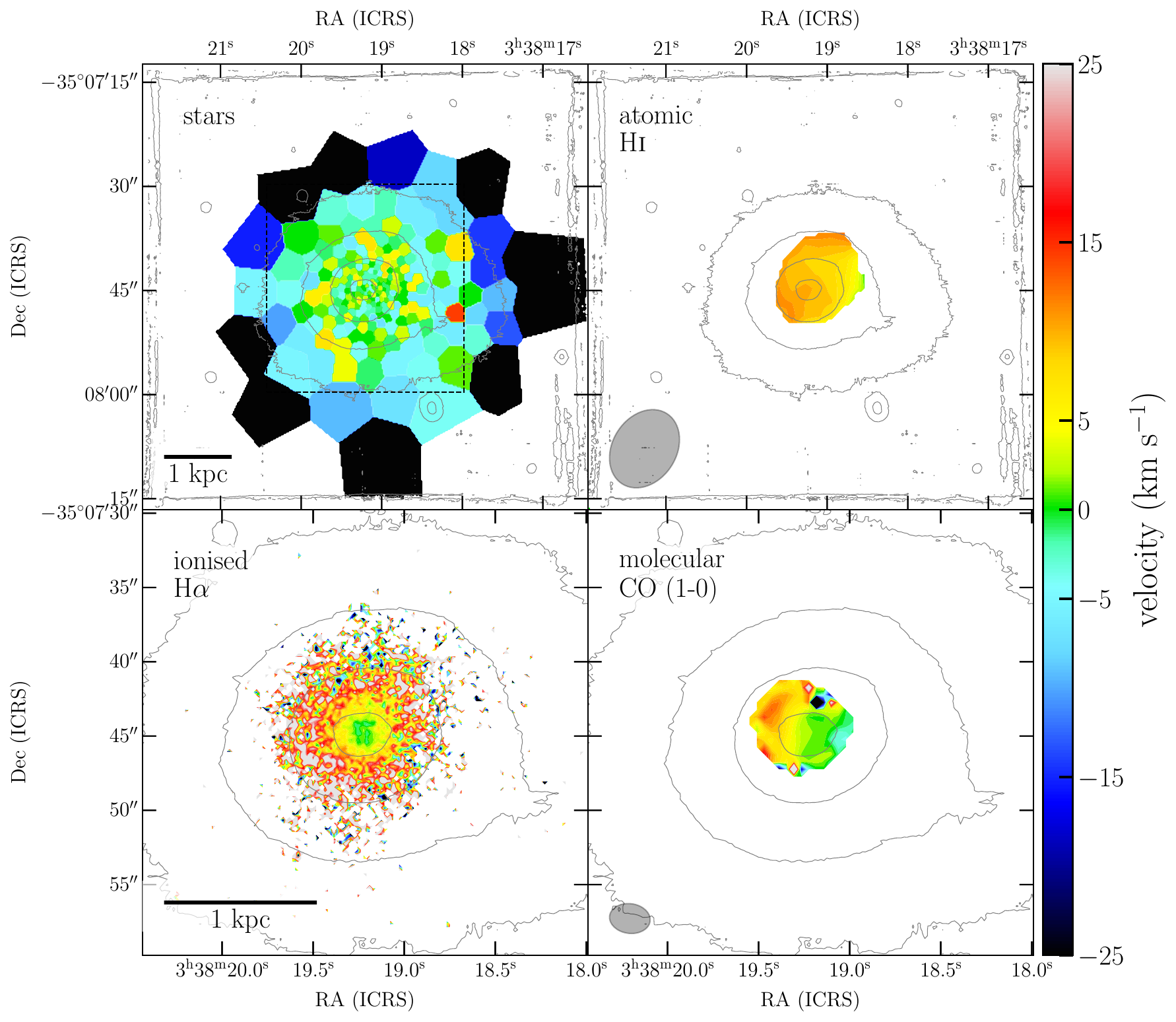}
        \caption{Velocity maps for FCC207.}
        \label{subfig:FCC207_velocity}
    \end{subfigure}
\end{figure*}

\begin{figure*}\ContinuedFloat     
    \begin{subfigure}{1\textwidth}
    \centering
        \includegraphics[width=0.8\textwidth]{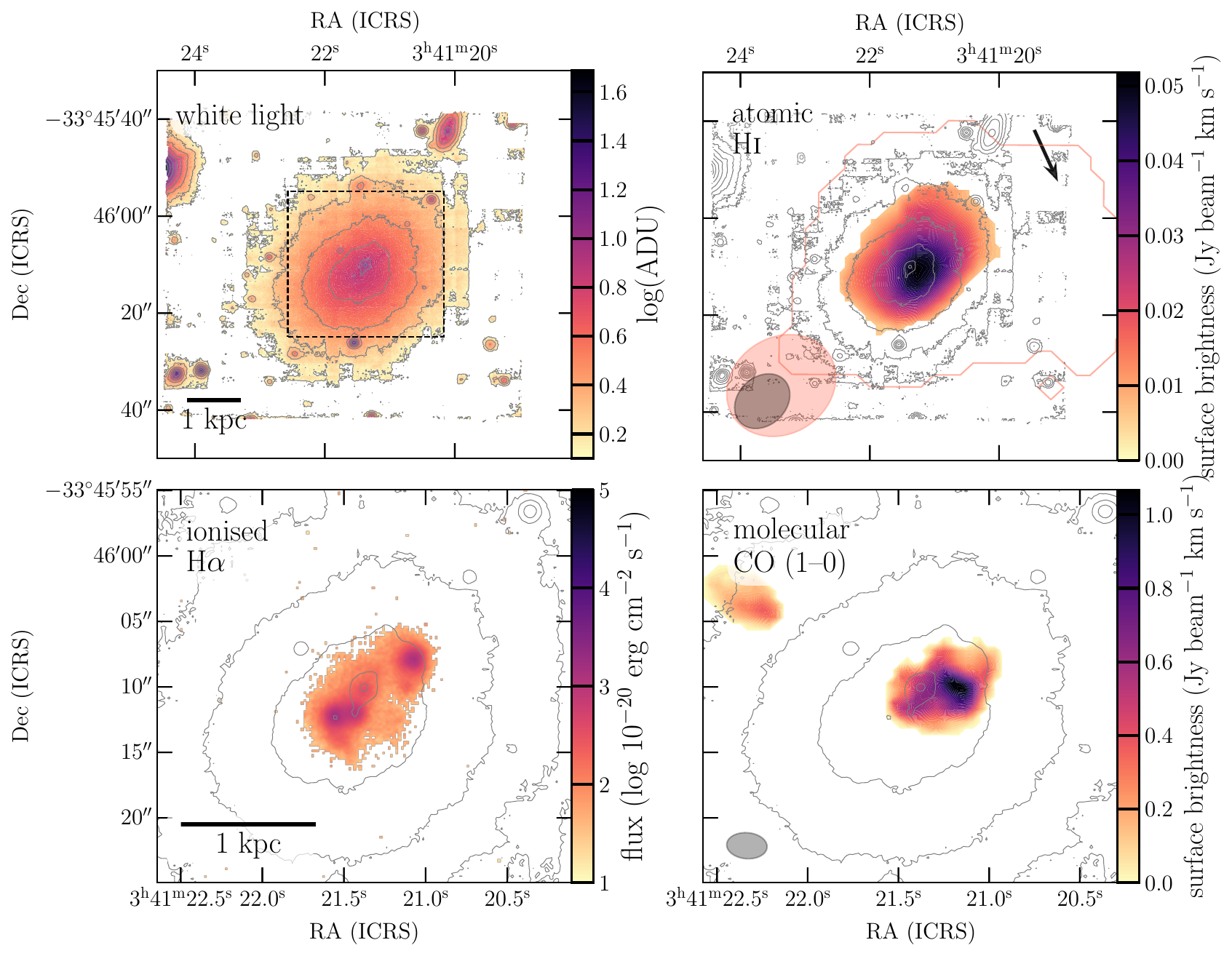}
        \caption{Intensity maps for FCC261.}
        \label{subfig:FCC261_intensity}
    \end{subfigure}
    \begin{subfigure}{1\textwidth}
        \centering
        \includegraphics[width=0.8\textwidth]{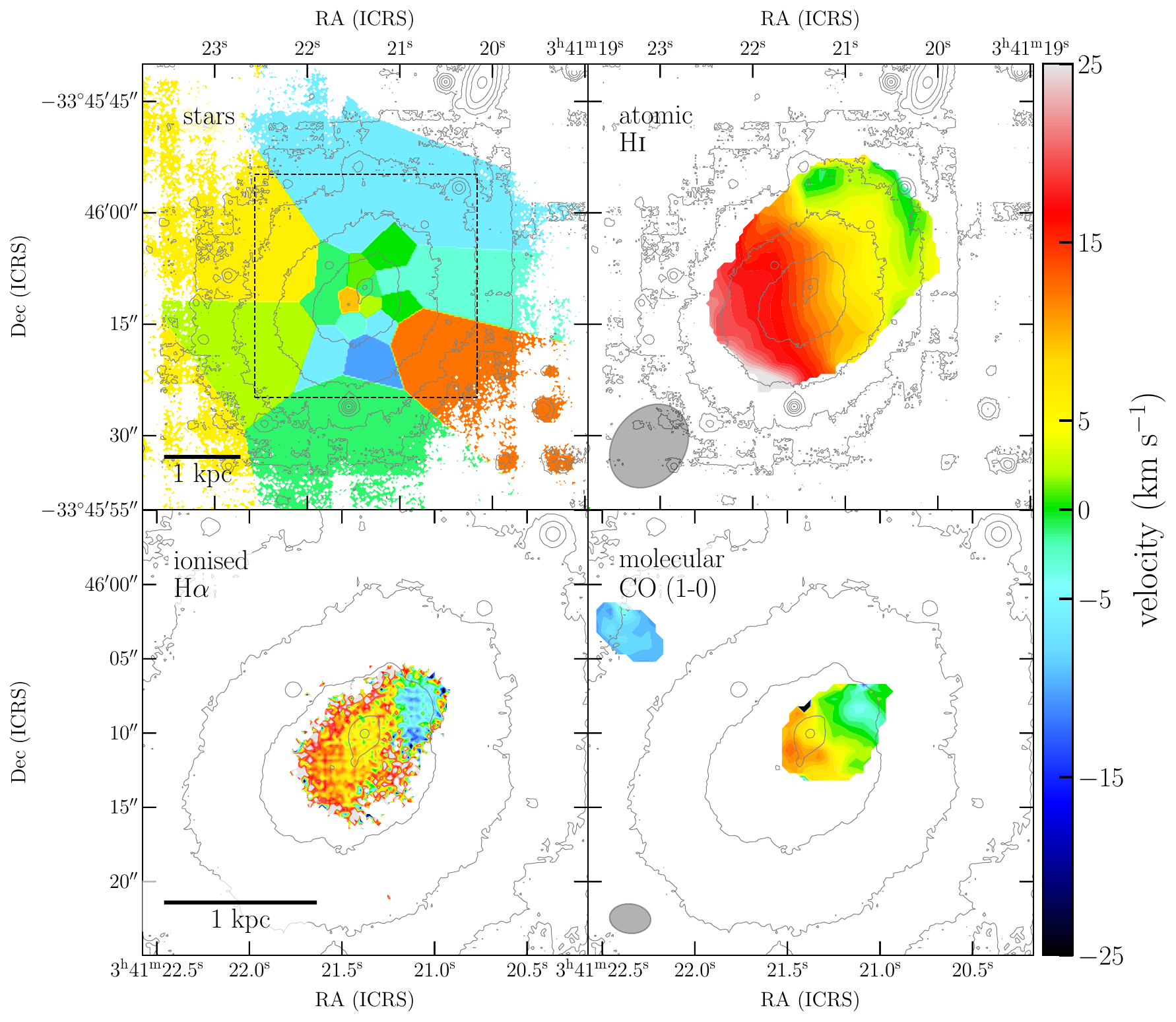}
        \caption{Velocity maps for FCC261.}
        \label{subfig:FCC261_velocity}
    \end{subfigure}
\end{figure*}

\begin{figure*}\ContinuedFloat     
    \begin{subfigure}{1\textwidth}
    \centering
        \includegraphics[width=0.8\textwidth]{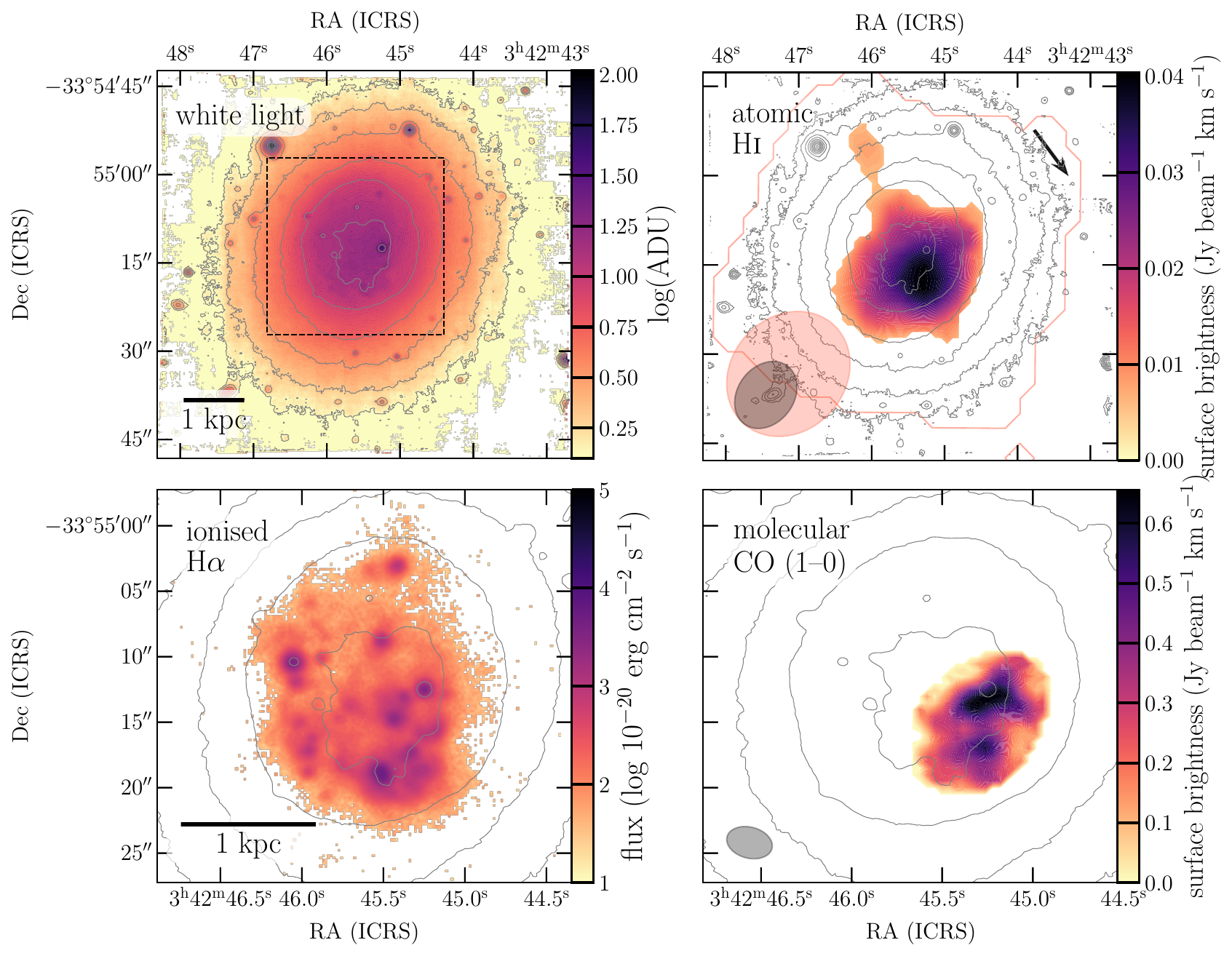}
        \caption{Intensity maps for FCC282.}
        \label{subfig:FCC282_intensity}
    \end{subfigure}
    \begin{subfigure}{1\textwidth}
        \centering
        \includegraphics[width=0.8\textwidth]{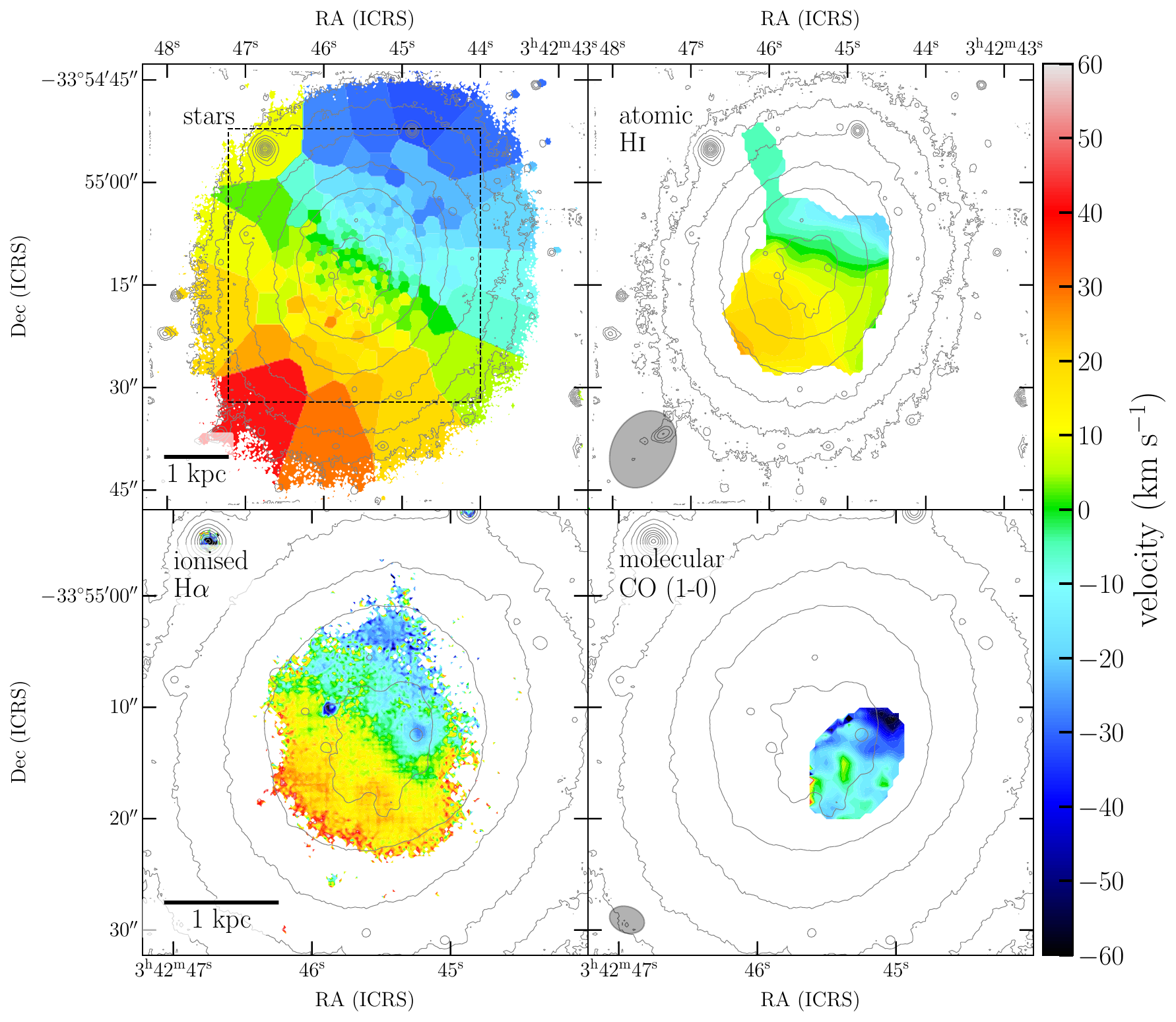}
        \caption{Velocity maps for FCC282.}
        \label{subfig:FCC282_velocity}
    \end{subfigure}
\end{figure*}

\begin{figure*}\ContinuedFloat    
    \begin{subfigure}{1\textwidth}
    \centering
    \vspace{-0.5cm}
        \includegraphics[width=0.8\textwidth]{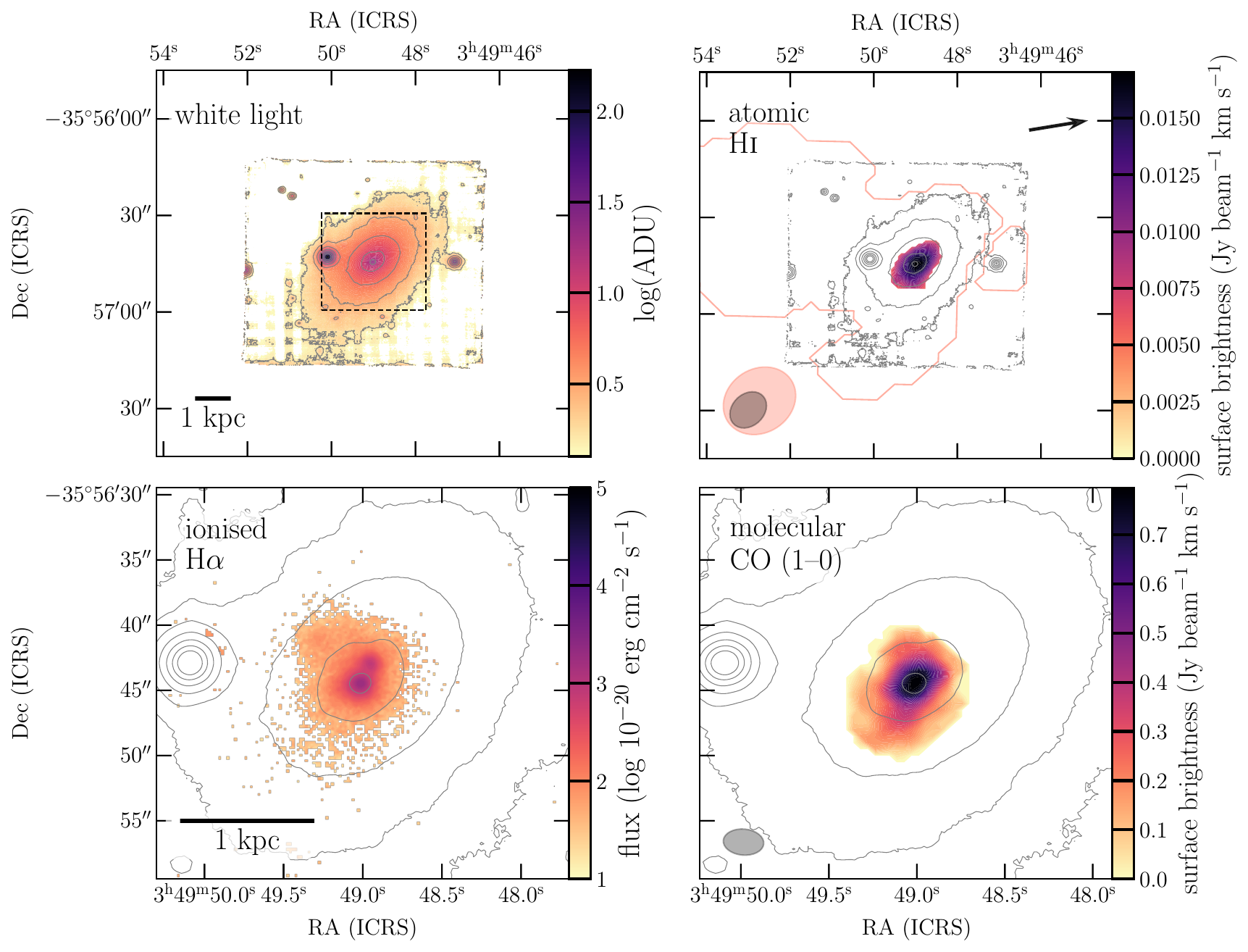}
        \caption{Intensity maps for FCC332.}
        \label{subfig:FCC332_intensity}
    \end{subfigure}
    \vspace{-0.5cm}
    \begin{subfigure}{1\textwidth}
        \centering
        \includegraphics[width=0.8\textwidth]{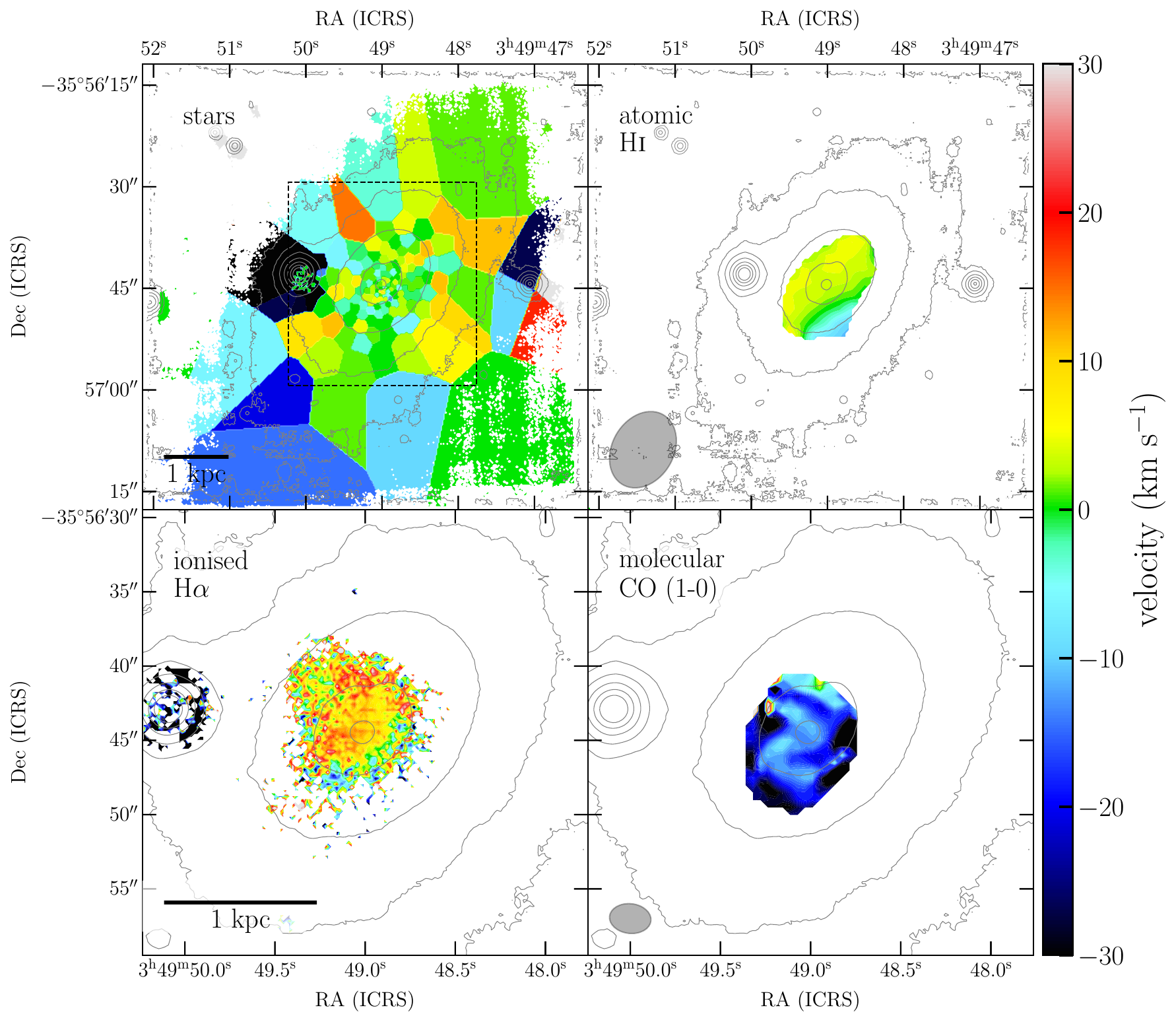}
        \caption{Velocity maps for FCC332.}
        \label{subfig:FCC332_velocity}
    \end{subfigure}
\end{figure*}

\begin{figure*}\ContinuedFloat 
    \begin{subfigure}{0.8\textwidth}
    \centering
        \includegraphics[width=0.915\textwidth]{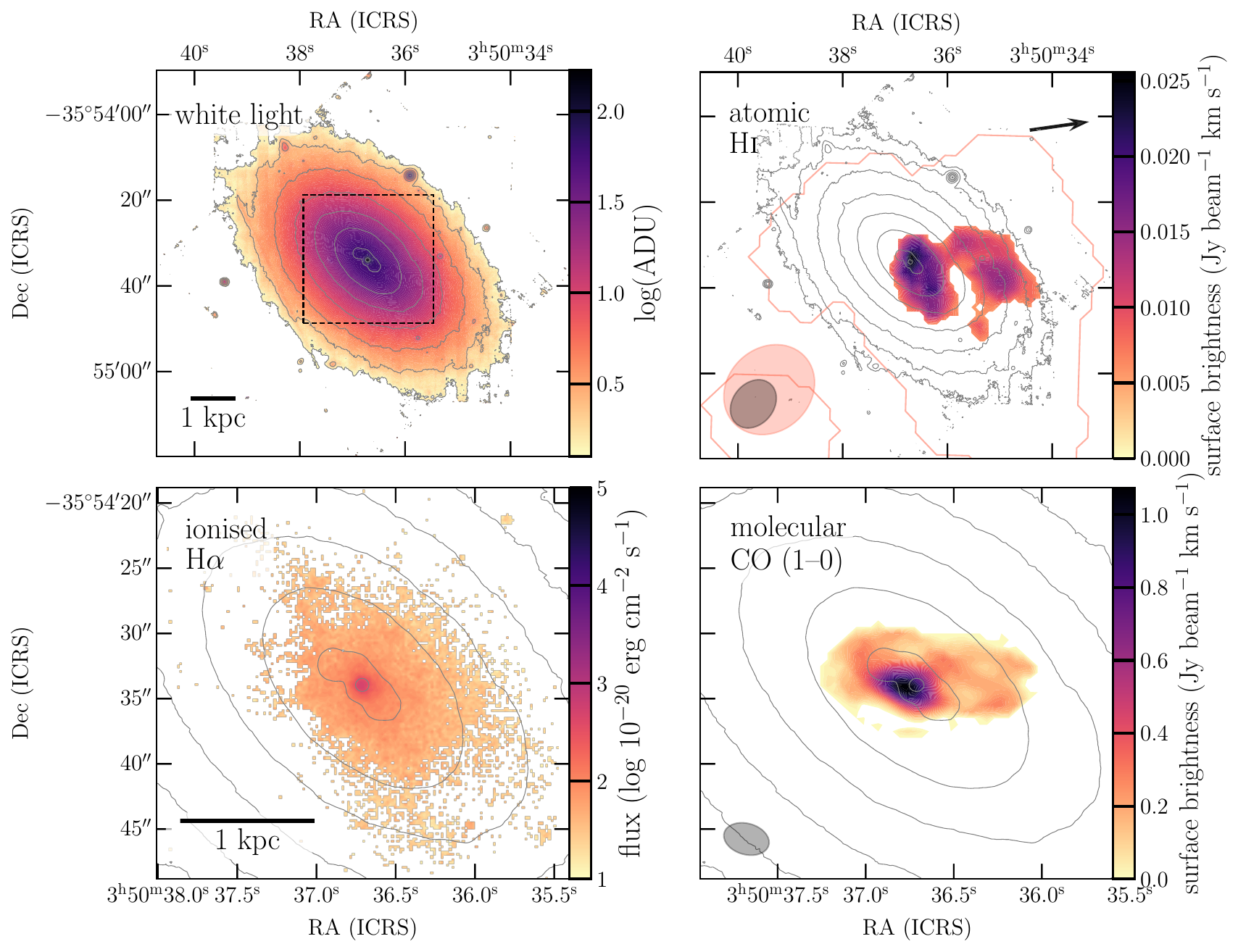}
        \caption{Intensity maps for FCC335.}
        \label{subfig:FCC335_intensity}
    \end{subfigure}
    \begin{subfigure}{0.8\textwidth}
        \centering
        \includegraphics[width=0.915\textwidth]{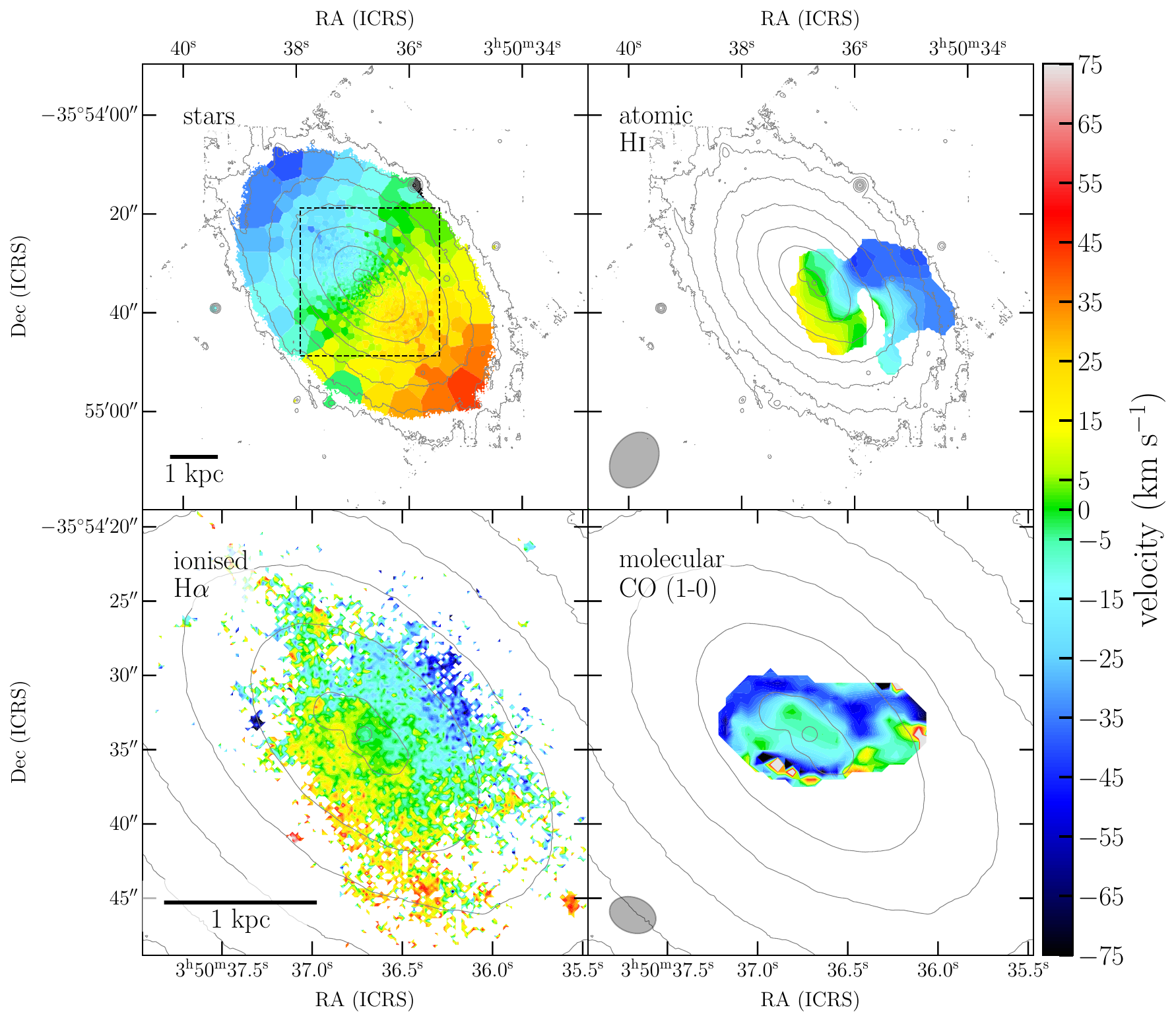}
        \caption{Velocity maps for FCC335.}
        \label{subfig:FCC335_velocity}
    \end{subfigure}
     \caption{Intensity (top panels) and velocity (bottom panels) maps of the stars, \HI, molecular gas, and ionised gas in the Fornax dwarfs. Top-left panels: white light image/stellar kinematics, respectively, top-right panels: \HI\ maps, \dtwo{bottom-left panels, ionised gas maps, and bottom-right panels: molecular gas maps.} The 11$^{\prime\prime}$ \HI\ maps are shown, with the outermost contour of the 22$^{\prime\prime}$ map shown in \dtwo{light-red}. The corresponding beams of the \HI\ and CO observations are shown in the respective bottom-left corners, and a scalebar is shown in the bottom-left corner of the ionised gas maps. \update{The arrow in the top-right corner of the \HI\ intensity panel indicates the projected direction to the cluster centre.} Contours of the MUSE white-light (in log-scale) image are overlaid on each panel for scale. \update{Additionally, the sizes of the bottom panels are indicated in the top-left panels with dashed rectangles.}}
     \label{fig:intensity_maps}
\end{figure*}

\begin{figure*}
	\centering
	\includegraphics[width=0.7\textwidth]{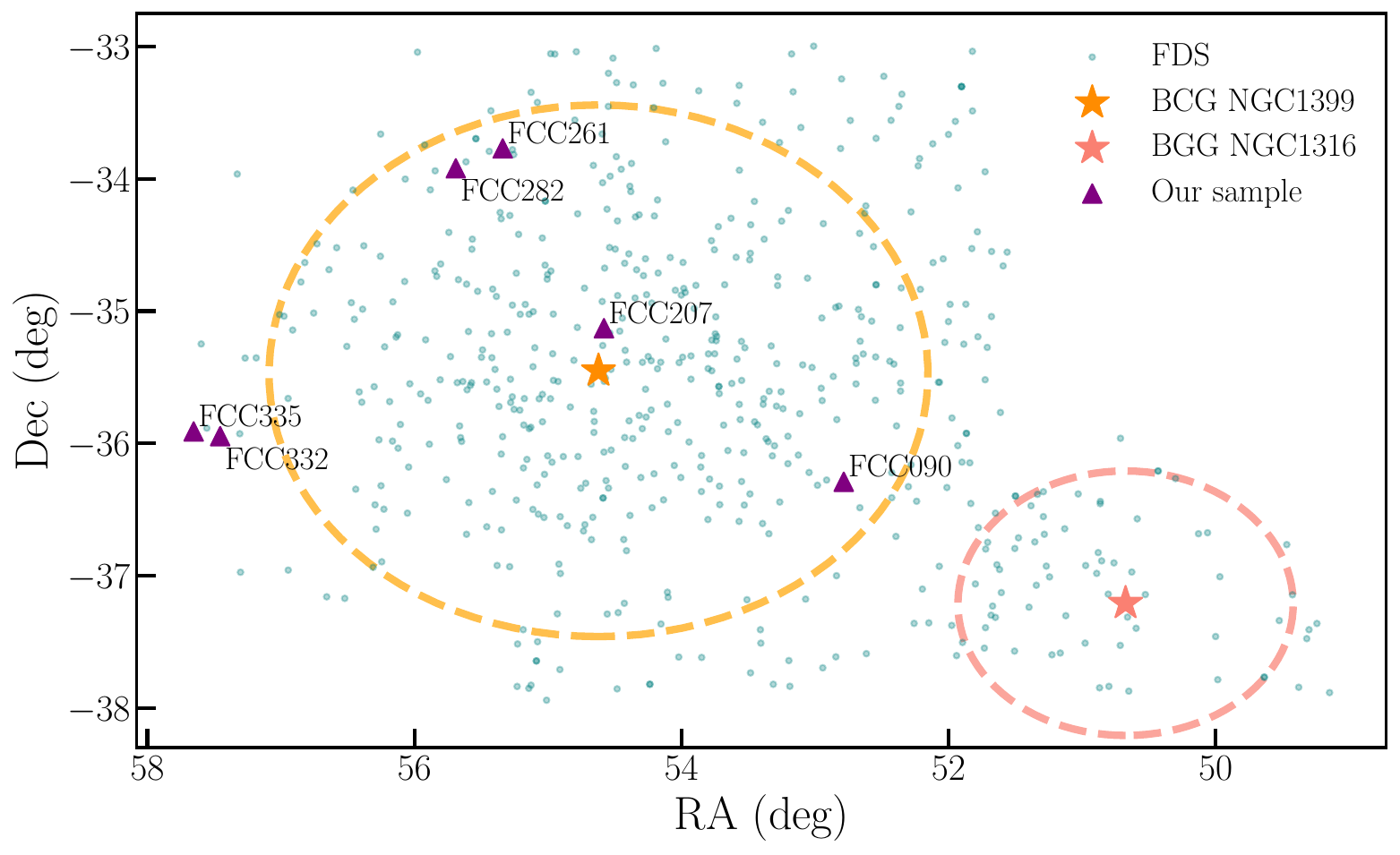}
	\caption{Map of the Fornax cluster. BCG NGC1399 is indicated with a yellow star, and the virial radius with a dashed circle in the same colour. BGG NGC1316 (Fornax A) and the infalling group's virial radius are indicated in a similar fashion. Satellite galaxies are taken from the FDS, and are shown as light blue markers. The dwarf galaxies in our sample are plotted as purple triangles, and have their names annotated. With exception of FCC207 the dwarf galaxies in our sample are located around or just outside the virial radius.}
	\label{fig:map}
\end{figure*}

\subsection{Locations within the cluster}
\dtwo{Figure \ref{fig:map} shows a map of the Fornax cluster, where cluster members, represented by small blue markers, are from the FDS. The brightest cluster galaxy NGC1399 is indicated with a yellow star, and the virial radius with a dashed yellow line. Similarly, the brightest group galaxy NGC1316 (Fornax A) of the infalling group is indicated with an orange star, and its virial radius with an orange dashed line. The six dwarf galaxies studied here are indicated with purple triangles. Most of them are located around the virial radius of the cluster core, \refrep{except} FCC207, which is close to the cluster centre in projection. FCC332 and FCC335 are \refrep{reasonably} close to each other on the sky \refrep{($\sim$70 kpc in projection)}, and have possibly interacted. This is discussed in \ref{subsub:FCC335}.}

\subsection{Description \& discussion of individual galaxies}

\begin{figure}
	\centering
	\includegraphics[width=0.47\textwidth]{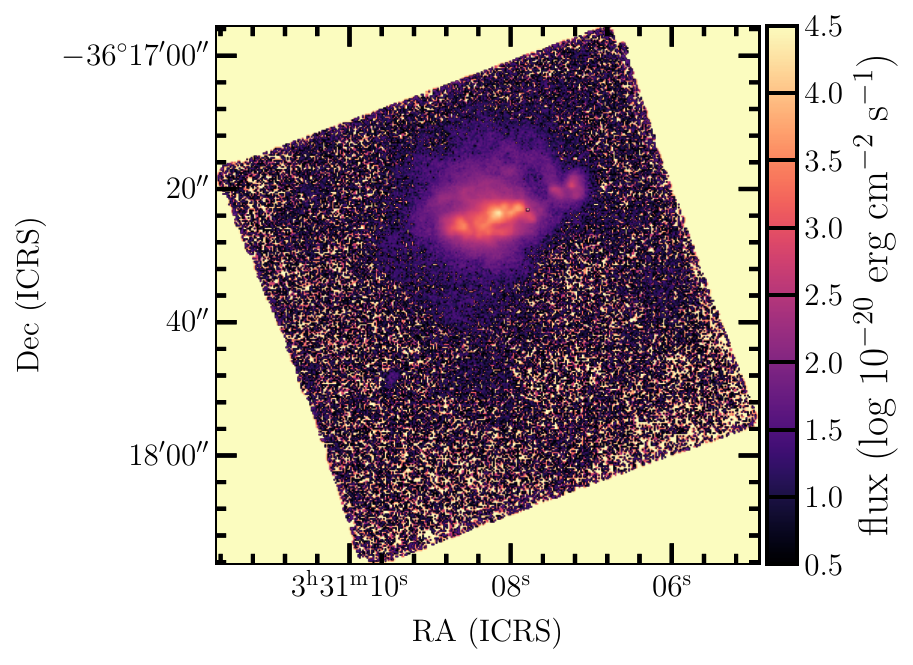}
	\caption{\dtwo{Observed (i.e. not de-reddened) map of the H$\alpha$ flux in FCC090, obtained by trimming and summing the MUSE cube around the H$\alpha$ line as described in \S \ref{subsub:FCC090}. Pixels with values below zero are masked. This map reveals an ionised gas feature towards the south-west of the galaxy that could be the result of a recent merger.}}
	\label{fig:Ha_map_fcc090}
\end{figure}

\label{sub:indiv_gals}
Since the main goal of this work is to provide an in-depth study of a small number of gas-rich dwarf galaxies in the Fornax cluster, we will start this section with a detailed description of the intensity and velocity maps of the various gas phases in each dwarf galaxy, and a discussion of \refrep{each} galaxy in the context of the Fornax environment. \\

\subsubsection{FCC090}
\label{subsub:FCC090}
FCC090 (Figures \ref{subfig:FCC090_intensity} and \ref{subfig:FCC090_velocity}) has an extended molecular gas tail, and disturbed ionised gas features that extend in the same direction, though further out. Its atomic gas is also disturbed and offset from the galaxy centre, towards the same side as the other two gas phases. However, its morphology is notably different, and the offset from the galaxy centre is more aligned with the projected direction towards the cluster centre. Furthermore, the areas with the highest molecular gas surface density are not co-spatial with those with the highest star formation rate surface density, as we might expect in a regular star-forming galaxy (see also \citealt{Zabel2019, Zabel2020}). 

\dtwo{By trimming the continuum-subtracted MUSE cube around the H$\alpha$ line and collapsing it over the spectral axis, more of the H$\alpha$ flux is recovered compared to the method described in \S \ref{subsub:ionised_gas_meas}. While these flux values cannot be used for any quantitative analysis, because the S/N in H$\beta$ is not sufficient to use the dereddening method described in \S \ref{subsub:ionised_gas_meas} (which relies on a measurement of the Balmer decrement), it shows some of the more diffuse ionised gas that is not visible in the maps in Figure \ref{fig:intensity_maps}. The resulting map for FCC090 is shown in Figure \ref{fig:Ha_map_fcc090}. It this reveals a faint ionised gas feature towards the south-west of the galaxy, connecting to the south and west sides of the main ionised gas reservoir. A similar feature is visible towards the north of the galaxy, though most of it is outside the MUSE FoV. This feature could be a trail of ionised gas left behind by a recent merger.}

From the stellar velocity map in the top-left panel of Figure \ref{subfig:FCC090_velocity} we can see that FCC090 is primarily supported by velocity disperson rather than rotation \update{(unless we are viewing the galaxy face-on, \dtwo{which is unlikely given its observed ellipticity})}, as was already shown by \citet{Iodice2019b}. The velocity maps of all three gas phases are irregular, although a hint of rotation can possibly (still) be seen in the centres of the H$\alpha$ \refrep{(}and CO\refrep{)} maps. Interestingly, the atomic and molecular gas are offset by $\sim$ 25~km~s$^{-1}$, with \HI\ being more redshifted. \update{To explore this velocity offset in more detail, we create position-velocity diagrams (PVDs) of the \HI\ and H$_2$ in FCC090 and the other objects in the sample, along the major axis of the \HI\ emission. These are shown in Appendix \ref{app:pvds}. The PVDs highlight that both gas phases are decoupled both spatially and in velocity in FCC090 (Figure \ref{subfig:PVD_FCC090}), suggesting that they are affected differently by the cluster environment, and possibly on different timescales.}

\update{FCC090 is located close to the virial radius in projection (Figure \ref{fig:map}). Since it still has significant amounts of gas and ongoing star formation (see Figures \ref{subfig:hi_h2_defs} and \ref{fig:SFMS}), it is likely on its first infall into the cluster. Since \HI\ is more susceptible to RPS than molecular gas, it could be that RPS is causing a \update{velocity} offset between the two phases. However, FCC090's velocity is higher than that of the BCG (see Figure \ref{fig:trumpet}), which means that it has a velocity component along the \dtwo{line of sight} that is moving away from us \dtwo{relative to} the cluster. Thus, we expect any ongoing RPS to result in a gas tail that is blueshifted with respect to the systemic velocity of the galaxy's stellar body. Since we are seeing the opposite here, it is unlikely that the velocity shift in \HI\ is caused by RPS. Therefore, \refrep{the most likely explanation is that this was} caused by a recent interaction with another galaxy.}

As discussed in \citet{Iodice2019a}, \refrep{who study} the optical data from the FDS \refrep{in depth}, FCC090 exhibits significant ``twisting'' in its stellar disc, i.e. the position angle of the isophotes varies by $\sim$60\textdegree\ between the centre and the outer radii. It also exhibits a tiny optical tail towards the \dtwo{east}, i.e. in the direction of the cluster centre in projection, and facing away from the gas tails \dtwo{(see appendix A.1 and figure C.1 in \citealt{Iodice2019a})}. Its inner regions are quite blue compared to the outer regions: \dtwo{it was identified as one of three blue outliers in a sample of 26 dwarf elliptical galaxies in the Fornax cluster studied by \citet{Hamraz2019}, with an unusually extended blue area. They conclude that FCC090 contains a ``large and extended amount of dust and young stellar populations''.} This aligns with the observation that star formation \dtwo{efficiency} is enhanced in \dtwo{its central regions} \citep{Zabel2020}. The colour structure in its inner region is very irregular (see figure C.1 in \citealt{Iodice2019a}). The authors suggest that FCC090 may have undergone a recent minor merger. This scenario could explain the disturbed and discrepant gas phases, the enhanced star formation in its centre (\S \ref{sub:res_ks} and Figure \ref{subfig:DT_FCC090}), \dtwo{and the ``trail'' of ionised gas visible in Figure \ref{fig:Ha_map_fcc090}}. FCC090 is still at the outskirts of the Fornax cluster, located in between the BCG and Fornax A in projection. Furthermore, \citet{Loni2021} suggest that FCC090 is part of a subgroup around NGC1365 that fell into the cluster simultaneously. Thus, \dtwo{it is also a possibility that FCC090 is influenced by the tidal field of the massive NGC1365 ($M_\star \sim 10^{11}\ \text{M}_\odot$, \citealt{Fuller2014}).} If it has undergone a recent merger, the galaxy it merged with was likely part of either the group around NGC1365, or that around Fornax A. FCC090 could be an example of a galaxy experiencing the effects of pre-processing as it is falling into the Fornax cluster. 

\subsubsection{FCC207}
\label{subsub:FCC207}
FCC207 is quite different from the other dwarf galaxies in the sample: while the white-light image (top left panel) shows some asymmetry, the various gas phases are relatively symmetric, centrally located, and truncated (Figure \ref{subfig:FCC207_intensity}). \dtwo{\citet{Rijcke2003a}, report a distorted nucleus in FCC207, which they find to be more elongated than the bulk of the galaxy, and describe as ``kidney-shaped''. They attribute this unusual shape to dust absorption towards the north of the nucleus, which is visible as a red patch in $B$-$R$ colour maps. They also report a slightly east-west elongated blue object towards the west of the nucleus, which is also visible in their H$\alpha$ narrow-band image. While we do see a feature towards the south-west of the nucleus in the MUSE whitelight image, it is not visible in our H$\alpha$ map. Instead, the H$\alpha$ disc looks somewhat flattened on the west-side of the nucleus.}

 \update{The \HI\ disc is unresolved, implying that its extent is comparable to or smaller than that of the molecular and ionise gas discs.} This means that FCC207 has lost a significant amount of its atomic gas. Indeed, with a \HI\ deficiency of \dtwo{$>$2} -- by far the largest \HI\ deficiency of the galaxies in the sample -- it contains less than 1\% of the amount of atomic gas than a typical field galaxy with the same stellar mass (see Table \ref{tab:props}). The remaining gas appears settled, \dtwo{though the molecular gas shows some asymmetry towards the north-east, and the unresolved \HI\ disc is offset from the galaxy centre towards the north-west.} There is very slow to no rotation in all three gas phases (Figure \ref{subfig:FCC207_velocity}). \dtwo{The \HI\ and CO are marginally offset, by $\sim$5-10 km s$^{-1}$ (as can be seen more clearly in the relative PVDs in Figure \ref{subfig:PVD_FCC207}), which is within the typical \HI\ Doppler broadening}. Like FCC090, FCC207 is a dwarf elliptical, and, \update{unless the galaxy is viewed face-on,} its stellar population is supported by turbulence rather than rotation.

Its truncated but relatively settled gas disc suggests that FCC207 is already in a more advanced stage environmental processing, having lost significant amounts of gas (\HI\ in particular, see also Figure \ref{subfig:hi_h2_defs}). FCC207 is the only dwarf galaxy in the sample that is close to the cluster centre both in projection and in phase space (Figures \ref{fig:map} and \ref{fig:trumpet}). This suggests that it has already spent more time in the cluster than the other objects, and environmental processes have had more time to remove its ISM, quenching the galaxy from the outside in. 

\dtwo{\citet{Rijcke2003a} find that the nucleus of FCC207 is comparatively blue ($B$-$R$ = 0.90 mag vs. 1.25 mag in the bulk of the galaxy), which is in agreement with with the presence of molecular gas and ongoing (albeit inefficient) star formation.} Indeed, the stellar population in FCC207 shows a strong age gradient of 5 Gyr between the centre and twice the effective radius ($2R_e$, \citealt{Koleva2009}). The authors of this work also find that FCC207 is nucleated, and has low central stellar metallicity \dtwo{(Fe/H] = -0.70 $\pm$ 0.05 dex within the central arcsec)}. Its inner regions show a constant star formation history (SFH) until recent epochs. They also harbour a young stellar population, as indicated by the Balmer and [O\textsc{iii}] \update{emission lines}, which are not observed around or beyond $R_e$. This implies that there has been no recent star formation outside the central regions of FCC207. This \dtwo{is} consistent with the observed gas distribution, and a scenario in which it has been quenched from the outside in, after having spent significant time in the cluster. \dtwo{Alternatively, the young, low-metallicity, gas-rich core of FCC207 could be the result of the accretion of fresh \HI\ through a minor merger or tidal interaction during an earlier stage of environmental processing, from which star formation ignited only after the gas settled in the central regions. In this case starvation of the \HI\ reservoir through ongoing star formation would have caused the H$_2$ mass to surpass the \HI\ mass, resulting in the gas composition observed today.}

\subsubsection{FCC261}
In FCC261 all three gas phases show asymmetries towards the north-west. Most notably, the \HI\ is offset from the optical centre, and forms a tail, which is most visible in the 22$^{\prime\prime}$ map. \update{This tail is not aligned with the direction towards the cluster centre, but closer to perpendicular with it. This \refrep{means} that it is unlikely to be the result of RPS.} \dtwo{There is a small offset between the bulk of the molecular and atomic gas, both spatially and in velocity (5 - 10 km s$^{-1}$, see also Figure \ref{subfig:PVD_FCC261}).} While the molecular and ionised gas phases extend towards the same direction, they are not fully co-spatial: the highest-density molecular gas is situated in between the regions with the bulk of the H$\alpha$ flux. The H\textsc{ii} regions towards the south-east of the galaxy centre do not coincide with \update{any of the detected} molecular gas. As a result, FCC261 lies below the Kennicutt-Schmidt relations from the literature, and the bulk of its molecular gas has long depletion times of $>$20 Gyr (see Figures \ref{subfig:ks_FCC261} and \ref{subfig:tdep_FCC261}). \dtwo{As expected for a dwarf elliptical galaxy,} FCC261 is also located well below the SFMS, with an SFR of $\sim$0.004 M$_\odot$ yr$^{-1}$ (Figure \ref{fig:SFMS}).

No rotation is seen in the stellar velocity field of FCC261, \update{which could either indicate a system that is supported by velocity dispersion, or projection effects if we are viewing the galaxy face-on}. \update{The \HI\ shows relatively regular rotation, although the velocity field is quite warped.} The velocity map of the ionised gas shows that the H\textsc{ii} regions towards the north-west are blueshifted compared to those towards the south-east, with a difference of $\sim$ 15 km s$^{-1}$. This is reflected in the velocity map of the molecular gas, showing that here the molecular gas and star forming regions are co-spatial and co-moving, as expected. 

FCC261 is located just inside the virial radius in projection (Figure \ref{fig:map}). While it is already quite \HI\ deficient, it still has \refrep{substantial} amounts of H$_2$ left (Figure \ref{subfig:hi_h2_defs}). Like the other dwarf galaxies, it is most likely on its first infall into the cluster. \update{The gas in this dwarf galaxy is clearly disturbed, but its morphologies are not compatible with RPS. FCC261 is not associated with any substructures or infalling groups, though it has several close neighbours on the sky (Figure \ref{fig:map}). Therefore, the most likely explanation for its gas asymmetries, velocity warps, and suppressed star formation, are tidal interactions with a nearby neighbour.}

\subsubsection{FCC282}
\update{FCC282 is one of the more active galaxies in the sample, located just over 1 $\sigma$ below the SFMS. It has efficient star formation, and is entirely located on or above the KS relations from the literature (Figure \ref{subfig:DT_FCC282}). It is one of two galaxies in the sample that is supported by rotation, and the H$\alpha$ map shows a significant number of \dtwo{star forming} regions in the central \dtwo{areas}. \dtwo{The most prominent} of these regions \dtwo{(those towards the south-west of the galaxy core)} have molecular gas associated with them, though in the locations of some them it remains undetected. This is likely the result of the detection limit of the ALMA observations. The \HI\ map shows an extended feature towards the north-east, and a similar but shorter feature on the opposite side of the galaxy centre. It is also offset from the galaxy centre in the direction towards the cluster centre.}

The \update{stars in} FCC282, which has been identified as an S0, \update{are} rotating fairly regularly, \update{though in the outskirts the redecing velocities appear to be higher than the approaching velocities, potentially indicating some instability in the stellar disc}. The ionised gas largely follows this rotation, \dtwo{though it exhibits a clear irregularity, likely indicating radial motions, or possibly a warp. The \HI\ also shows some irregularity in addition to rotation: its rotation axis is under an angle compared to that of the stars, and there seems to be a radial change in position angle.} Finally, the velocities of the molecular gas are very disturbed, and the molecular gas \update{is blueshifted with respect to} the atomic gas. \update{This is highlighted in Figure \ref{subfig:PVD_FCC282}, where we can see that the molecular gas that spatially coincides with the \HI\ emission is blueshifted by $\sim$30 km s$^{-1}$ with respect to the bulk of the \HI.}

FCC282 is \refrep{located} in the same area of the cluster as FCC261 (i.e. in the north-east, close to the virial radius), but it has a negative rather than positive velocity compared to the BCG (Figures \ref{fig:map} and \ref{fig:trumpet}). Although it has a feature in the opposite direction, the bulk of the atomic gas is shifted in the direction towards the cluster centre. This is incompatible with what we would expect from RPS. As the galaxy approaches the cluster core, the effects of RPS will become stronger as the density of the ICM and the infall velocities increase. If RPS does not play a \dtwo{major} role in displacing the gas in FCC261 yet, it is expected to play a more significant role in the future. FCC261 does not have any near neighbours on the sky. This means that its disturbed features have to be the result of a past tidal interaction with another galaxy, or an ongoing tidal interaction with the cluster potential. \dtwo{However, the latter is unlikely, as FCC282 has a stellar mass similar to NGC1427A (both have $M_\star \sim10^9$ M$_\odot$, see Table \ref{tab:props} and \citealt{Loni2021}), which was estimated to be well outside the tidal radius of Fornax, despite being much closer to the cluster centre in projection \citep{Serra2024}.} 

\subsubsection{FCC332}
\label{subsub:FCC332}
FCC332 has a large \dtwo{\HI\ feature towards} its north-east side. This feature is seen in the cubes with resolutions of $\geq 22^{\prime\prime}$, but resolved out in the \dtwo{11}$^{\prime\prime}$ cube. \update{This is also true for the \HI\ features towards the south-east and west side of the galaxy in the 22$^{\prime\prime}$ map. To obtain a better understanding of these three features, we show the the 22$^{\prime\prime}$ velocity field in Figure \ref{fig:vel_FCC332_22arcsec}.} This map shows that the cloud in the north-east is offset in velocity from the \HI\ in the galaxy disc by $\sim$50 km s$^{-1}$, \update{which is significantly more than the typical \HI\ rotation velocity in these dwarfs, and the velocity dispersion of the \HI\ in the disc of FCC332, which is $<$10 km s$^{-1}$. Since FCC332 is blueshifted with respect to the BCG, a redshifted tail would be consistent with RPS. \dtwo{\refrep{However,} the morphology at $22^{\prime\prime}$ and large velocity difference between the disc and the ``tail'' look somewhat different from ``classic'' examples RPS galaxies (e.g. \citealt{Chung2009,Poggianti2017, Moretti2020, Brown2021})}. 
FCC332 is located in the very outskirts of the cluster and has no close neighbours on the sky (Figure \ref{fig:map}), thus ruling out an ongoing tidal interaction as the explanation for this external cloud. A third possible explanation would be that the \HI\ cloud originates from free-floating gas that is being accreted onto the galaxy, though it is unclear how likely it is for such gas to be available in this region of the cluster. The H$\alpha$ map also shows an extention in the direction of the large \HI\ cloud to the north-east, implying that some of this gas is forming stars. This means that either the gas is forming stars as it is being stripped, or the gas has been accreting onto the galaxy for long enough for star formation to have commenced.}

\update{The other two \HI\ features, the cloud on the west side of the disc and the extension towards the south, both have velocities similar to the systemic velocity of the stars and the bulk of the \HI\ in the disc. This is consistent with a scenario in which these features originate from the galaxy disc. The \HI\ feature towards the south side of the galaxy is also seen in the CO and H$\alpha$ maps, and the morphologies of the three phases suggest that this gas is being stripped from the galaxy. This feature is located at a $\sim$45\textdegree\ angle from the \HI\ cloud described above, which means that it is unlikely for both features to be caused by RPS. The most probable explanation is that a past tidal interaction is responsible for the stripping (and possibly accretion) of the anomalous gas, possibly aided by ongoing RPS.} 

In Figure \ref{fig:bpt} we show maps with the BPT classifications for each pixel in FCC332 (left-hand panel, and FCC335 in the right-hand panel). The orange pixels indicate areas where the ionisation is dominated by star formation. The purple and dark-purple pixels indicate regions where the ionisation is dominated by radiation from an \dtwo{AGN-like sources (Seyfert-like or LINER-like, respectively)}, or other sources of ionisation (such as shocks or old stars). The pink pixels indicate regions where the ionisation is due to a combination of star formation and other sources of ionisation (``composite'' or ``transition'' regions). The majority of the ionised gas in FCC332 is classified as composite. A small area in the north-east is classified as ionised by star formation, and the extension towards the north-west is mostly classified as Seyfert (or other sources of ionisation). There is no known AGN in FCC332, and although it is not impossible for a galaxy of $M_\star \sim 10^{8.6}\ \text{M}_\odot$ to harbour one, the distribution of the ionised gas and its ionisation source classification make it seem more plausible that the gas is ionised by shocks. It is clear that FCC332 is strongly affected by its environment (as discussed above). Furthermore, the region that is classified as ``AGN-like'' corresponds to the gas that is being stripped from the galaxy. The area that is classified as ``SF'' overlaps with where the bulk of the molecular gas is (Figures \ref{subfig:FCC332_intensity} and \ref{subfig:DT_FCC332}), and the remaining areas are classified as ``composite''. Therefore, it is more likely that the regions that are classified as ``AGN-like/LINER-like'' are ionised as a result of shocks from interaction with the ICM, and/or from a past tidal interaction. Similar effects are seen in galaxies that are known to be undergoing RPS, where significant fractions of the ionised gas, in particular the stripped, more diffuse ionised gas at the outskirts, show LINER-like line ratios \citep{Poggianti2019, Tomicic2021}. Alternatively, the gas could be ionised by radiation from the ICM. However, since FCC332 is located outside the virial radius (Figure \ref{fig:map}), it is unlikely that the gas in this object is affected by radiation from the ICM more than the galaxies that are closer to the cluster centre.

The \dtwo{optical shape of FCC332 is} supported by turbulence, and no rotation is seen in any of the gas phases. \update{Moreover, there is a 10 - 20 km~s$^{-1}$ velocity difference between the atomic and the molecular gas, and between the ionised gas and the stars. The former can be seen more clearly in the PVD in Figure \ref{subfig:PVD_FCC332}.} FCC332 also has low star formation activity, although the star formation in the galaxy centre is still relatively efficient (Figures \ref{fig:SFMS}, \ref{subfig:KS_FCC332}, and \ref{subfig:DT_FCC332}). While there are multiple possible explanations for what exactly is happening to FCC332, it is clearly heavily affected by its environment as it is entering the cluster. 

\subsubsection{FCC335}
\label{subsub:FCC335}
FCC335 is the \update{other} galaxy in the sample with a rotating stellar disc. Its \HI\ is displaced from the galaxy centre, and shows features towards the west and south-west, \update{as well as a separate \dtwo{clump} on the south-east side}. \dtwo{The 40$^{\prime\prime}$ \HI\ moment 0 map, shown in Figure \ref{fig:sb_FCC335_40arcsec}, reveals a larger feature pointing east\refrep{,} which is not visible in the higher-resolution maps, thus showing that FCC335 has multiple \HI\ features pointing in different directions.} The ionised gas \update{follows these features (except for the \dtwo{clump})}, and the molecular gas also shows a gas tail towards the west of the galaxy. The gas-phase morphologies, combined with the regular stellar disc, remind of RPS. However, the velocity fields of the three gas phases are \update{unusual}. The \HI\ and H$\alpha$ show a gradient \update{(rotation?)} that is perpendicular to the rotation axis of the stellar disc, \dtwo{and agrees remarkably well between both phases compared to the velocity fields of these phases in the remainder of the sample}. The velocity field of the molecular gas is disturbed, and does not show any rotation or gradient. \dtwo{It is also offset in velocity with respect to the atomic gas (see Figure \ref{subfig:PVD_FCC335}).} It is unclear what \refrep{caused} the velocity axis of the gas to be perpendicular to that of the stars. \update{It is possible FCC335 has undergone an interaction with another galaxy which has disturbed its gas velocities \dtwo{(possibly FCC332, see below)}, and/or it has accreted gas through a minor merger before entering the cluster, \dtwo{as is usually the explanation for this phenomenon in higher-mass galaxies}. FCC335 is located in the very outskirts of the cluster (Figure \ref{fig:map}) close to FCC332 on the sky \update{(70 kpc in projection)}.}

\update{A somewhat similar effect is seen in Fornax galaxy \mbox{ESO358-063}, where the \HI\ tail, which the authors attribute to RPS, exhibits a velocity gradient that does not align with that of the disc, but is more or less perpendicular to it \citep[fig. 10]{Serra2023}. However, in this case, the atomic gas in the disc is still regularly rotating along the major axis, following the stars. Furthermore, the molecular gas in this galaxy is rotating regularly, and shows no signs of a tail \citep[fig. B13]{Zabel2019}. However, ESO358-063 is a more massive galaxy ($M_\star \sim 10^{10}$ M$_\odot$), as are most galaxies which exhibit indisputable signatures of RPS. It is possible that RPS manifests itself differently in dwarf galaxies, where a larger fraction of the gas reservoir is expected to be affected by it, \dtwo{and thus cannot be ruled out as the cause of} the misalignment of gas in these galaxies}. 

In Figure \ref{subfig:FCC335_bpt} we can see that the gas in FCC335 is ionised by a combination of star formation and AGN/other sources in the central regions, and by AGN/other sources only in the outskirts. The ``composite'' area overlaps with where the bulk of the molecular and ionised gas are (Figures \ref{subfig:FCC335_intensity} and \ref{subfig:DT_FCC335}), so most of the ionised gas in FCC335 is ionised by a combination of star formation and other sources. The regions classified as \dtwo{``Seyfert-like''} or \dtwo{``LINER-like''} correspond to the stripped gas in the outskirts. Similarly to FCC332, it seems most likely that this gas is ionised as a result of shocks from interaction with the ICM or with another galaxy.

FCC335 is quite deficient both in atomic and molecular gas. While all the molecular gas in this galaxy has ionised gas emission associated with it, there are steep gradients in its \refrep{measured} star formation efficiency, with depletion times at both ends of the CO ``disc'' exceeding 10 Gyr (see Figure \ref{subfig:DT_FCC335}). Overall, star formation in FCC335 is quite inefficient, with the vast majority of its resolution elements lying below the KS relations from the literature (Figure \ref{subfig:KS_FCC335}). \update{Keeping in mind that the gas is ionised predominantly by other sources than star formation, this means that \refrep{it} is even more inefficient than is indicated in this Figure, in which the star formation rate surface density should be interpreted as an upper limit (and hence depletion times as lower limits). This is another indication that the ISM in this galaxy is affected by its environment, \refrep{limiting} its ability to form stars.} \\

\dtwo{While they are currently $\sim$70 kpc apart in projection, it is possible that FCC332 and FCC335 have interacted. Both galaxies are among the most irregular objects in the sample, showing multiple tails/features in multiple gas phases that are difficult to explain with RPS alone, and some of which face one another in projection. Both objects are located at the outskirts of the cluster, where the effects of the cluster potential are estimated to be negligible, and have no detected nearby neighbours. They contain gas that is ionised almost exclusively as a result of mechanisms other than star formation, as well as unusual dust features. It is not unlikely that, similarly to NGC1427A \citep{Serra2024}, a combination of interactions between these dwarf galaxies and ongoing RPS is shaping their ISM.}

\begin{figure*}
    \begin{subfigure}{0.65\textwidth}
    \centering
        \includegraphics[width=\textwidth]{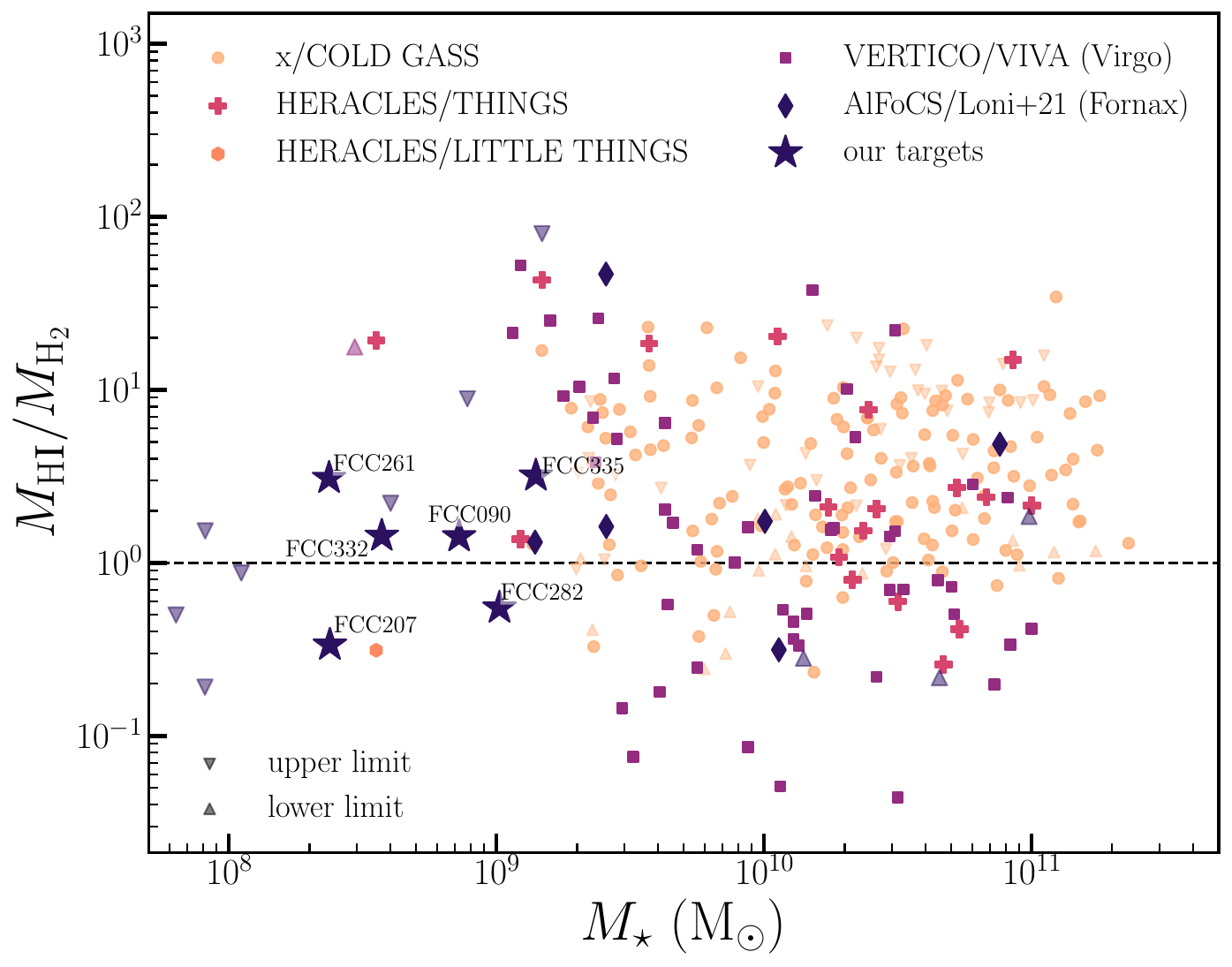} \vspace{-3mm} 
        \caption{Global \HI-to-H$_2$ mass ratios of the sample, compared to field galaxies from x/COLD GASS (yellow markers), HERACLES and LITTLE/THINGS (pink diamond markers and orange marker, respectively), spiral galaxies in the Virgo cluster from VERTICO/VIVA (purple square markers) and other \dtwo{mass} galaxies in the Fornax cluster from AlFoCS/\update{\citet[dark purple, diamond markers]{Loni2021}}. \HI/H$_2$ ratios of the Fornax dwarfs cover a range similar to that in field galaxies and remaining galaxies in the Fornax cluster, while Virgo spirals can have ratios much smaller than 1.}
    \label{subfig:hi_over_h2}
    \end{subfigure}
   
    \begin{subfigure}{0.6\textwidth}
        \centering
        \vspace{-1mm}
        \includegraphics[width=\textwidth]{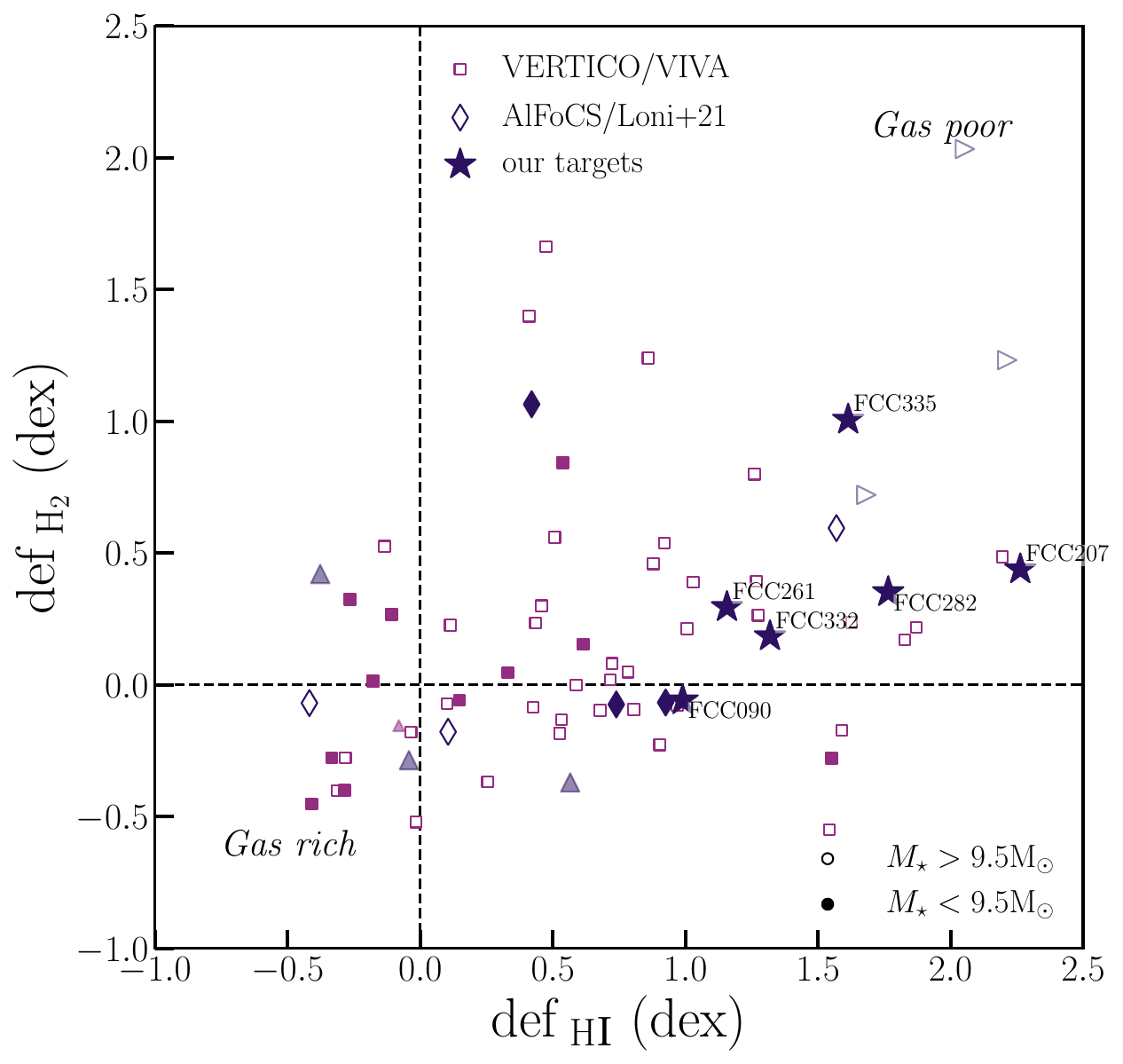}
        \caption{Comparison of \HI\ and H$_2$ deficiencies between the galaxies in the sample (dark-purple stars), the remaining Fornax galaxies (dark-purple diamonds), and spiral galaxies in the Virgo cluster (purple squares). \update{Dwarf galaxies ($M_\star < 9.5 \text{M}_\odot$) are represented by filled markers, whereas higher-mass galaxies ($M_\star > 9.5 \text{M}_\odot$) are indicated with empty markers.} The Fornax dwarfs are among the most \HI- and H$_2$-deficient galaxies.}
    \label{subfig:hi_h2_defs}
    \end{subfigure}
     \caption{Comparison of the relative molecular and atomic gas contents of our sample of dwarf galaxies in relation to field galaxies, spiral galaxies in the Virgo cluster, and remaining galaxies in the Fornax cluster.}
     \label{fig:glob_HI_H2}
\end{figure*}

\begin{figure*}
	\centering
	\includegraphics[width=0.7\textwidth]{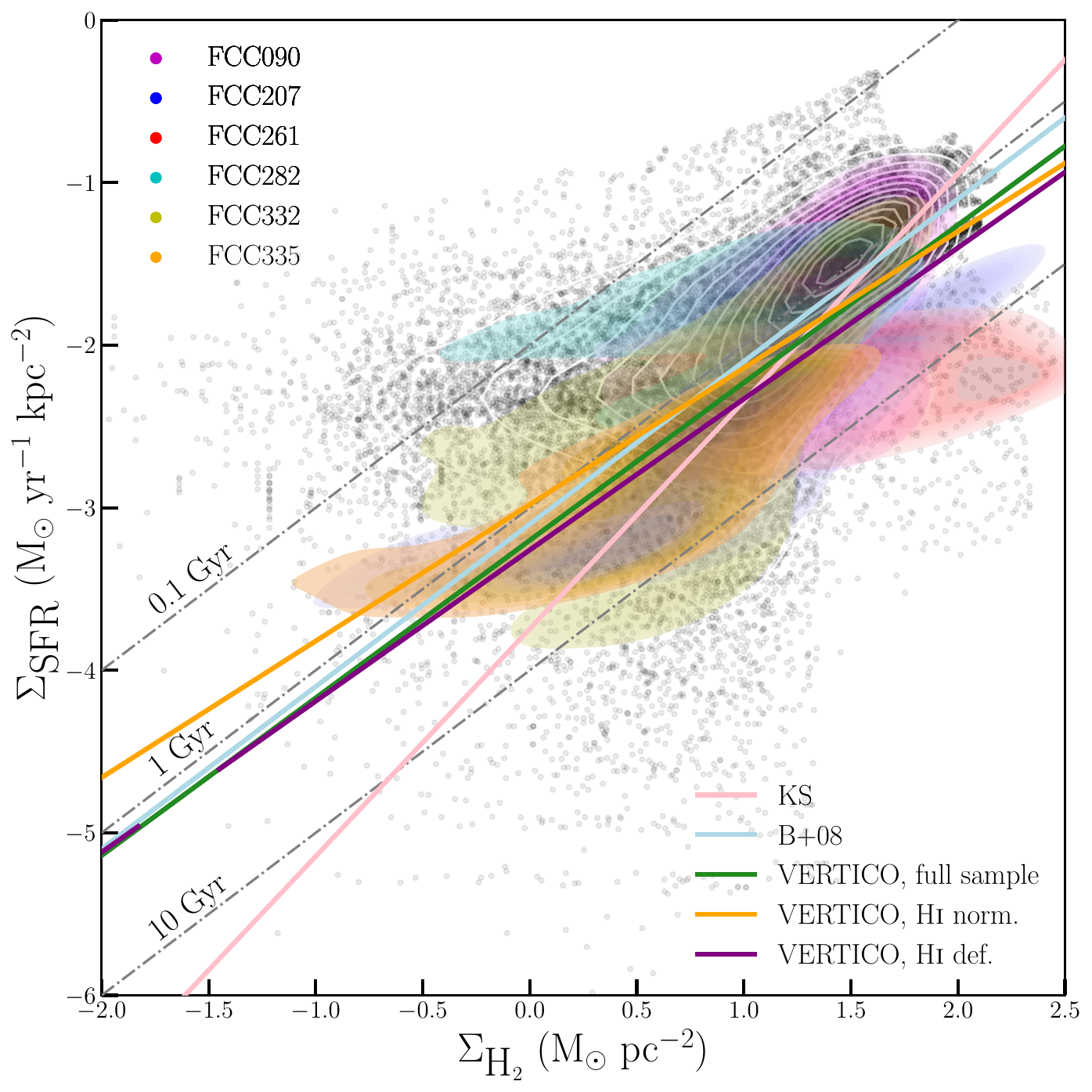}
	\caption{Resolved Kennicutt-Schmidt relation (rKS) of the dwarf galaxies in our sample, shown as coloured KDEs, compared to \dtwo{(r)KS} relations from the literature. The rKS for the remaining, \dtwo{higher-mass} galaxies in the Fornax cluster is represented by small grey markers, with the corresponding KDE overplotted as white contours. Relations from the literature shown are the original (unresolved) Kennicutt-Schmidt relation \citep{Kennicutt1998}, in pink, the relation from \citet{Bigiel2008}, in blue, and various rKS relations of spiral galaxies in the Virgo cluster from \citet{Donaire2023}: the full VERTICO sample in green, \HI-normal VERTICO galaxies in yellow, and \HI-deficient VERTICO galaxies in purple. While FCC090 and FCC282 occupy an area similar to that of the remaining Fornax galaxies \update{(figure adapted from \citealt{Zabel2022})}, the other dwarf galaxies lie below the various relations from the literature, and thus have relatively inefficient star formation. \refrep{Note that the H$\alpha$ in FCC335 is almost entirely due to ionisation sources other than star formation, and its location along the $\Sigma_\text{SFR}$ axis strictly serves as an upper limit. The same is true for FCC332, in which most of the spaxels are classified as ``composite'' (see also Figure \ref{fig:bpt}).}}
	\label{fig:rKS}
\end{figure*}

\subsection{The \HI\ and H$_2$ content of the Fornax dwarfs}
In Figure \ref{fig:glob_HI_H2} we show a comparison of the global \HI\ and H$_2$ contents of our sample of Fornax dwarfs, and compare them with those in field galaxies, galaxies in the Virgo cluster, and remaining galaxies in the Fornax cluster. In Figure \ref{subfig:hi_over_h2} we show global \dtwo{\HI-to-H$_2$ mass fractions} as a function of stellar mass. Our Fornax dwarfs are shown as dark-blue stars, while the remaining Fornax galaxies are represented by dark-blue diamonds. Other comparison samples are x/COLD GASS, shown as dark yellow/orange markers, HERACLES/(LITTLE) THINGS, shown as pink plus-signs (the single LITTLE THINGS detection is indicated with an orange marker), and the VERTICO/VIVA (spiral galaxies in Virgo), \update{sample} is shown as purple squares. For all samples upper and lower limits are represented by semi-transparent downward/upward triangles (respectively) in the same shade as the detections in the corresponding sample. 

There is significant spread in $M_\text{\HI} / M_{\text{H}_2}$ among field galaxies, though only a \update{small} fraction of the x(COLD )GASS sample has fractions of $M_\text{\HI} / M_{\text{H}_2} < 1$ (i.e. the vast majority of galaxies have more \HI\ than H$_2$). This is different in the Virgo cluster, with a significant fraction of the VERTICO/VIVA sample lying around or below the $M_\text{\HI} / M_{\text{H}_2} = 1$ line. This suggests more efficient removal of \HI\ compared to H$_2$ by environmental processes in the Virgo cluster (see also \citealt{Brown2021, Zabel2022}). The dwarf galaxies in our sample are scattered around the $M_\text{\HI} / M_{\text{H}_2} = 1$ line, which means that their H$_2$ mass is similar to their \HI\ mass. \update{While other galaxies in the Fornax cluster span a somewhat larger range of \HI-to-H$_2$ mass ratios, with fractions mostly $>1$, the majority of them are also close to $M_\text{\HI} / M_{\text{H}_2} = 1$ (though statistics are small). This indicates that, similarly to the Virgo galaxies discussed above, in the environmentally-stripped Fornax galaxies \HI\ is removed in significant quantities before H$_2$ (see also \citealt{Loni2021}). \dtwo{Additionally, enhanced conversion from \HI\ to H$_2$ could play a role in areas where the ISM is compressed and the density increased. Such areas with enhanced H$_2$ surface densities have been reported in studies of the effects of RPS on the ISM in cluster galaxies, both in observational studies (e.g. \citealt{Mok2017, Cramer2020, Cramer2021, Roberts2022}) and simulations (e.g. \citealt{Bekki2014, Steinhauser2016, Tonnesen2019}). Other studies of the cold ISM in cluster galaxies (including galaxies undergoing RPS) have found no evidence for enhanced H$_2$ surface densities or \HI-to-H$_2$ conversions (e.g. \citealt{Zabel2022, Watts2023}). Since most dwarfs in the sample are H$_2$ deficient with suppressed SFRs and SFEs, and because RPS is not \refrep{likely to be }the main cause of the observed morphological disturbances, it is unlikely for this to be the main explanation for the low $M_{\text{H}\textsc{i}}/M_{\text{H}_2}$ values. However, in objects such as FCC090, which has a global SFR close to the main sequence and enhanced SFEs in its disc, increased gas densities, for example as a result of the hypothesised past minor merger, could potentially offset some of the molecular gas stripping.}}

In Figure \ref{subfig:hi_h2_defs} we show respective \HI\ and H$_2$ deficiencies (compared to field galaxies with the same stellar mass, see \S \ref{sub:deficiencies}) for the Fornax and Virgo samples. Markers are similar to those in Figure \ref{subfig:hi_over_h2}, \update{while we split the samples into galaxies with \dtwo{log $\left( M_\star / \text{M}_\odot \right) > 9.5$} (open markers) and galaxies with \dtwo{log $\left( M_\star / \text{M}_\odot \right) < 9.5$} (filled markers) to aid comparison between galaxies of similar stellar mass}. In the Virgo cluster there is a weak relation between \HI- and H$_2$-deficiency, where more \HI-deficient galaxies are also increasingly H$_2$ deficient (albeit to a lesser extent, see also \citealt{Zabel2022}). Fornax galaxies appear to follow this trend, with moderately \HI-deficient Fornax galaxies having H$_2$ content comparable to that of field galaxies with similar stellar mass, and highly \HI-deficient galaxies being moderately H$_2$ deficient (though numbers for the Fornax cluster are smaller). Most dwarf galaxies are both \HI\ and H$_2$ deficient. Exceptions are FCC090 and FCC332, which have \HI\ deficiencies of def\textsubscript{\HI} $>$ 1, but \update{H$_2$ masses in line with those expected given their stellar masses}.

\subsection{The resolved Kennicutt-Schmidt relation}
\label{sub:res_ks}
In the previous section we have seen that the majority of Fornax dwarfs are deficient in \HI\ and, to a lesser extent, in H$_2$. While they lie below the SFMS, they are not yet completely quenched (Figure \ref{fig:SFMS}). This gives rise to the question whether these dwarfs still follow \dtwo{star-formation relations $\left( \Sigma_{\text{SFR}} - \Sigma_{\text{gas}} \right)$ from the literature}, and form fewer stars purely because of their lack of cold gas, or if their disturbed gas reservoirs \refrep{also form stars less efficiently}. In Figure \ref{fig:rKS} we show the resolved, \dtwo{molecular} Kennicutt-Schmidt relations \dtwo{($\Sigma_{\text{SFR}} - \Sigma_{\text{H}_2}$, henceforth referred to as ``rKS'')} for the dwarfs in our sample, using coloured kernel density estimate (KDE) plots. The combined rKS relation for the remaining galaxies in the Fornax cluster is shown as grey dots with corresponding KDE overlaid as white contours. Lines of constant depletion times (0.1, 1, and 10 Gyr), are shown as grey dash-dotted lines. We also show various trends from the literature, with the original Kennicutt-Schmidt relation, \dtwo{which relates $\Sigma_{\text{SFR}}~-~\Sigma_{\text{H}_2+\text{H}\textsc{i}}$ \citep{Schmidt1959, Kennicutt1998},} plotted as a pink line, the \dtwo{rKS} relation from \citet{Bigiel2008} as a blue line, and three different \dtwo{rKS} relations for the Virgo cluster from \citet{Donaire2023}: the green line represents the entire (detected) VERTICO sample, consisting of 49 spiral galaxies in the Virgo cluster, the yellow line represents \HI\ ``normal'' VERTICO galaxies, i.e. galaxies apparently unaffected by environment, with a \HI-deficiency $< 0.3$ dex, and the purple line represents \HI-deficient VERTICO galaxies, i.e. galaxies with \HI-deficiencies $> 0.3$ dex. The spread and uncertainties on the slopes of these literature relations are omitted for the sake of readability, but are described as follows. The uncertainty in the slope of the original KS relation is $\pm$0.15, and the spread in $\Sigma_\text{SFR}$ is $\sim$0.1 dex. For the relation from \citet{Bigiel2008} the uncertainty in the slope is $\pm$0.2, and the spread 0.2 dex. The relations for the Virgo cluster have uncertainties in the slope of $\pm$0.07, and typical spreads of 0.42 dex.

FCC090 and FCC282 have quite efficient star formation, and overlap with the various KS relations from the literature, as well as more massive galaxies in the Fornax cluster. The other four dwarf galaxies, on the other hand, have significantly less efficient star formation, and are located well below the various literature relations in the $\Sigma_{\text{H}_2} - \Sigma_\text{SFR}$ plane. \update{Since not all H$\alpha$ detected in FCC332 and FCC335 is ionised as a result of star formation (see \S \ref{subsub:sfr_sd}), their $\Sigma_\text{SFR}$s should be considered upper limits. This means that in reality they would be shifted downwards in Figure \ref{fig:rKS}, meaning that their depletion times are longer than what is indicated here and in Figures \ref{subfig:DT_FCC332} and \ref{subfig:DT_FCC335}, thus moving them further away from the literature relations. Of the literature relations, these four galaxies are closest to that derived for \HI\ deficient galaxies in the Virgo cluster.} \citet{Donaire2023} attribute the suppressed star formation efficiency in these \HI\ deficient galaxies to environmental effects. \update{Both RPS and tidal interactions are expected to decrease the density of the gas while increasing turbulence, resulting in less efficient star formation (though RPS may increase the density of the gas in the centre of the leading side of the galaxy, thus having the opposite effect in these regions, e.g. \citealt{Mok2017})}. The dwarf galaxies in the VERTICO sample (galaxies with $M_\star < 9.5\ \text{M}_\odot$: NGC4064, NGC4294, NGC4299, NGC4351, NGC4383, NGC4532, NGC4561, and NGC4713, see \citealt{Brown2021}) have relatively short depletion times, typically varying from $\sim$0.1 to $\sim$1 Gyr, with most resolution elements in most dwarfs being closer to 0.1 Gyr \citep[figure 5]{Donaire2023}. This is different from \refrep{this work's result, which indicates that} most dwarf galaxies have prolonged depletion times. With one exception, the low-mass VERTICO galaxies are classified as being similar to field galaxies, or as only having started experiencing environmental effects recently, based on their \HI\ content \citep{Yoon2017}. This could mean that the more crowded Fornax environment is more efficient at removing the ISM from dwarf galaxies, or it could be a selection effect, where only \HI-rich dwarf galaxies are observed in the Virgo cluster. \dtwo{The ongoing ViCTORIA (Virgo Cluster multi Telescope Observations in Radio of Interacting galaxies and AGN) project, which is blindly mapping the \HI\ in the Virgo cluster, out to the virial radius, down to ~$\sim 2\times 10^{19}$ cm$^{-2}$ at 3$\sigma$, 30$^{\prime\prime}$ and 25 km s$^{-1}$ resolution (\citealt{Boselli2023}, de Gasperin et al. in prep.), will provide more insight into this.}

There is significant variation in the rKS relation, not only between, but also within galaxies, with depletion times in a single galaxy ranging from several Gyr to over 10 Gyr between the various resolution elements. KS relations and depletion time maps for the individual galaxies are shown in Figure \ref{fig:DT_plots}. The more efficiently star forming FCC090 and FCC282 also have a significant amount of resolution elements with less efficient star formation (see Figures \ref{subfig:KS_FCC090}, \ref{subfig:DT_FCC090}, \ref{subfig:KS_FCC282}, and \ref{subfig:DT_FCC282}). This is in agreement with the results from \citet{Donaire2023}, who also find substantial differences in star formation efficiency between and within galaxies in the Virgo cluster.

\begin{figure*}
    \begin{subfigure}{0.49\textwidth}
    \centering
        \includegraphics[width=\textwidth]{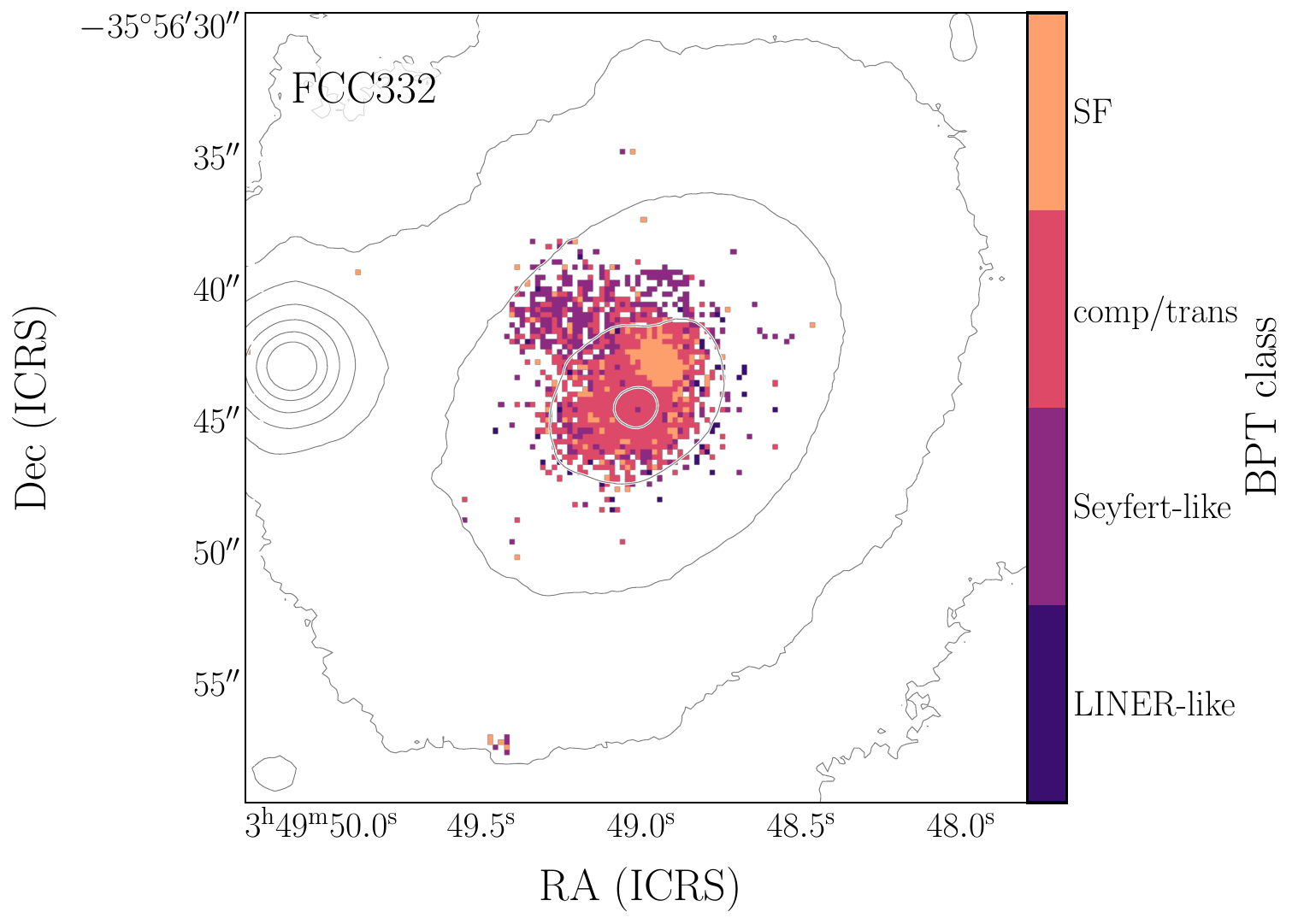}
        \caption{BPT map for FCC332.}
        \label{subfig:FCC332_bpt}
    \end{subfigure}
    \begin{subfigure}{0.49\textwidth}
        \centering
        \includegraphics[width=\textwidth]{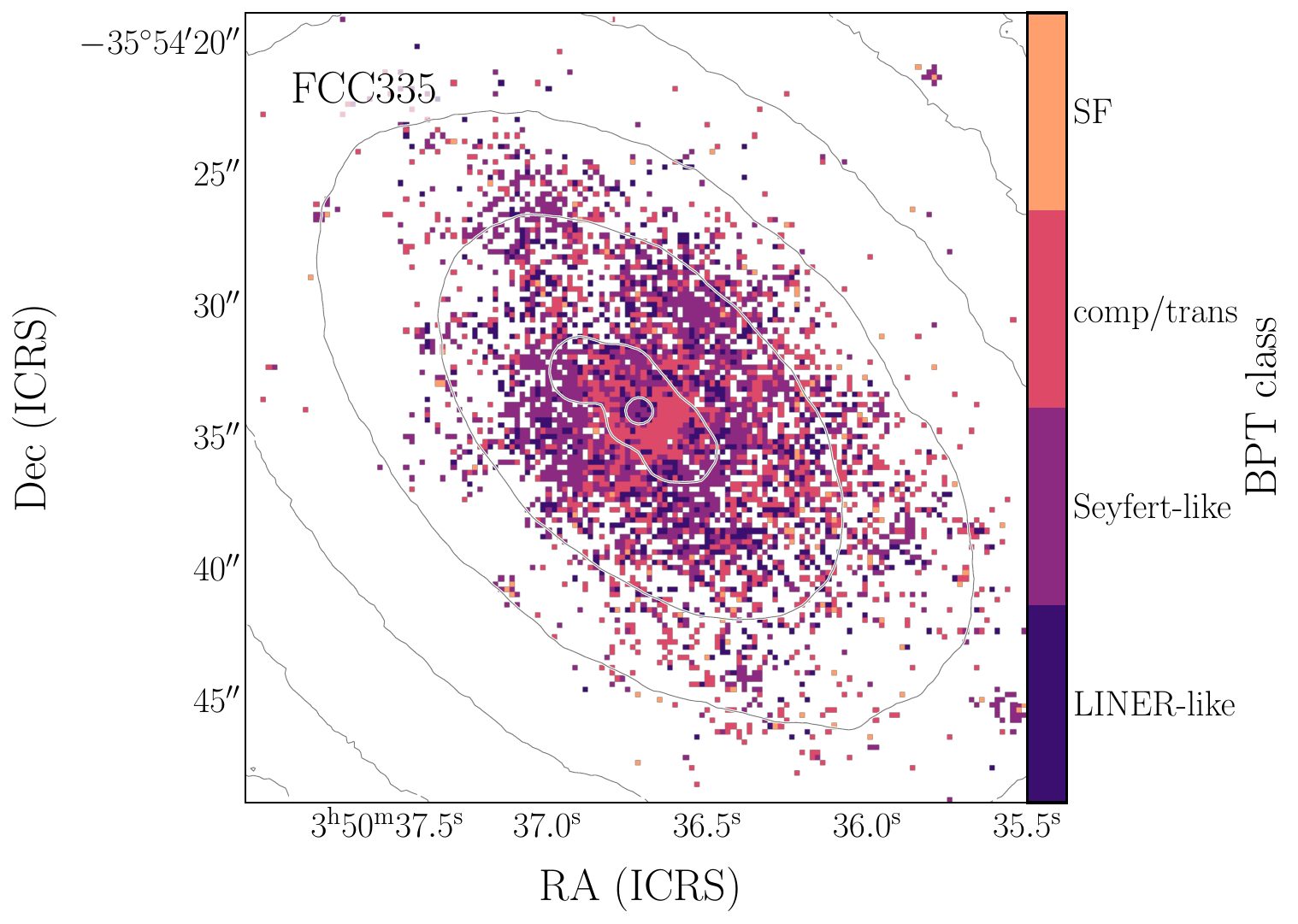}
        \caption{BPT map for FCC335.}
        \label{subfig:FCC335_bpt}
    \end{subfigure}
    \caption{Maps of the BPT classification of the ionised gas for FCC332 (left-hand panel) and FCC335 (right-hand panel), the two galaxies in the sample whose gas ionisation \dtwo{is} due to other reasons than star formation alone. Orange pixels indicate where the ionisation is due to star formation, purple and dark-purple where it is classified as \dtwo{AGN-like} (Seyfert or LINER-like, respectively) or other sources, and pink pixels indicate ``composite'' or ``transition'' regions where the ionisation is caused by a combination of star formation and AGN or other sources. Both sub-figures are shown on the same scale as the bottom panels in Figure \ref{fig:intensity_maps}, and the same contours corresponding to the MUSE white-light images are overlaid. While in the traditional BPT diagram the corresponding line ratios are classified as ``\dtwo{Seyfert-like}'' or ``\dtwo{LINER-like}'', it is more likely that this gas is ionised as a result of shocks from interaction with the ICM (RPS) or tidal interactions.}
    \label{fig:bpt}
\end{figure*}

\section{Discussion}
\label{sec:discussion}
All six \refrep{dwarf galaxies} in the sample have disturbed morphologies and kinematics in the three gas phases studied here (atomic, molecular, and ionised, Figure \ref{fig:intensity_maps}). While these disturbances look quite different between galaxies \refrep{both} in terms of morphology and velocity field, there are \refrep{several aspects} they have in common. For example, the \HI\ ``disc'' is often displaced from the galaxy centre, and the molecular gas often has the most irregular velocity field. While the three gas phases usually have features extending in the same direction, their peak surface density regions are co-spatial in only one of the dwarf galaxies (FCC207). This results in long depletion times for the molecular gas, with significant variation within galaxies. In addition to spatial offsets, the majority of the dwarfs in the sample show offsets in velocity between \dtwo{all three gas phases (most notably in FCC090 and FCC332)}. These spatial and velocity offsets suggest that both gas phases are displaced on different timescales, with the more loosely bound and extended \HI\ being displaced more efficiently than the more tightly bound and centrally located H$_2$. \dtwo{Alternatively, they are decoupled and affected differently \refrep{(but simultaneously)} by the ongoing environmental processes.} Most of the dwarf galaxies in the sample are ellipticals whose stellar velocities are supported by turbulence rather than rotation, making it difficult to assess whether their stellar velocities are also perturbed. Exceptions are FCC282 and FCC335, both of which have regularly rotating stellar discs, without any clear evidence for disturbances. RPS is a hydrodynamical mechanism, which disturbes the various components of the ISM, while leaving the stellar disc intact. RPS could explain some of the one-sided features observed in the sample. 

\dtwo{\cite{Kleiner2023} provide \HI\ moment 1 maps of the 17 dwarf galaxies located in the central $\sim 2.5 \times 4\ \text{deg}^2$ in which \HI\ was detected with MeerKAT (their appendix A). The majority of these \HI-detected dwarf galaxies are late-types with regularly rotating \HI\ discs ($\sim$ 11/17), while some have disturbed \HI\ reservoirs with no clear rotation ($\sim$ 6/17, including FCC090 and FCC207, which are also studied in this work). This shows that the six CO-detected dwarf galaxies studied here are part of a sub-population that has particularly disturbed ISM.} Since all galaxies except FCC207 are located around the virial radius or outside of it, and have detectable ISM and ongoing star formation (albeit at lower rates than isolated galaxies with comparable stellar masses), we suspect that these dwarf galaxies are on their first infall into the Fornax cluster. \dtwo{As such, they still have detectable molecular gas, while they are also already actively being processed by the cluster environment, having started their transition from active to passive. The disturbed features in most of these objects are not consistent with RPS (alone), and some galaxies show evidence of ongoing or recent interactions. This is consistent with the picture that dwarf galaxies interact as they fall into the Fornax cluster, as was also found in a detailed study of NGC1427A (log $\left( M_\star / \text{M}_\odot \right) \sim 9.3$, \citealt{Serra2024}).} FCC207 stands out from the rest of the sample in several ways, which could indicate that it has already spent more time in the cluster. In summary, it is the only galaxy in the sample with a relatively relaxed but truncated ISM (see \ref{subsub:FCC207}). In Figure \ref{fig:glob_HI_H2}, we can see that FCC207 is the most \HI\ deficient galaxy in the sample (in agreement with its truncated, unresolved \HI\ disc), and has more H$_2$ than \HI.

\HI\ deficiencies in the dwarfs in the sample are quite substantial, with the lowest value being $\sim$1 dex, and the highest $\sim$2 dex. While most dwarfs are somewhat H$_2$ deficient (FCC090 and FCC332 are exceptions), these H$_2$ deficiencies are less significant. In this aspect the dwarf galaxies in Fornax do not stand out from their higher-mass counterparts or from the majority of galaxies in the Virgo cluster, suggesting that the gas removal process occurs in a similar way (i.e. the removal of significant amounts of \HI\ before the molecular gas reservoirs are reduced by comparable amounts. \dtwo{\citet{Kleiner2023} found that the majority of \HI-detected late-type dwarf galaxies (LTDs) follow the luminosity-\HI\ relation (see their figure 5), while the majority of dwarf galaxies in their sample has no detectable \HI. The three \HI-detected early-type dwarf galaxies (ETDs), on the other hand, lie $>3 \sigma$ below this relation. The \refrep{vast} majority of galaxies in our sample are ETDs, so the \HI\ deficiencies found here are in agreement with this result. Indeed, two galaxies that are included in both samples, FCC090 (classified as an LTD by \citealt{Kleiner2023}), and FCC207, have \HI\ deficiencies that agree with their position in the $M_{r^\prime} - M_\textsc{Hi}$ plane. However, the CO-detected, rotation-supported FCC282 and FCC335 also have substantial \HI-deficiencies, which means their \HI\ content is closer to that of the ETDs than the LTDs described in \citet{Kleiner2023}. This could imply that they represent a rare population of dwarf galaxies that is rapidly losing their \HI\ to the cluster environment as they transition from LTDs to ETGs.}

Some values of M$_{\text{H}_2}$ (and thus def$_{\text{H}_2}$) differ from those published in earlier work \citep{Zabel2019, Zabel2021}. This is because here we use the H$\alpha$ weighted global metallicity to calculcate X\textsubscript{CO} \citep{Accurso2017}, whereas in previous work we used the mass-metallicity relation to estimate it, as direct metallicity measurements were not available for the majority of the sample. While most changes are minor, the exception is FCC090, whose X\textsubscript{CO} (and thus M$_{\text{H}_2}$) was estimated to be $\sim 4 \times$ higher than previously derived. As a result, it is no longer considered H$_2$ deficient, and it has shifted much closer to the Kennicutt-Schmidt relation compared to \citet{Zabel2020}. The difference between both X\textsubscript{CO} estimates suggests that FCC090 lies below the mass-metallicity relation, resulting in an underestimation of its X\textsubscript{CO} in past work.

\section{Summary}
\label{sec:summary}
\update{In this work we have compared the molecular, atomic, and ionised gas in six dwarf galaxies in the Fornax cluster in detail, using recently obtained MUSE (four objects) and MeerKAT data, in combination with archival ALMA and MUSE (two objects) data. We have studied resolved maps of the distributions and velocities of the three gas phases, as well as the resolved Kennicutt-Schmidt relation, and integrated measurements of their \HI\ and H$_2$ masses and deficiencies. We have analysed each galaxy in detail, taking into consideration additional information such as their location relative to the SFMS, location in the cluster, and location in phase-space, for a comprehensive picture of their place within the cluster. The main conclusions from this work are as follows:}
\begin{itemize}
\item \update{All six dwarfs have disturbed ISM, with all three gas phases having irregular morphologies and velocity fields. While all three typically show features in the same general direction, they tend to be decoupled both spatially and in velocity. This means that different phases of the cold ISM are affected differently by environmental processes. The exception is FCC207, whose ISM is more settled and more truncated than in the other objects.}
\item While some of the observed features remind of RPS, all galaxies show \refrep{additional} characteristics that are inconsistent with it, such as very disturbed velocity fields or tails that are inconsistent with the direction of the RPS wind. While RPS may contribute to some of the gas removal, we conclude that tidal interactions and minor mergers likely play a role in the environmental processing of these galaxies. \dtwo{This is consistent with the picture that dwarf galaxies interact as they fall into the Fornax cluster \refrep{(or even before, as a result of ``pre-processing'')}, after which they may be \refrep{shaped} further by RPS.}
\item The dwarfs are quite deficient in \HI\ (1 $\lesssim$ def$_{\text{\HI}}$ $\lesssim$ 2 dex) and moderately deficient to normal in H$_2$ (0 $\lesssim$ def$_{\text{H}_2}$ $\lesssim$ 1 dex), suggesting that \HI\ is removed in significant quantities before H$_2$. This is consistent with what is observed in higher-mass Fornax galaxies and spiral galaxies in the Virgo cluster.
\item Most objects in the sample \update{(4/6)} have suppressed star formation and increased molecular gas depletion times, varying from several up to tens of Gyr. Thus, most \dtwo{of these} dwarfs have started to lose molecular gas, and, \update{additionally,} the remaining gas is losing its ability to form stars, likely because of turbulence and/or diffuseness as a result of environmental mechanisms.
\item \update{Exceptions to this are FCC090 and FCC282, which are closer to the SFMS and have rKS relations that are more in agreement with those from the literature. Possibly their SFE has not yet been affected by the ongoing environmental processes, or, alternatively, they could be (temporarily) enhancing the SFE in parts of the galaxy, offsetting the suppression of it in other parts.}
\end{itemize}

\update{The still substantial but highly disturbed ISM in the gas-rich dwarf galaxies studied in this work, along with their suppressed star formation rates and, in most cases, SFEs, suggest that these objects are on their first infall into the Fornax cluster, and are in the process of being transformed from actively star-forming to passive and quenched. This shows that dwarf galaxies are very susceptible to environmental processes, even in the outskirts of the relatively poor Fornax cluster. \refrep{It is quite possible that several of them have experienced pre-processing.} The nature of the disturbances, in particular the spatial and velocity offsets between the three gas phases, and their disturbed velocity fields, are inconsistent with RPS alone. Therefore, we suggest that tidal interactions must play a role in transforming these galaxies, possibly aided by RPS.}

\section{Data availability}
\dtwo{The MUSE observations used in this work are available from the ESO archive (programme IDs 0104.A-0734, 296.B-5054, 098.B-0239, 094.B-0576, 097.B-0761, and 096.B-0063). The ALMA observations are available from the ALMA archive (project ID 2015.1.00497.S), and the MeerKAT data are available at \url{https://sites.google.com/inaf.it/meerkatfornaxsurvey/data}.}

\section*{Acknowledgements}
NZ is supported through the South African Research Chairs Initiative of the Department of Science and Technology and National Research Foundation.

The MeerKAT Fornax Survey has received funding from the European Research Council (ERC) under the European Union’s Horizon 2020 research and innovation programme (grant agreement no. 679627; project name FORNAX).

\dtwo{This research has made use of the NASA/IPAC Extragalactic Database (NED), which is operated by the Jet Propulsion Laboratory, California Institute of Technology, under contract with the National Aeronautics and Space Administration.}

\dtwo{This work made use of Astropy:\footnote{http://www.astropy.org} a community-developed core Python package and an ecosystem of tools and resources for astronomy \citep{Astropy2013, Astropy2018, Astropy2022}.}

\dtwo{This research made use of APLpy, an open-source plotting package for Python \citep{Robitaille2012}.}

\dtwo{This work made use of SAOImage DS9 \citep{Joye2003}, TOPCAT \citep{Taylor2005}, Matplotlib \citep{Hunter2007}, Numpy \citep{Harris2020}, SciPy \citep{2020SciPy}, lifelines \citep{Davidson-Pilon2019}, and reproject: Python-based astronomical image reprojection \citep{Reproject2020}.}




\bibliographystyle{mnras}
\bibliography{References} 



\clearpage

\appendix

\section{Additional figures}
\label{app:add_figs}
In this section we include additional figures, providing context as to the status of the dwarfs in our sample within the Fornax cluster. We show their positions on the SFMS (Figure \ref{fig:SFMS}) and where they are located in the Fornax cluster, both in projection (Figure \ref{fig:map}) and in phase space (Figure \ref{fig:trumpet}). 

\subsection{The SFMS}
\begin{figure*}
	\centering
	\includegraphics[width=0.7\textwidth]{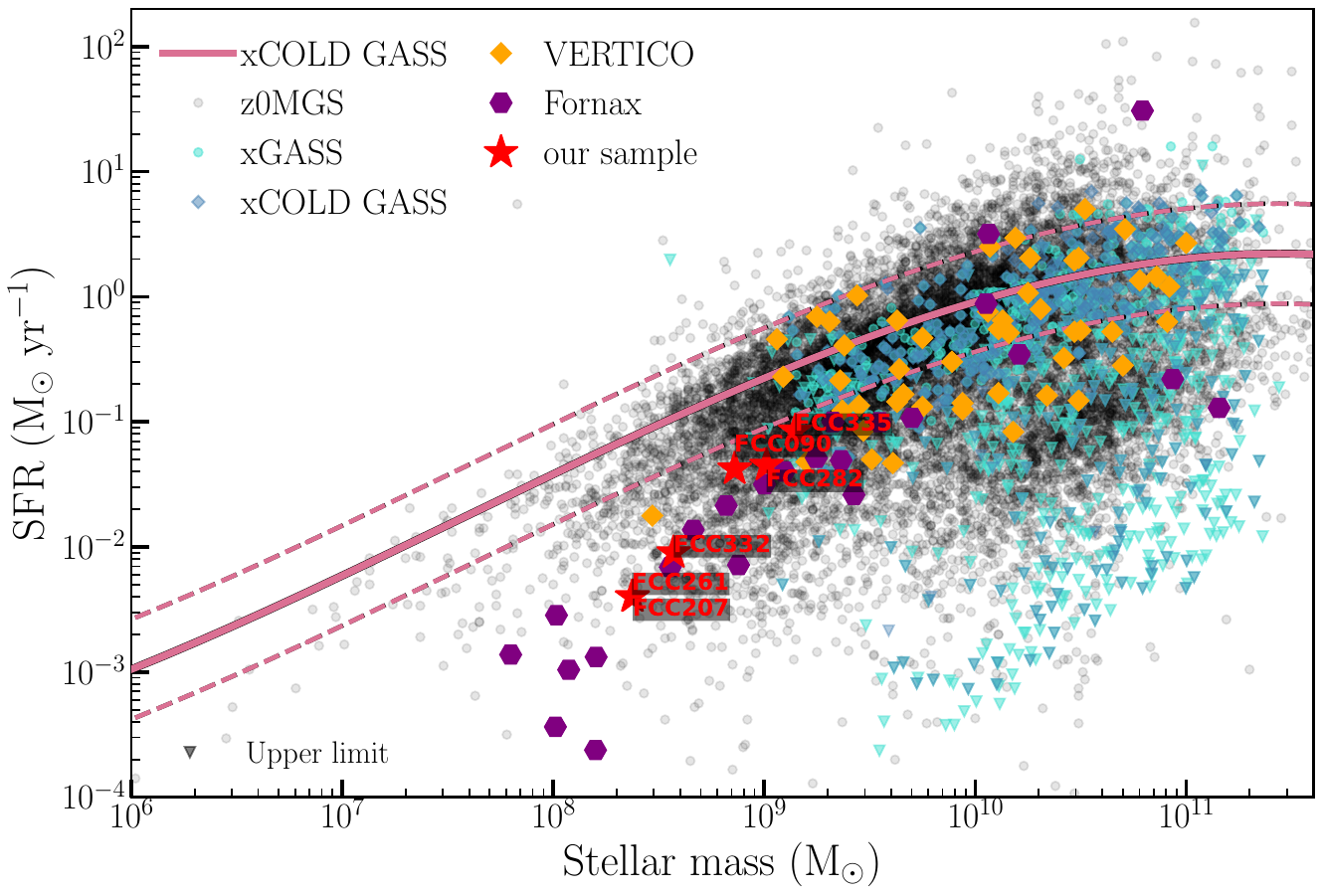}
	\caption{Positions of our targets, shown as red stars, with respect to the SFMS from xCOLD GASS \citep{Saintonge2017}, represented by the pink line (dashed lines show the 1$\sigma$ spread in the relation). For comparison, data from the z0MGS (small grey markers, \citealt{Leroy2019}), xGASS (small light-blue markers, \citealt{Catinella2018}), VERTICO (yellow diamonds, \citealt{Brown2021}), and \dtwo{Fornax galaxies from \citet[][purple hexagonals]{Loni2021}} are included. The dwarf galaxies in our sample lie systematically below the SFMS, \dtwo{and follow the trend of the Fornax cluster}.}
	\label{fig:SFMS}
\end{figure*}

Figure \ref{fig:SFMS} shows the positions of our dwarf galaxies (represented by red stars and annotated) on the SFMS from xCOLD GASS \citep{Saintonge2017}. Individual galaxies from the xCOLD GASS sample are shown as small dark blue diamonds. For comparison, we also show the positions of additional field galaxies from the z0MGS (small grey markers, \citealt{Leroy2019} and xGASS (small blue markers, \citealt{Catinella2018}), as well as spiral galaxies in the Virgo cluster from VERTICO (\citealt{Brown2021}, represented by yellow diamons) and other galaxies in the Fornax cluster shown as purple hexagonals. The dwarfs in our sample are positioned systematically below the SFMS from xCOLD GASS, although most are close to the lower 1$\sigma$ confidence interval. Thus, while our targets have suppressed SFRs, most still have significant ongoing star formation. FCC261 and FCC332 are furthest removed from the SFMS.

\subsection{Phase space diagram}
\begin{figure*}
	\centering
	\includegraphics[width=0.6\textwidth]{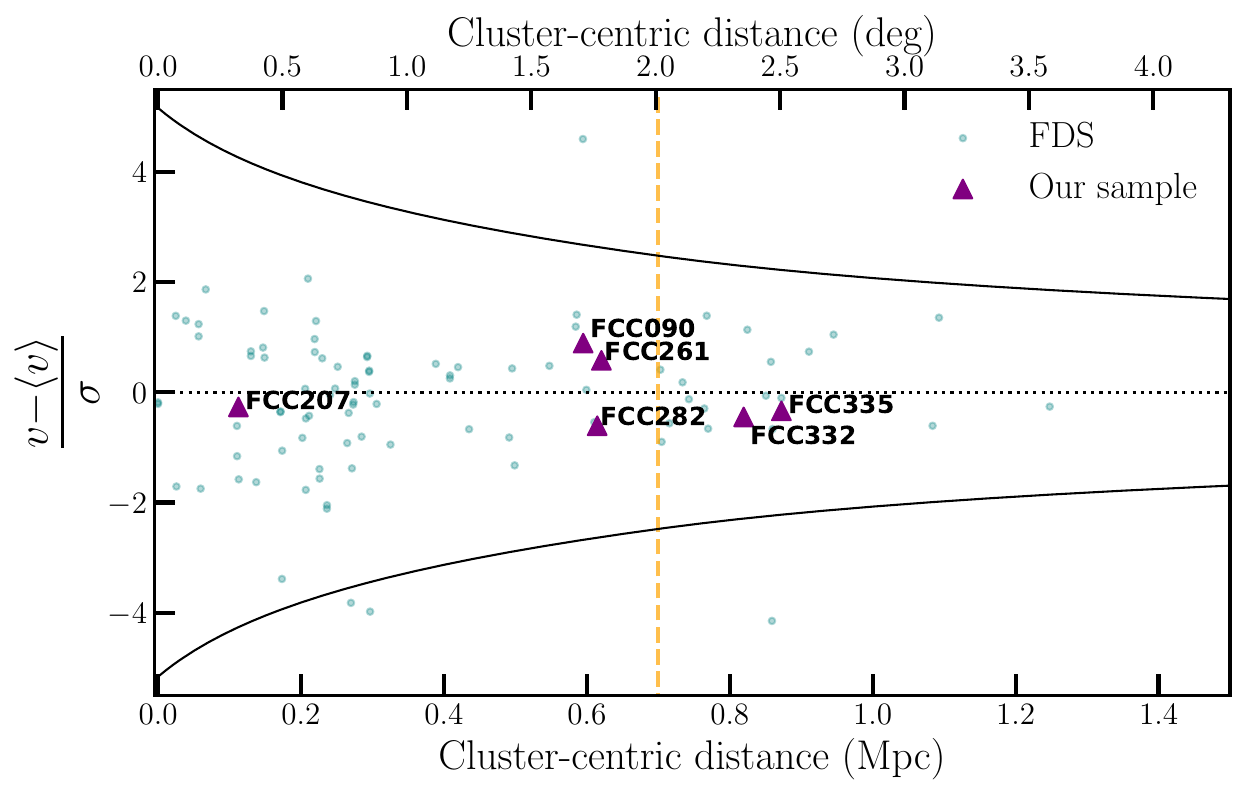}
	\caption{Phase space diagram of the Fornax cluster. \dtwo{The solid black lines represent the escape velocities of the cluster. Remaining} markers and lines are the same as in Figure \ref{fig:map}. $\left\langle v \right\rangle = 1493$ km s$^{-1}$ and $\sigma$ = 374 km s$^{-1}$ \citep{Drinkwater2001a}. The dwarf galaxies in our sample are close to $\left\langle v \right\rangle$.}
	\label{fig:trumpet}
\end{figure*}

Figure \ref{fig:trumpet} shows the locations of the dwarf galaxies in phase space. Lines and markers are the same as in Figure \ref{fig:map}. Velocity information comes from the Two Micron All-Sky Survey (2MASS) Redshift Survey (2MRS, \citealt{Huchra2012}) and the Two-degree-Field (2dF) Galaxy Redshift Survey (2dFGRS, \citealt{Colless2001}). FCC207's position and velocity are close to that of the BCG. The other dwarfs in the sample are scattered around the virial radius and also have velocities relatively close to that of the BCG. 

\begin{figure*}
	\centering
	\includegraphics[width=0.65\textwidth]{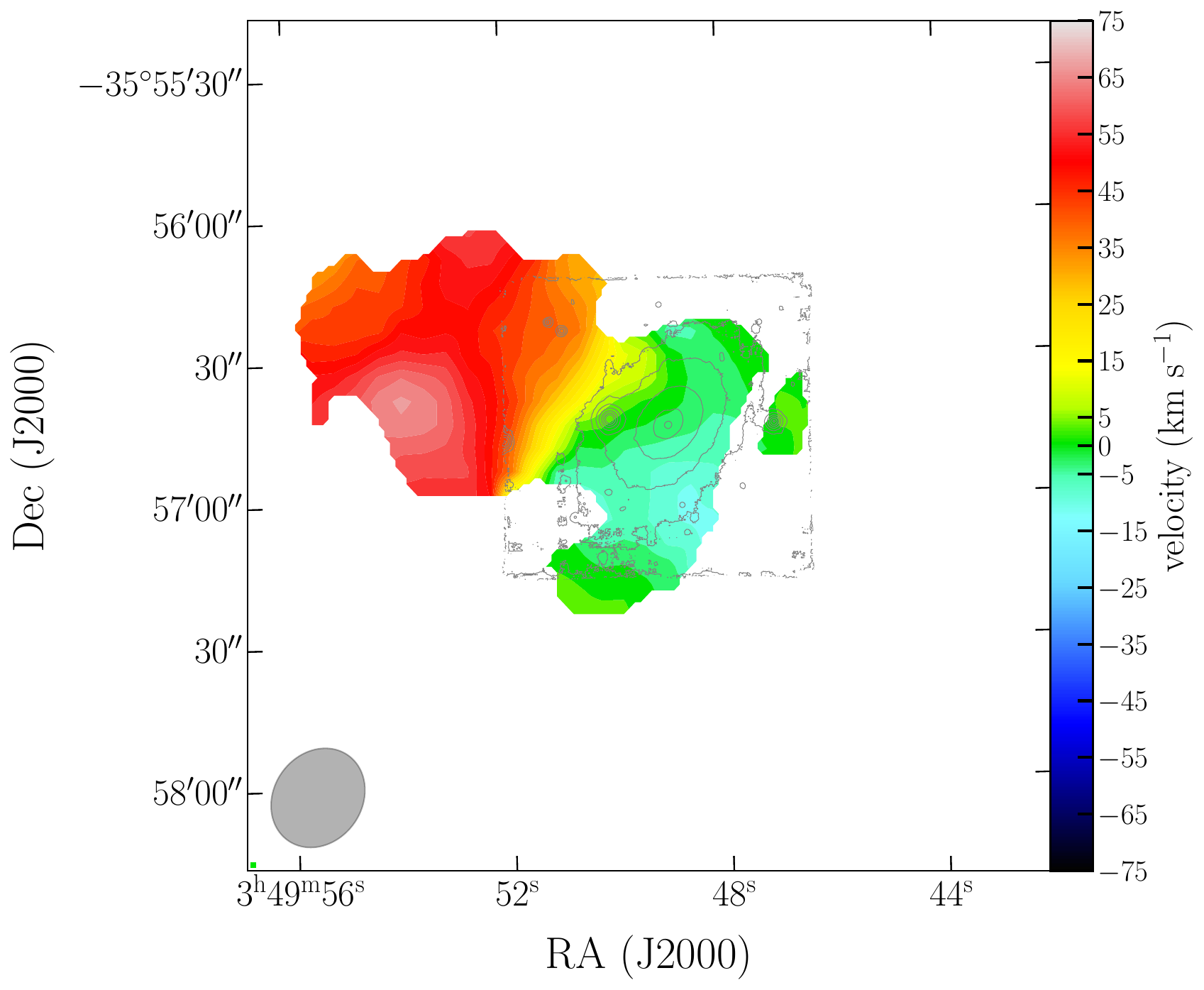}
	\caption{Moment 1 map of the \HI\ in FCC332 at 22$^{\prime\prime}$ resolution. The systemic velocity and contours are identical to those in Figure \ref{subfig:FCC332_velocity}. The large \HI\ cloud towards the north-east of the galaxy is offset in velocity from the galaxy disc by $\sim$ 50 km s$^{-1}$.}
	\label{fig:vel_FCC332_22arcsec}
\end{figure*}

\begin{figure*}
	\centering
	\includegraphics[width=0.65\textwidth]{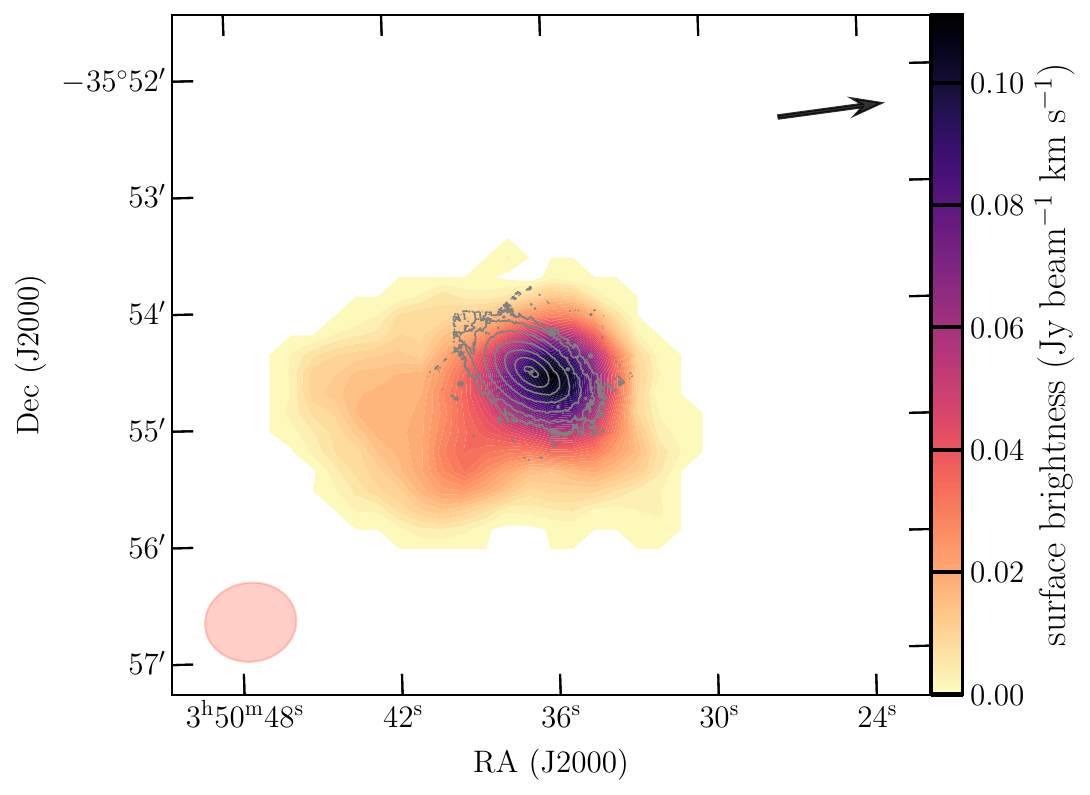}
	\caption{\dtwo{Moment 0 map of the \HI\ in FCC335 at 40$^{\prime\prime}$ resolution. The contours and arrow in the (projected) direction of the cluster centre are identical to those in Figure \ref{subfig:FCC335_intensity}, and the 40$^{\prime\prime}$ beam is shown in the lower-left corner. This lower-resolution image reveals a tail in the (south-)east direction that is not visible in the higher-resolution (11$^{\prime\prime}$ and 22$^{\prime\prime}$) images, showing that FCC335 has several \HI\ features in multiple directions.}}
	\label{fig:sb_FCC335_40arcsec}
\end{figure*}


\subsection{Multi-wavelength PVD analysis}
\label{app:pvds}
\begin{figure*}
    \begin{subfigure}{0.36\textwidth}
    \centering
        \includegraphics[width=\textwidth]{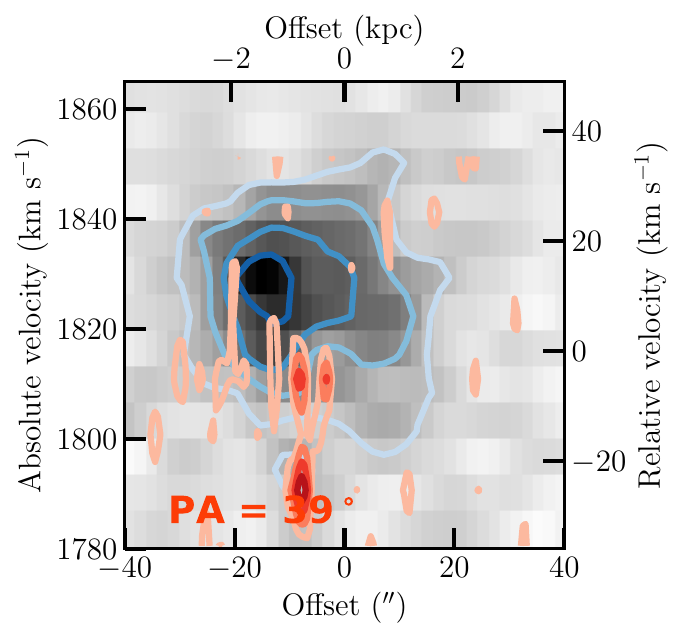}
        \caption{\HI\ (greyscale, blue) and CO (red) PVDs for FCC090.}
    \label{subfig:PVD_FCC090}
    \end{subfigure} \hspace{10mm}
    \begin{subfigure}{0.36\textwidth}
        \centering
        \includegraphics[width=\textwidth]{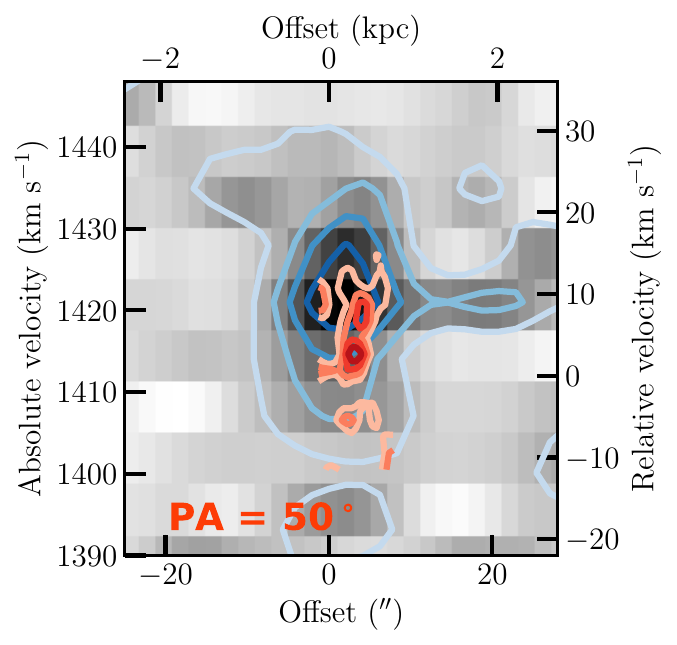}
        \caption{\HI\ (greyscale, blue) and CO (red) PVDs for FCC207.}
    \label{subfig:PVD_FCC207}
    \end{subfigure}
    
    \begin{subfigure}{0.41\textwidth}
    \centering
        \includegraphics[width=\textwidth]{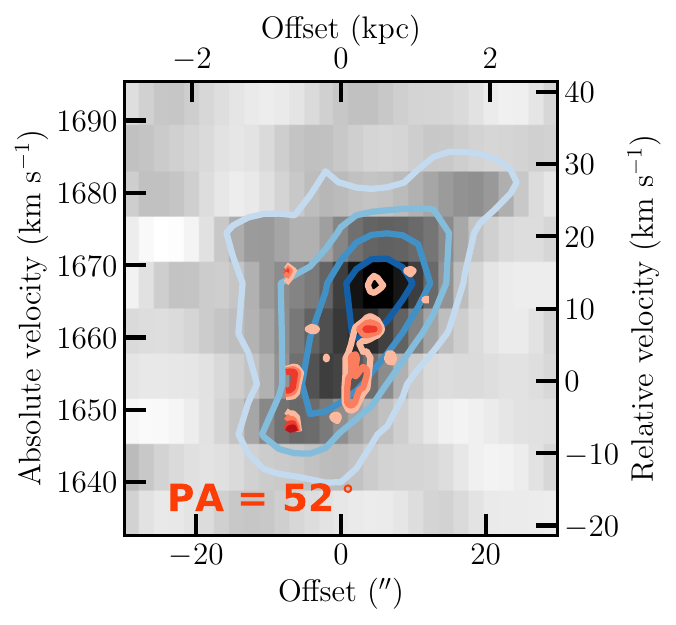}
        \caption{\HI\ (greyscale, blue) and CO (red) PVDs for FCC261.}
    \label{subfig:PVD_FCC261}
    \end{subfigure} \hspace{10mm}
    \begin{subfigure}{0.34\textwidth}
        \centering
        \includegraphics[width=\textwidth]{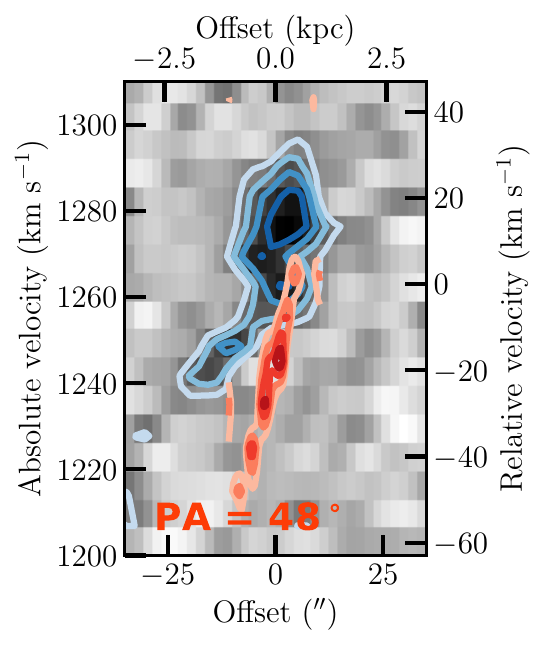}
        \caption{\HI\ (greyscale, blue) and CO (red) PVDs for FCC282.}
    \label{subfig:PVD_FCC282}
    \end{subfigure}

    \begin{subfigure}{0.32\textwidth}
    \centering
        \includegraphics[width=\textwidth]{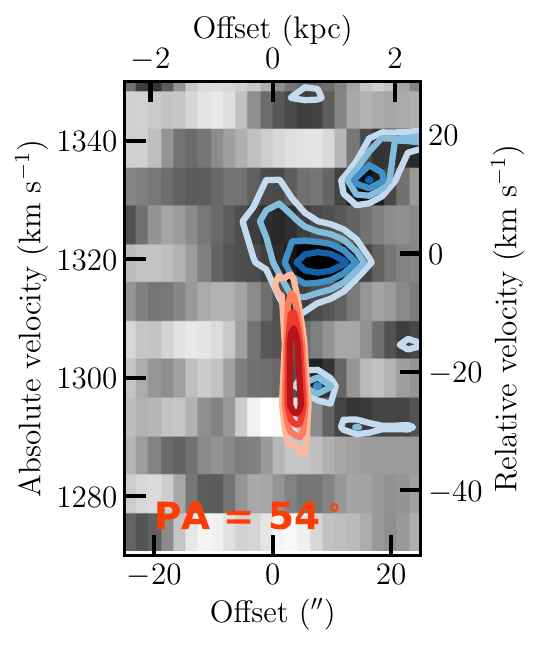}
        \caption{\HI\ (greyscale, blue) and CO (red) PVDs for FCC332.}
    \label{subfig:PVD_FCC332}
    \end{subfigure} \hspace{10mm}
    \begin{subfigure}{0.33\textwidth}
        \centering
        \includegraphics[width=\textwidth]{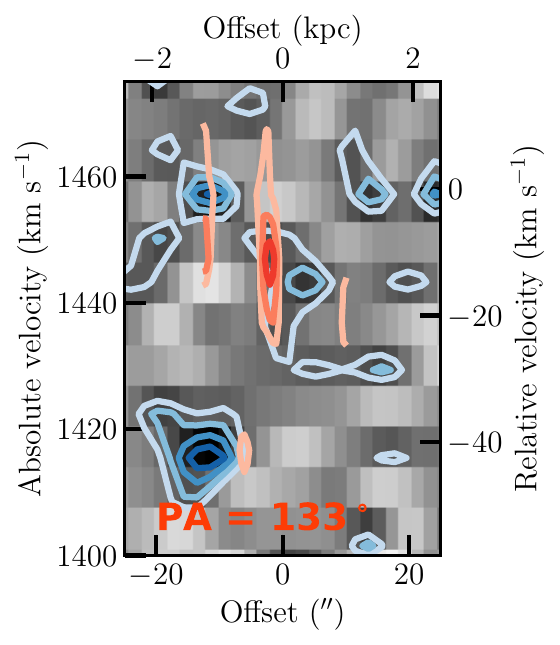}
        \caption{\HI\ (greyscale, blue) and CO (red) PVDs for FCC335.}
    \label{subfig:PVD_FCC335}
    \end{subfigure}
    \caption{Position-velocity diagrams (PVDs) of the \HI\ (greyscale and blue contours, \dtwo{5 contours between 0.15 mJy/b and the maximum value in the PVD}) and CO (red contours \dtwo{5 contours between 4 mJy/b and the maximum value in the PVD}) in the galaxies in the sample, \dtwo{with the slice positioned in such a way that it covers as much of the high-density \HI\ as possible (the corresponding position angles, measured clockwise from 12 o'clock, are indicated in each panel). The relative velocities are shown with respect to the systemic velocity of the stars.} These PVDs highlight the sometimes significant offsets between the \HI\ and CO emission both spatially and in velocity.}
    \label{fig:pvds}
\end{figure*}

\update{To further analyse the offsets between the atomic and molecular gas phases observed in Figure \ref{fig:intensity_maps}, we create and compare their position-velocity diagrams along the major axis of the \HI\ emission. These are shown in Figure \ref{fig:pvds}, where the greyscale background and corresponding blue contours show the \HI\ PVD, and the red contours show that of the CO emission. In some galaxies, most notably FCC090 and FCC282, but also in FCC261 and FCC332, the molecular gas is clearly offset from the atomic gas, both spatially and in velocity. This could be an indication of both gas phases being affected by environment in different ways and/or on different timescales.}

\clearpage

\section{Individual Kennicutt-Schmidt + $\lowercase{t}_\text{\lowercase{dep}}$ plots}
\label{app:ks_dt}
\dtwo{In Figure \ref{fig:DT_plots} we show the rKS relations for the dwarfs in the sample individually (left-hand panels), and corresponding depletion time maps (right-hand panels). In the left-hand panels the colours of the markers indicate the local density of the data, with lighter colours indicating a higher-density of data points. Similarly to Figure \ref{fig:rKS}, the original Kennicutt-Schmidt relation (which relates $\Sigma_{\text{SFR}}$ and $\Sigma_{\text{H}_2 + \text{H}\textsc{i}}$) is shown in pink, and the rKS relation from \citet{Bigiel2008} is shown in blue. Dash-dotted lines of constant depletion times are also shown. In the right-hand panels, maps of the corresponding depletion times are overlaid on optical $g$-band images from the FDS, whose logarithmic contours are shown in white. The extent of the H$\alpha$ emission is indicated by an orange contour, while the extent of the CO emission is shown by a blue contour. In areas where H$\alpha$ is detected, but CO remains undetected, the upper limit on the depletion time is shown, using the 3$\sigma$ upper limit on the $\Sigma_{\text{H}_2}$, calculated following \citet{Zabel2019}. Note that in Figures \ref{subfig:KS_FCC332}, \ref{subfig:DT_FCC332}, \ref{subfig:KS_FCC335}, and \ref{subfig:DT_FCC335} all H$\alpha$ emission is assumed to be related to star formation, which is not accurate, as can be seen in Figure \ref{fig:bpt} (\S \ref{subsub:sfr_sd}). Therefore, the $\Sigma_\text{SFR}$ and corresponding depletion times in these galaxies should be treated as upper limits.}

\dtwo{There is a variety in depletion times both between and within individual galaxies. In FCC090 and FCC207 the asymmetric molecular gas ``tails'' have significantly longer depletion times than the H$_2$ in the discs. In other galaxies however, such as FCC261 (Figures \ref{subfig:ks_FCC261} and \ref{subfig:tdep_FCC261}), there is no one-sided gradient in depletion times, but it decreases towards the outskirts of the CO disc, and is higher in the central region. In FCC282 (Figures \ref{subfig:KS_FCC282} and \ref{subfig:DT_FCC282}) CO is only detected in a relatively small part of the star-forming areas. This is likely due to detection limits. In FCC332 (Figures \ref{subfig:KS_FCC332} and \ref{subfig:DT_FCC332}) no CO is detected in the extension towards the north-east, while in the extension towards the south depletion times are significantly longer than in the disc, especially keeping in mind that the contribution of star formation towards ionising the gas in this area is minimal.}

\begin{figure*}
    \begin{subfigure}{0.44\textwidth}
    \centering
        \includegraphics[width=\textwidth]{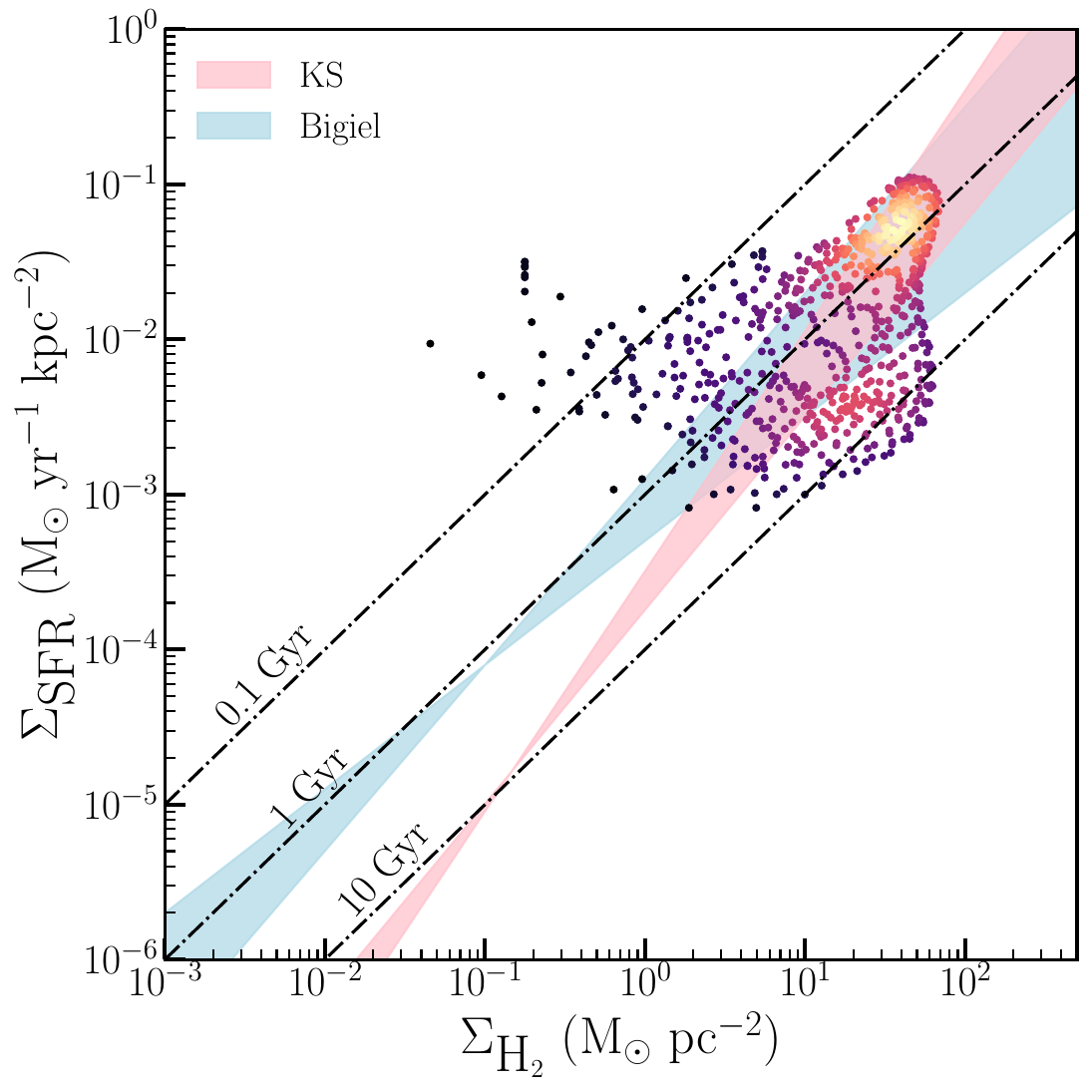}
        \caption{Kennicutt-Schmidt relation for FCC090. The colours of the markers indicate the density of the data points. The pink shaded region indicates the relation from \citet{Kennicutt1998}, and the blue shaded region that from \citet{Bigiel2008}.}
    \label{subfig:KS_FCC090}
    \end{subfigure}
    \begin{subfigure}{0.54\textwidth}
        \centering
        \includegraphics[width=\textwidth]{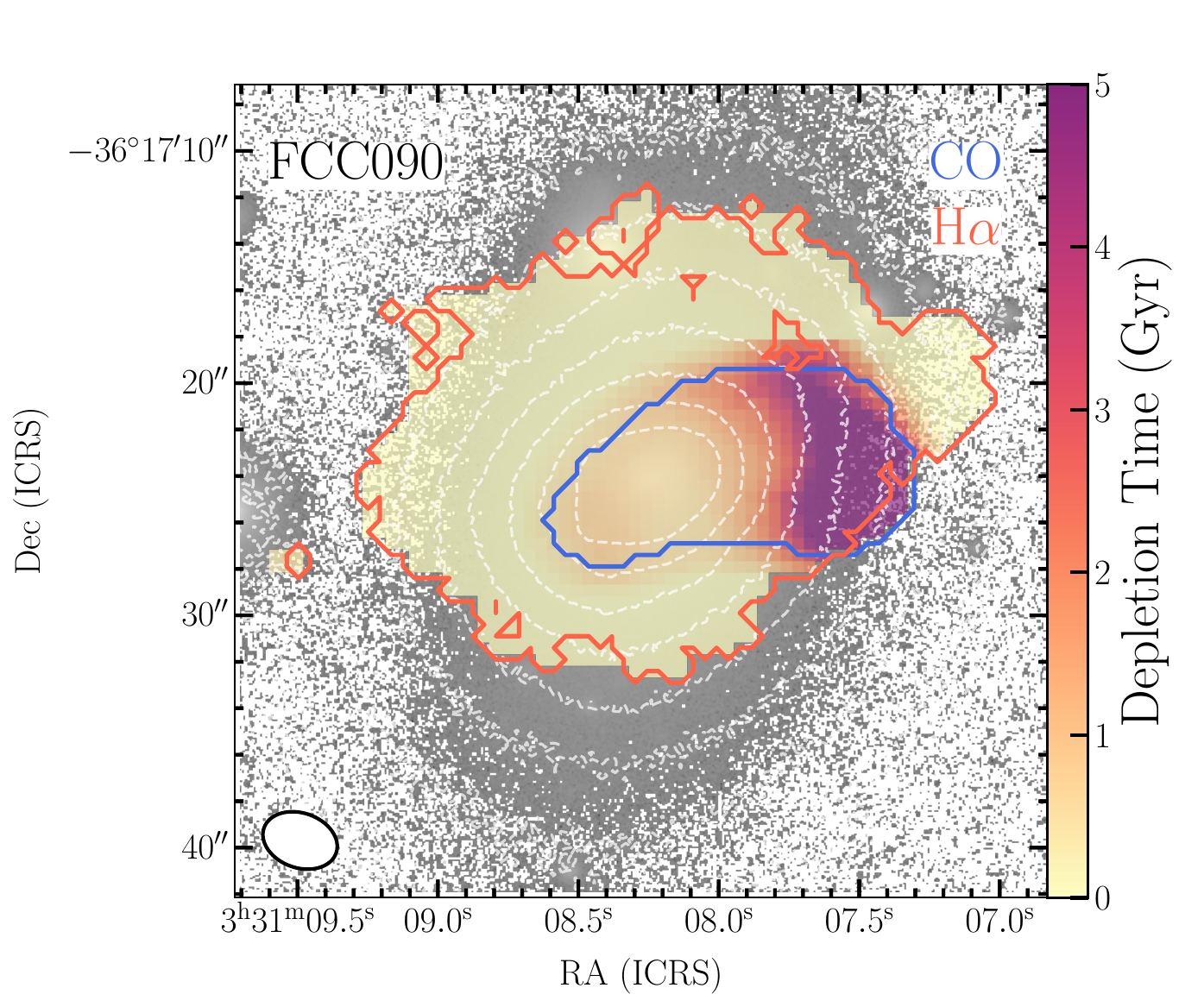}
        \caption{The depletion time map for FCC090, corresponding to the Kennicutt-Schmidt relation in Figure \ref{subfig:KS_FCC090}. The optical $g$-band image from the FDS in logaritmic scale, with corresponding contours shown as dashed white lines. The extent of the detected CO is shown with a blue contour, and the extent of the detected H$\alpha$ is indicated with an orange contour. The beam of the ALMA observations is shown in the bottom-left corner.}
    \label{subfig:DT_FCC090}
    \end{subfigure}
\end{figure*}

\begin{figure*}\ContinuedFloat 
    \begin{subfigure}{0.43\textwidth}
    \centering
        \includegraphics[width=\textwidth]{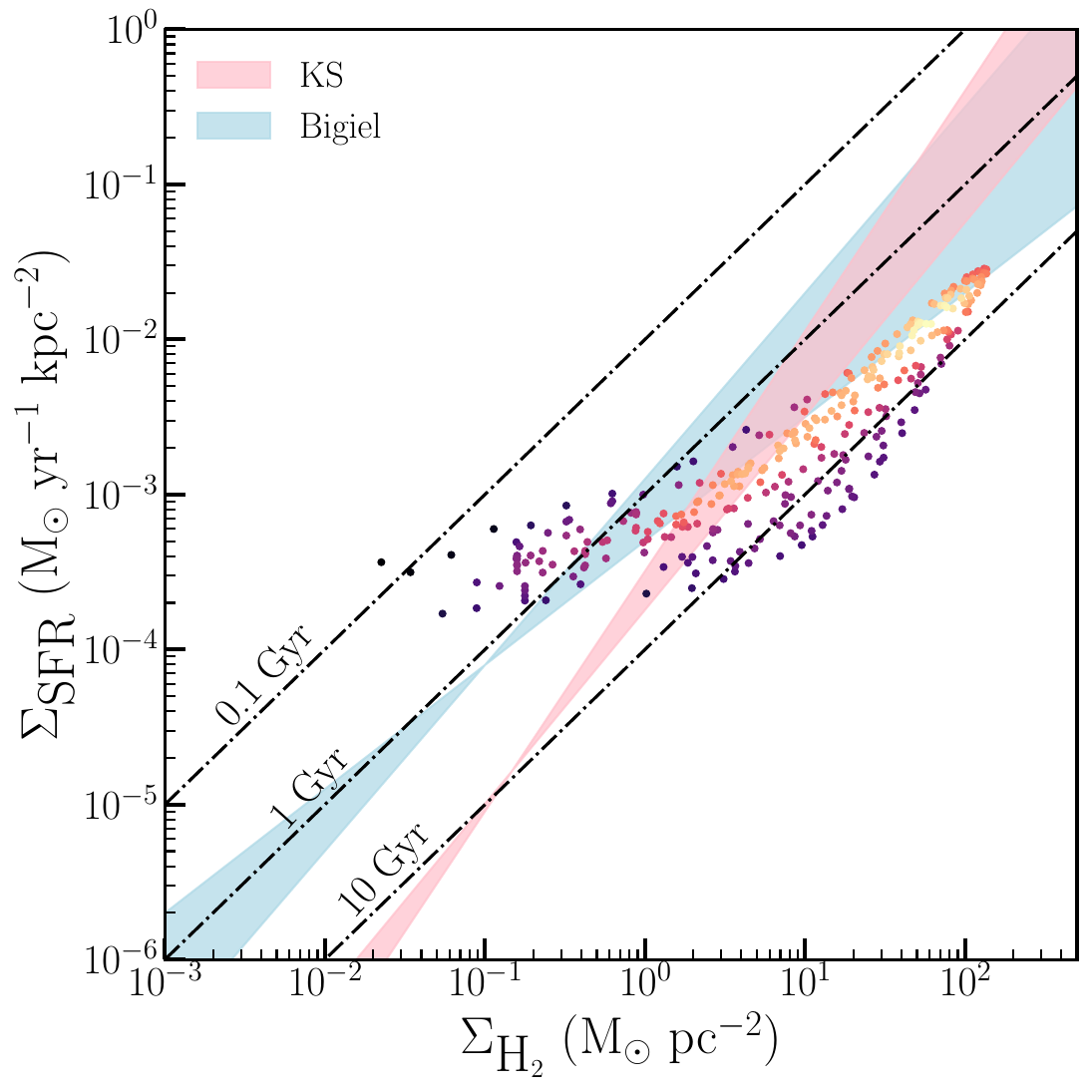}
        \caption{Same as Figure \ref{subfig:KS_FCC090}, but for FCC207.}
    \end{subfigure}
    \begin{subfigure}{0.55\textwidth}
        \centering
        \includegraphics[width=\textwidth]{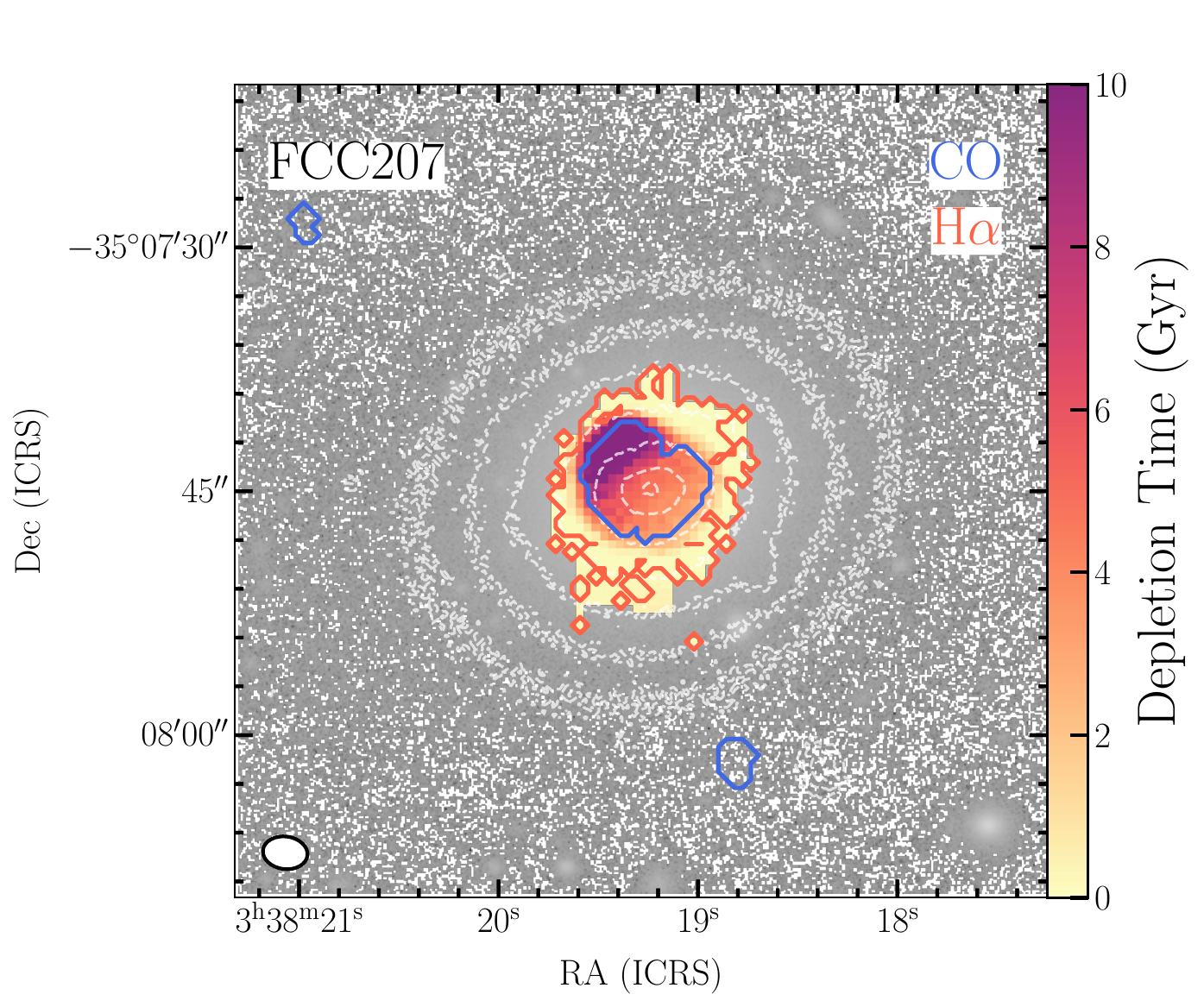}
        \caption{Same as Figure \ref{subfig:DT_FCC090}, but for FCC207.}
    \end{subfigure}
\end{figure*}

\begin{figure*}\ContinuedFloat     
    \begin{subfigure}{0.43\textwidth}
    \centering
        \includegraphics[width=\textwidth]{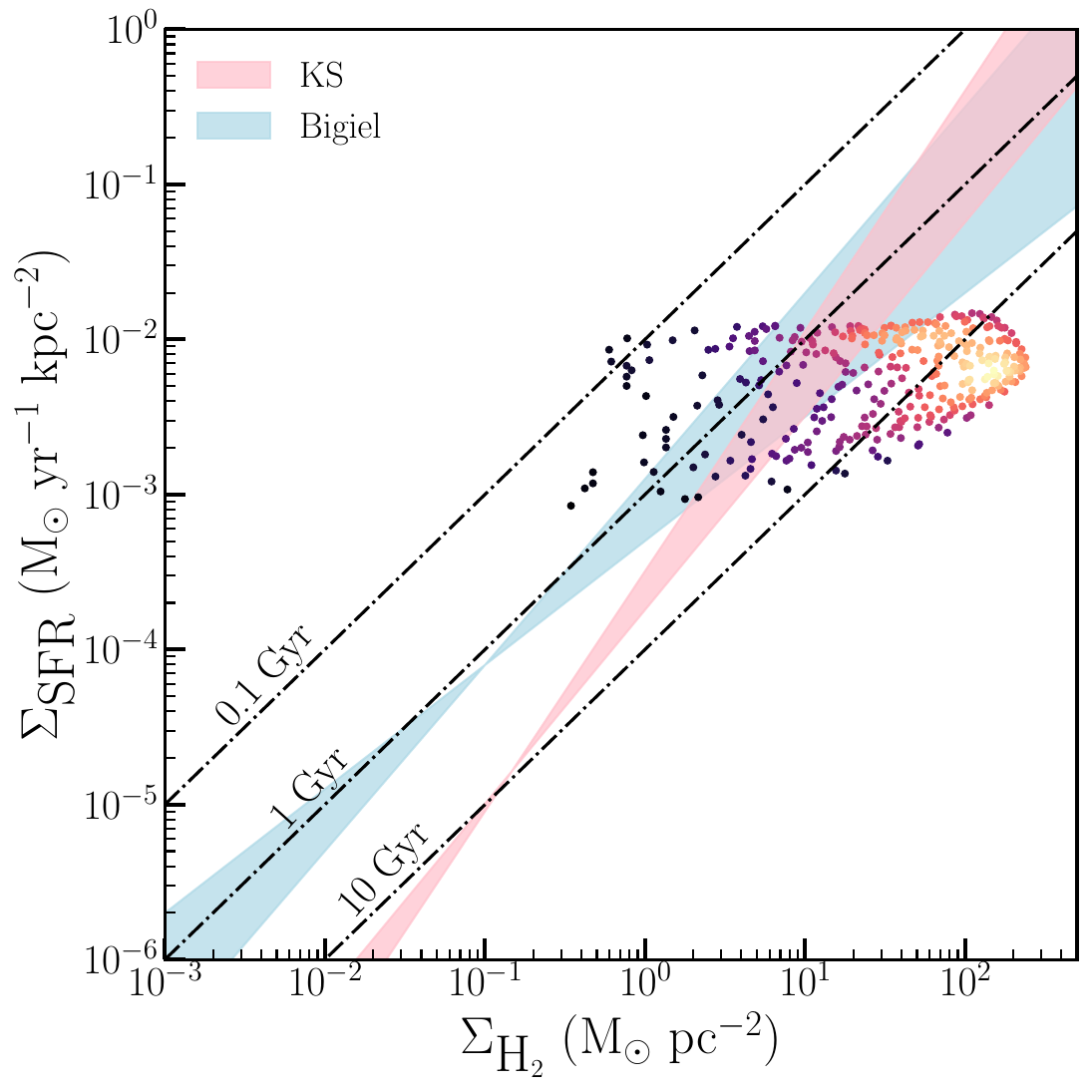}
        \caption{Same as Figure \ref{subfig:KS_FCC090}, but for FCC261.}
        \label{subfig:ks_FCC261}
    \end{subfigure}
    \begin{subfigure}{0.56\textwidth}
        \centering
        \includegraphics[width=\textwidth]{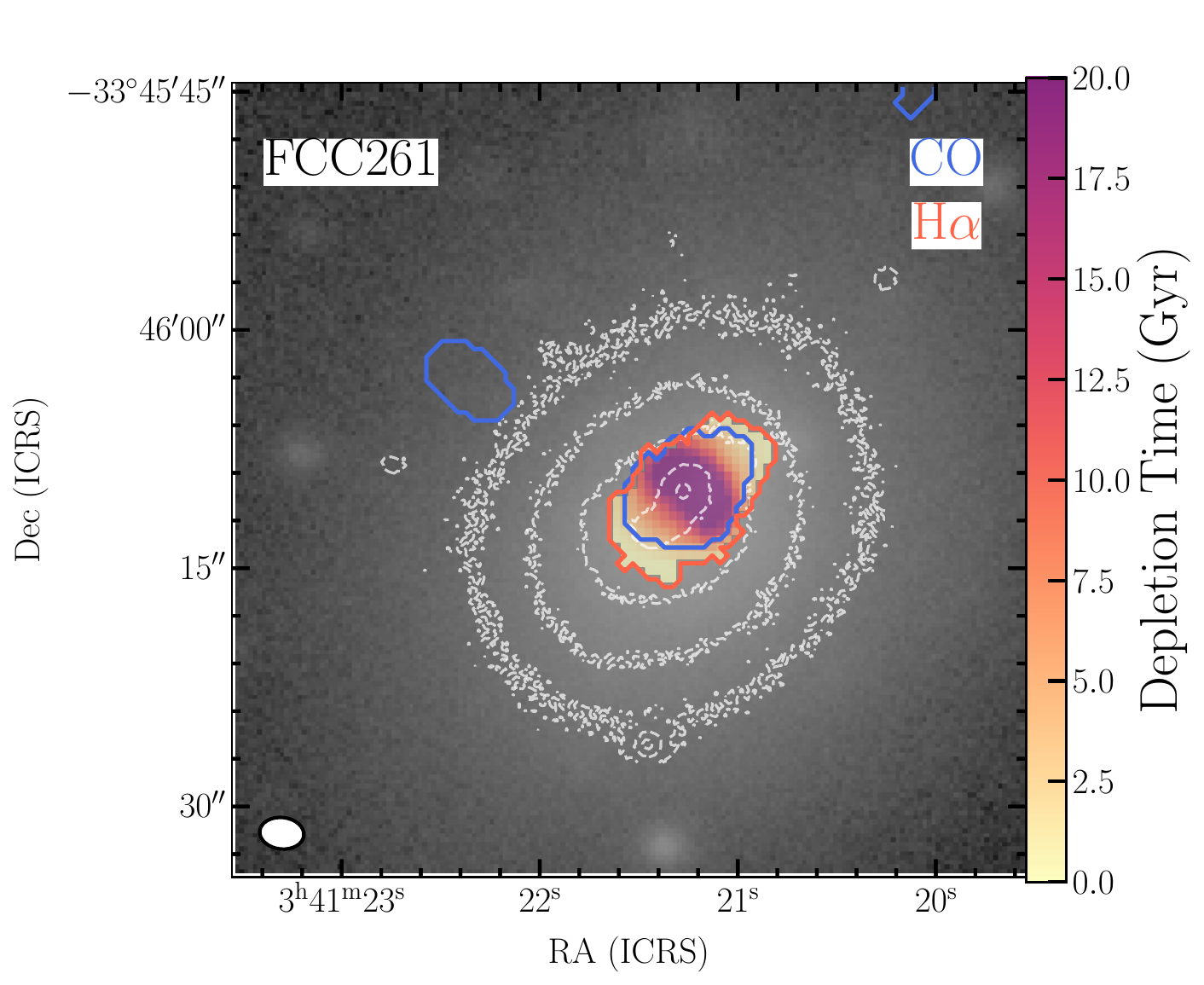}
        \caption{Same as Figure \ref{subfig:DT_FCC090}, but for FCC261.}
        \label{subfig:tdep_FCC261}
    \end{subfigure}
\end{figure*}

\begin{figure*}\ContinuedFloat     
    \begin{subfigure}{0.42\textwidth}
    \centering
        \includegraphics[width=\textwidth]{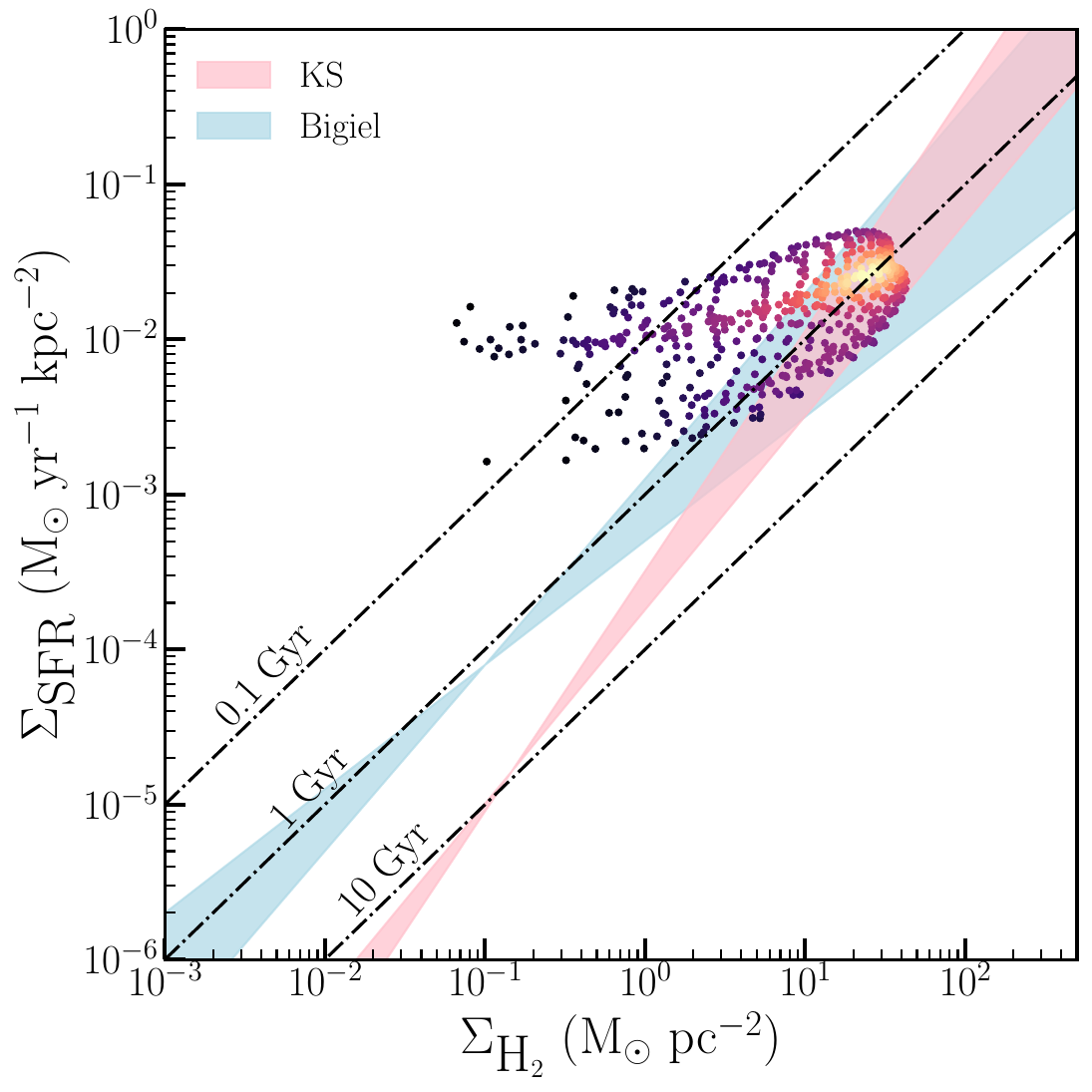}
        \caption{Same as Figure \ref{subfig:KS_FCC090}, but for FCC282.}
        \label{subfig:KS_FCC282}
    \end{subfigure}
    \begin{subfigure}{0.54\textwidth}
        \centering
        \includegraphics[width=\textwidth]{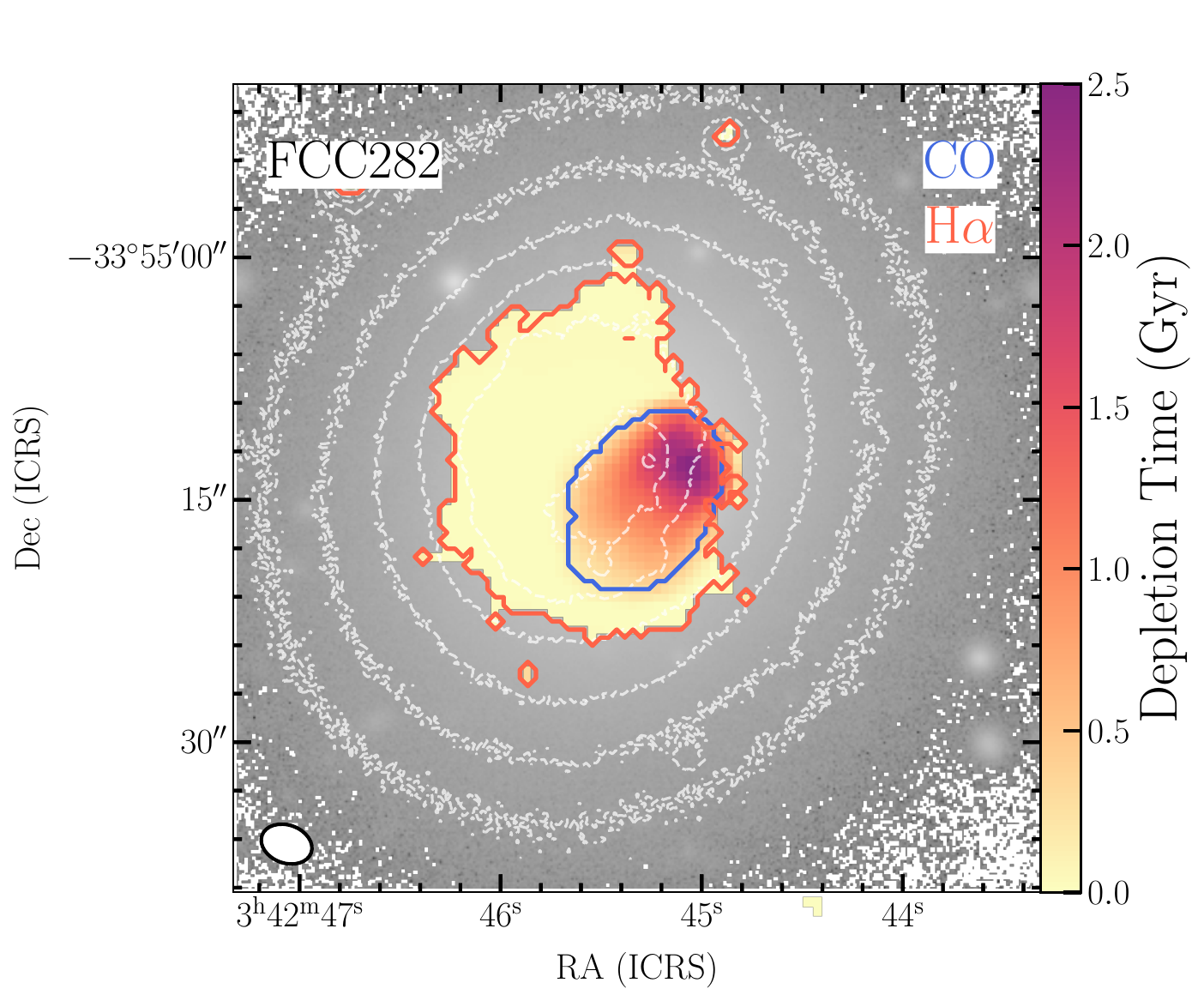}
        \caption{Same as Figure \ref{subfig:DT_FCC090}, but for FCC282.}
        \label{subfig:DT_FCC282}
    \end{subfigure}
\end{figure*}

\begin{figure*}\ContinuedFloat    
    \begin{subfigure}{0.42\textwidth}
    \centering
    \vspace{-0.5cm}
        \includegraphics[width=\textwidth]{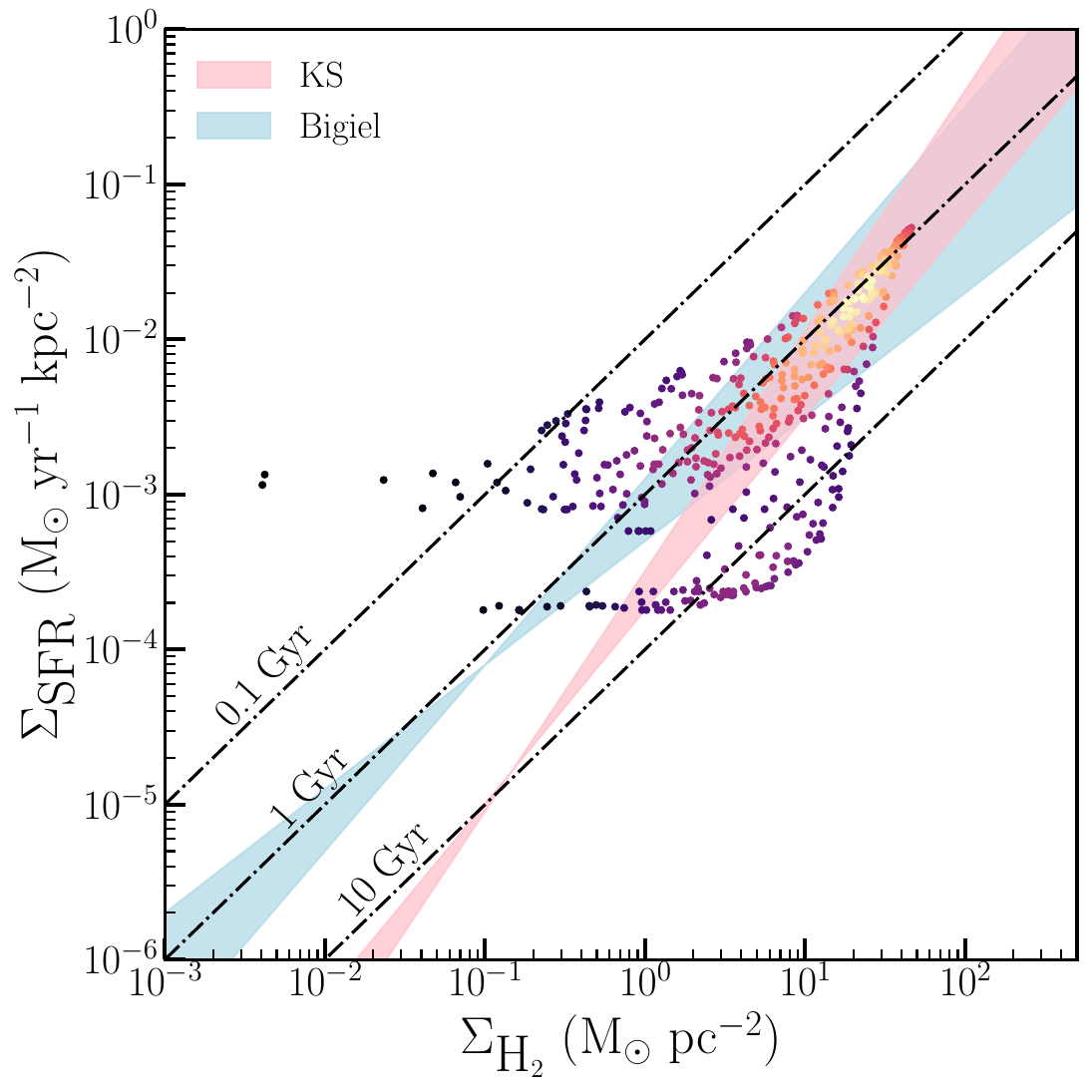}
        \caption{Same as Figure \ref{subfig:KS_FCC090}, but for FCC332. \refrep{Note that there are regions with significant contributions from ionisation sources other than O-stars in this galaxy (Figure \ref{fig:bpt}), and the $\Sigma_{\text{SFR}}$ values in these regions should be treated as upper limits, and caution should be used when interpreting this Figure.}}
        \label{subfig:KS_FCC332}
    \end{subfigure}
    \vspace{-0.5cm}
    \begin{subfigure}{0.56\textwidth}
        \centering
        \includegraphics[width=1\textwidth]{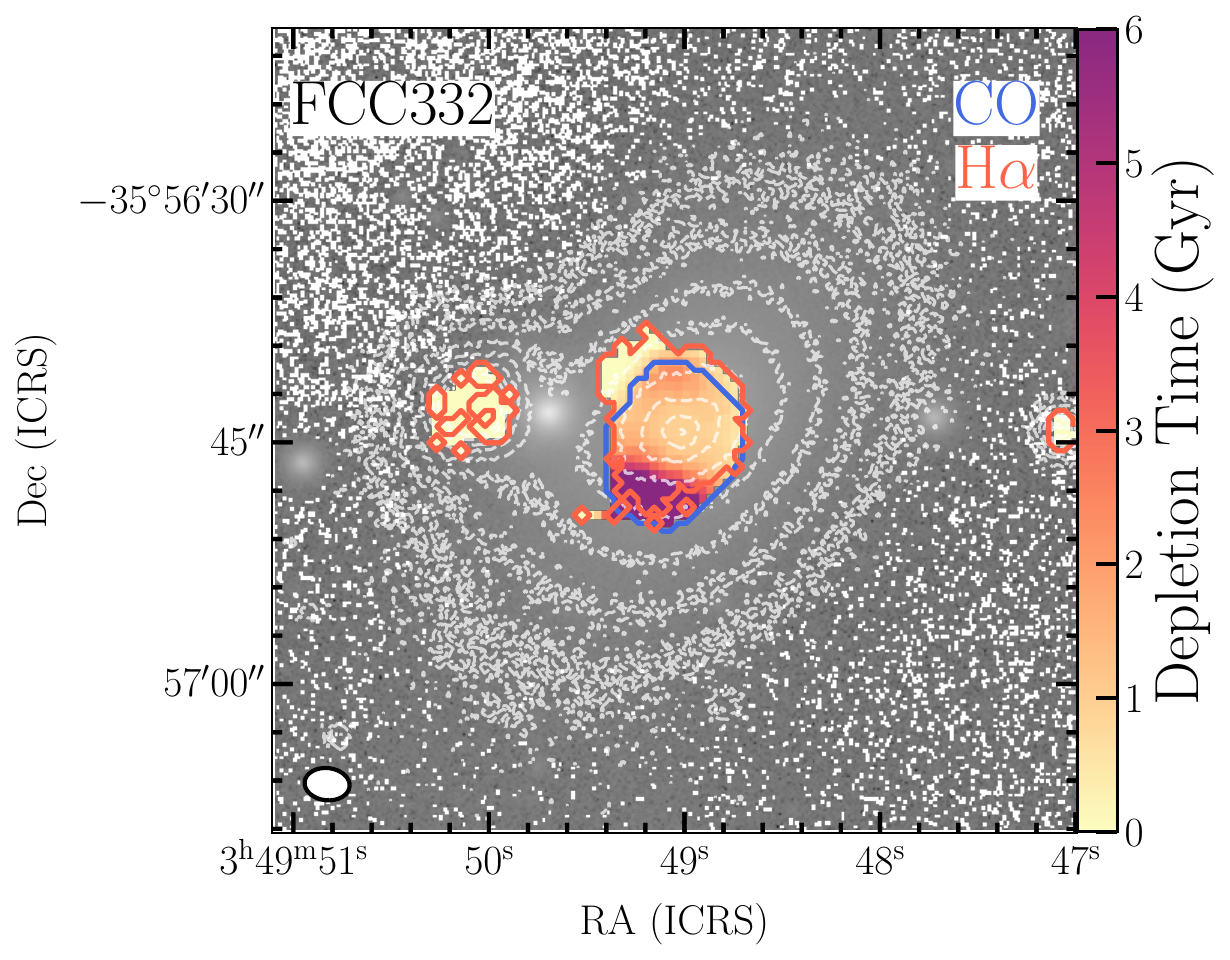}
        \caption{Same as Figure \ref{subfig:DT_FCC090}, but for FCC332. \refrep{Note that there are regions with significant contributions from ionisation sources other than O-stars in this galaxy (Figure \ref{fig:bpt}), and the $\Sigma_{\text{SFR}}$ values in these regions should be treated as upper limits, and caution should be used when interpreting this Figure.}}
        \label{subfig:DT_FCC332}
    \end{subfigure}
\end{figure*}

\begin{figure*}\ContinuedFloat 
    \begin{subfigure}{0.42\textwidth}
    \centering
        \includegraphics[width=\textwidth]{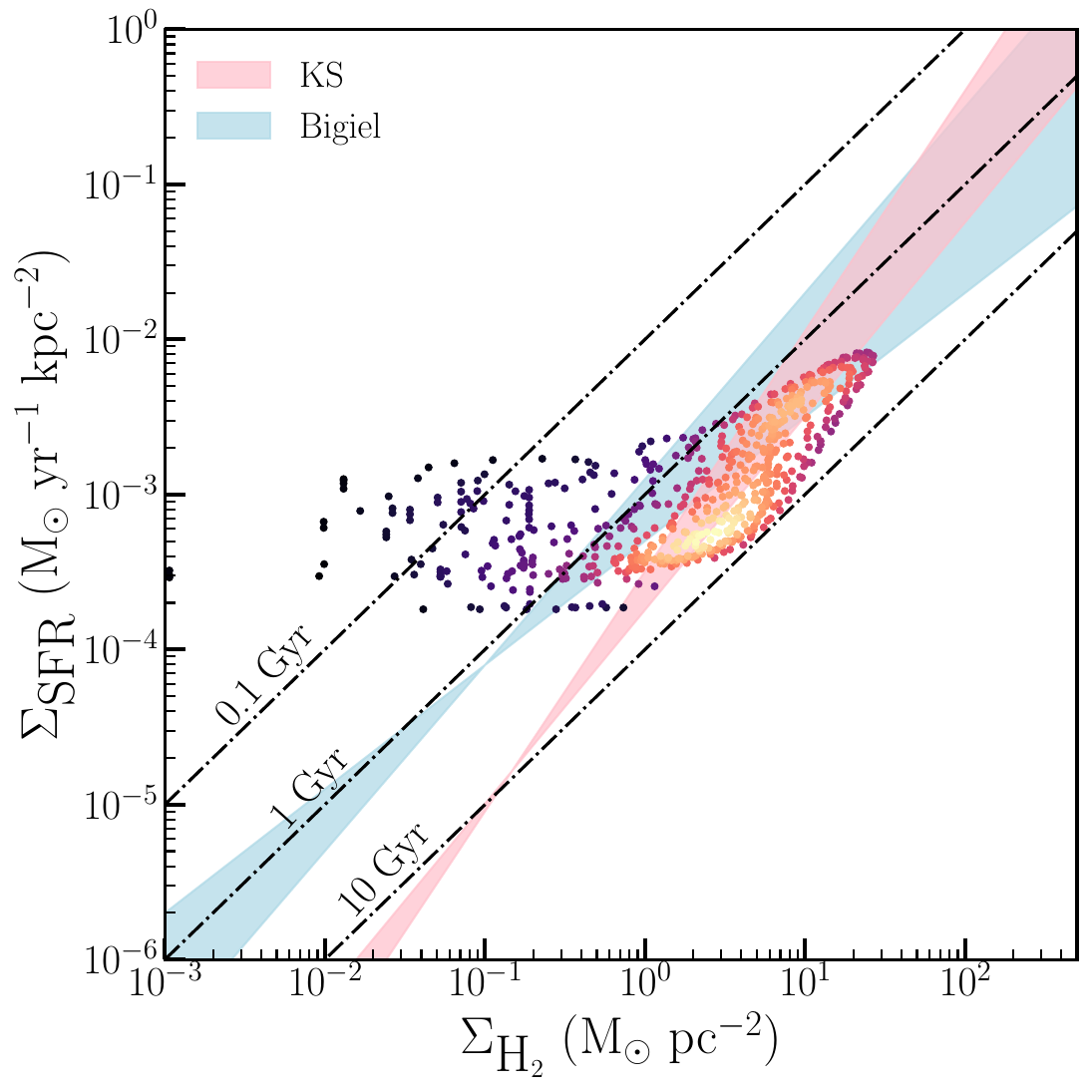}
        \caption{Same as Figure \ref{subfig:KS_FCC090}, but for FCC335. \refrep{Note that none of the ionised gas in this galaxy is the result of star formation alone (Figure \ref{fig:bpt}). The $\Sigma_{\text{SFR}}$ shown here are therefore strictly upper limits, and caution should be used when interpreting this Figure.}}
        \label{subfig:KS_FCC335}
    \end{subfigure}
    \begin{subfigure}{0.56\textwidth}
        \centering
        \includegraphics[width=\textwidth]{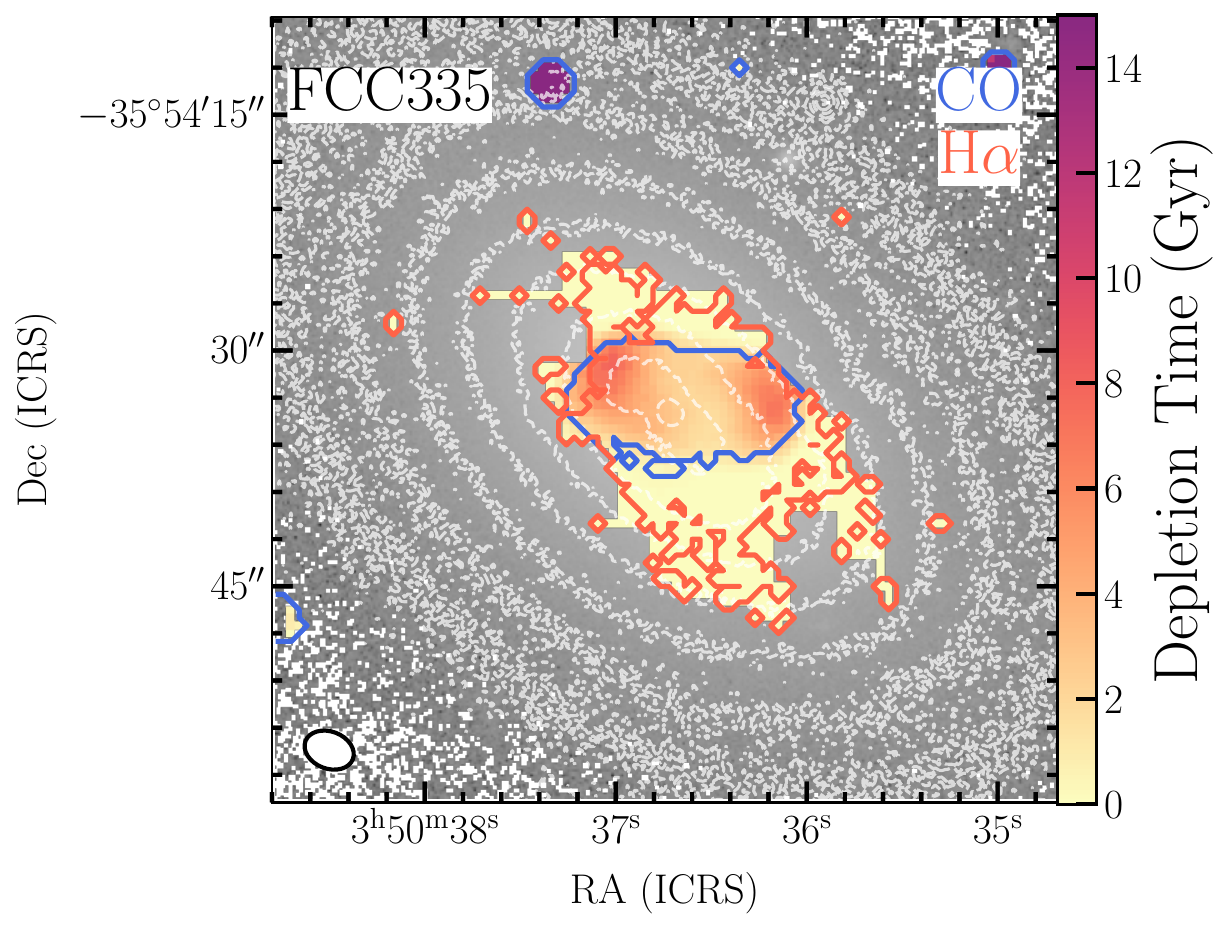}
        \caption{Same as Figure \ref{subfig:DT_FCC090}, but for FCC335. \refrep{Note that none of the ionised gas in this galaxy is the result of star formation alone (Figure \ref{fig:bpt}). The depletion times shown here are therefore strictly lower limits, and caution should be used when interpreting this Figure.}}
        \label{subfig:DT_FCC335}
    \end{subfigure}
     \caption{Individual Kennicutt-Schmidt relations (left-hand panels) and corresponding depletion time maps (right-hand panels) for the six dwarf galaxies in the sample. All dwarfs span a substantial area in the $\Sigma_{\text{H}_2} - \Sigma_\text{SFR}$ plane, and show spatial variation in their depletion times, where in some cases star formation is more efficient in the galaxy centre, while in other cases it is more efficent towards the outskirts, or there is an overall gradient.}
     \label{fig:DT_plots}
\end{figure*}

\subsection{BPT classification maps}
\label{app_sub:bpt_maps}
\refrep{In this section we show the BPT classifications of the remaining galaxies, which are not shown in Figure \ref{fig:bpt}. The emission from the four galaxies shown here is the result of star formation, with only a small number of spaxels around the edges classified as composite or AGN-like.}

\begin{figure*}
    \begin{subfigure}{0.49\textwidth}
    \centering
        \includegraphics[width=\textwidth]{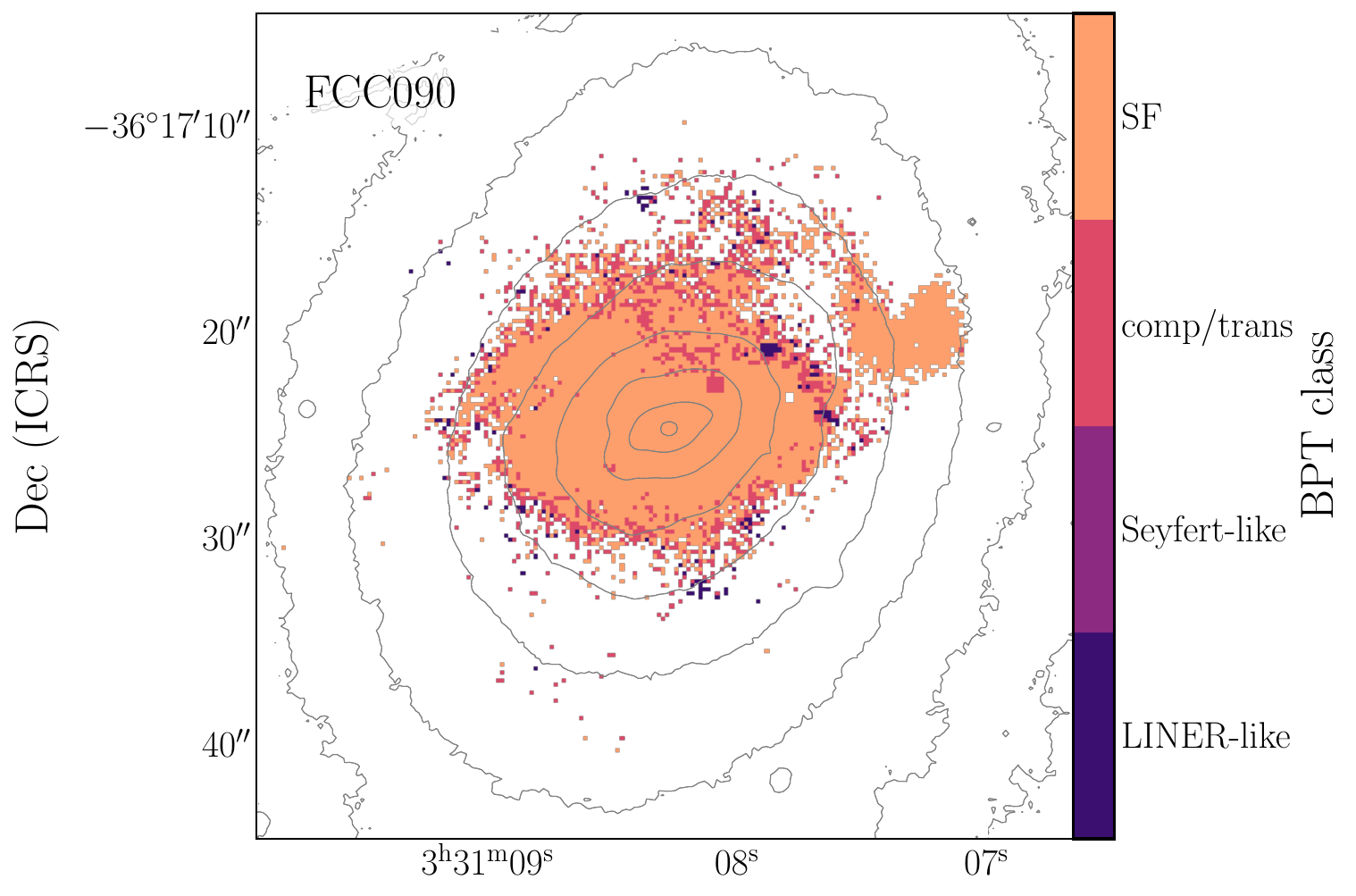}
        \caption{Map of the BPT classification for the ionised gas map of FCC090.}
        \label{subfig:FCC090_bpt}
    \end{subfigure}
    \begin{subfigure}{0.49\textwidth}
        \centering
        \includegraphics[width=\textwidth]{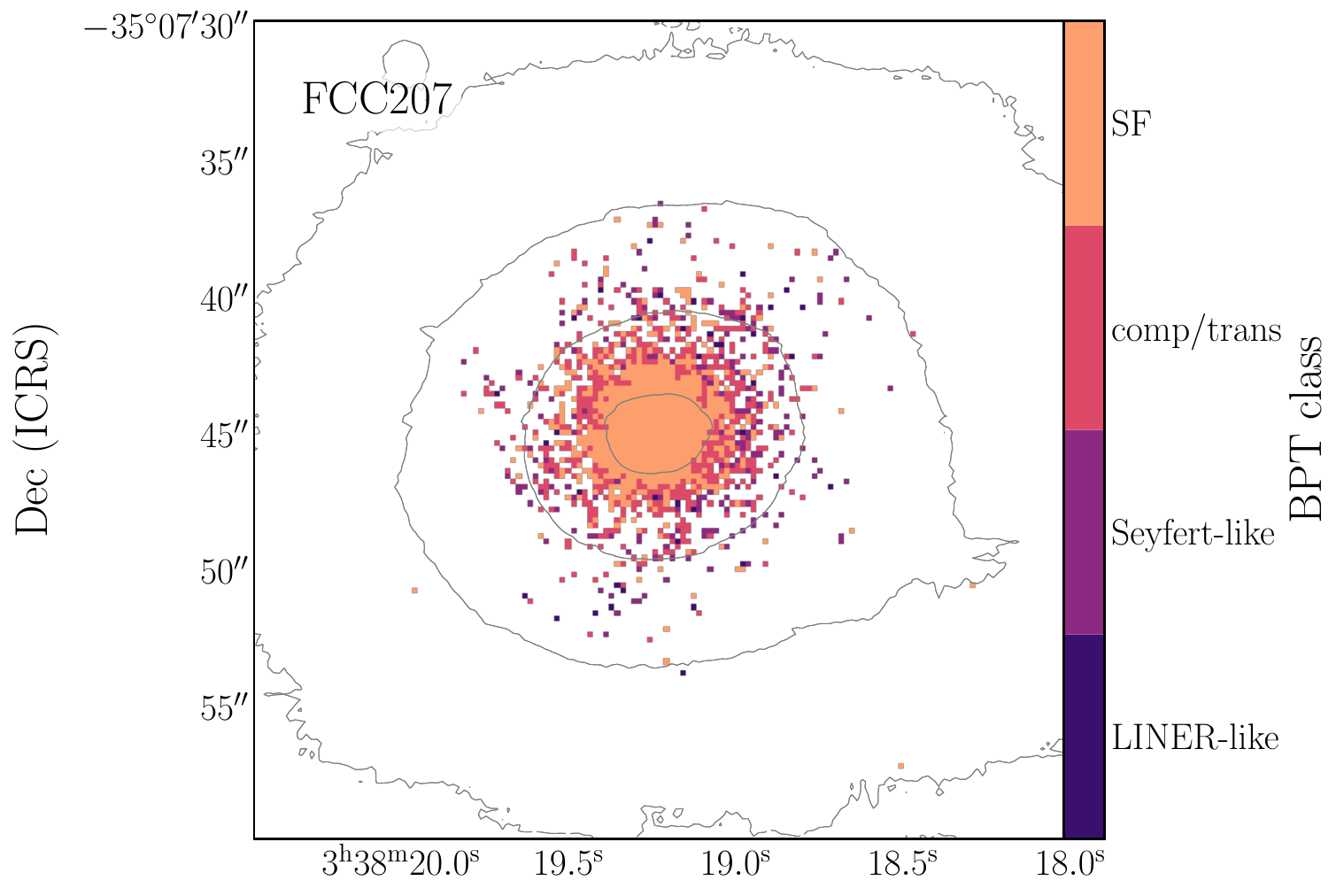}
        \caption{Map of the BPT classification for the ionised gas map of FCC207.}
        \label{subfig:FCC207_bpt}
    \end{subfigure}
    
    \vspace{5mm}
    
       \begin{subfigure}{0.49\textwidth}
    \centering
        \includegraphics[width=\textwidth]{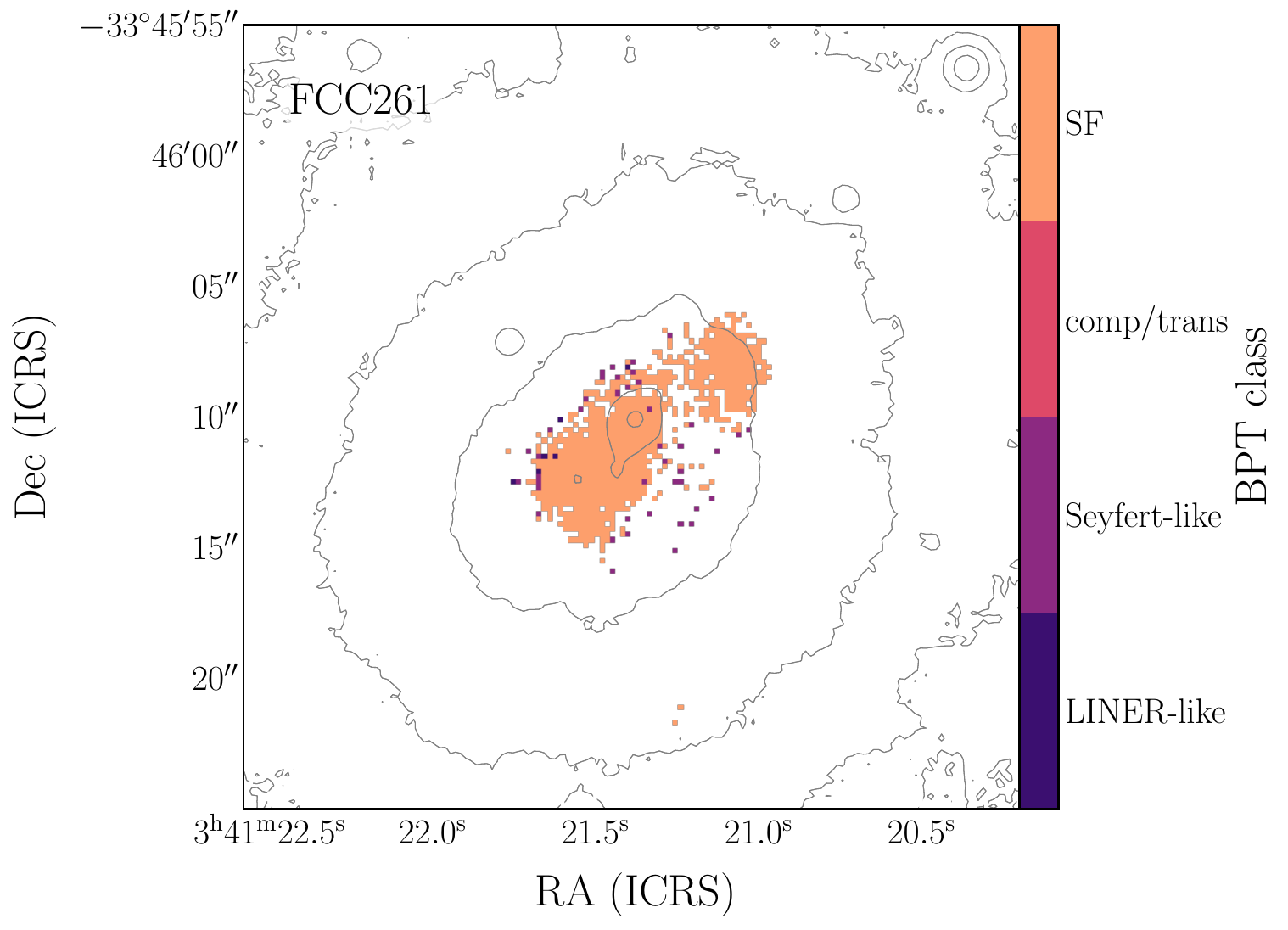}
        \caption{Map of the BPT classification for the ionised gas map of FCC261.}
        \label{subfig:FCC261_bpt}
    \end{subfigure}
    \begin{subfigure}{0.49\textwidth}
        \centering
        \includegraphics[width=\textwidth]{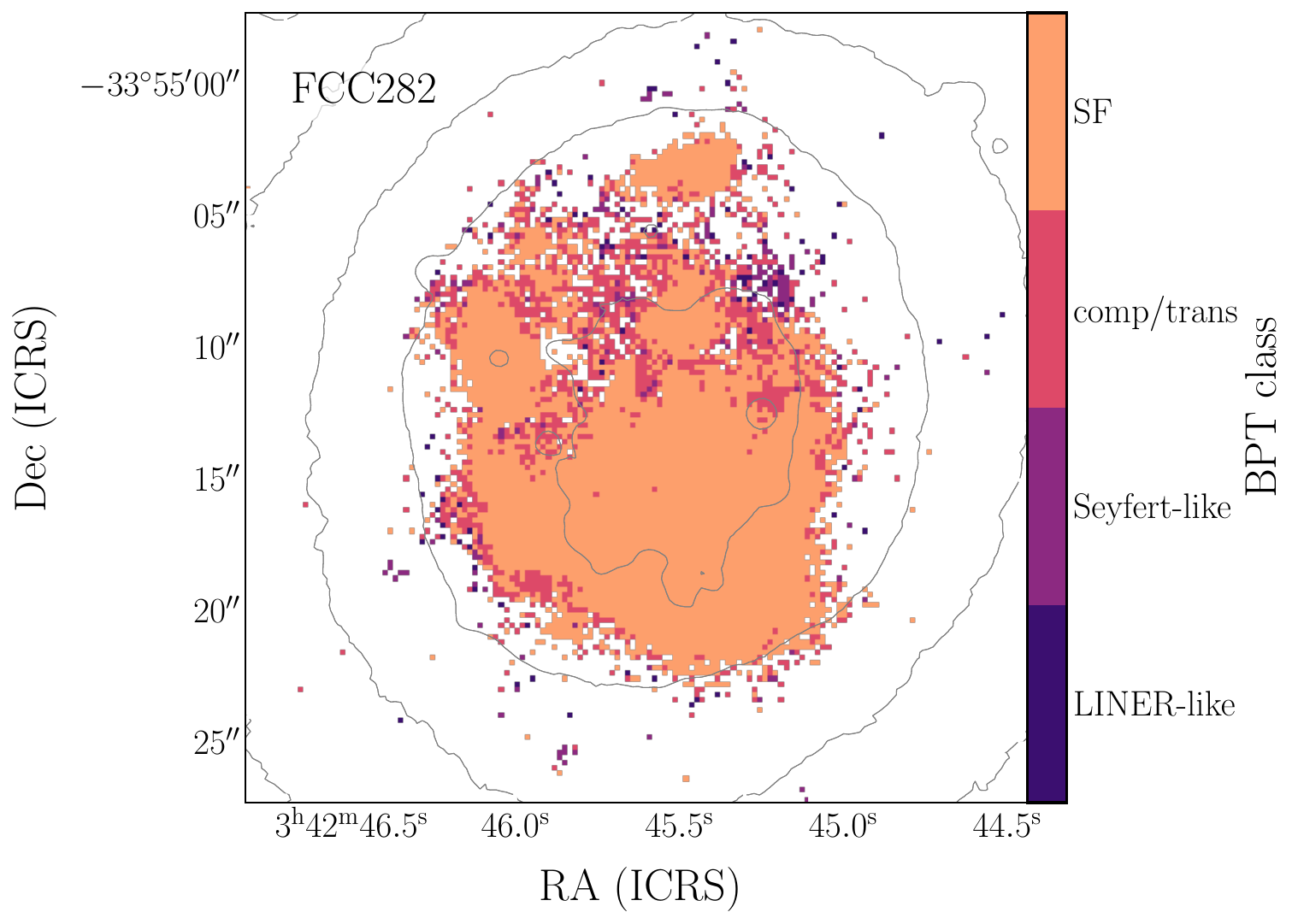}
        \caption{Map of the BPT classification for the ionised gas map of FCC282.}
        \label{subfig:FCC282_bpt}
    \end{subfigure}
     \caption{\update{Maps of the BPT classification for the galaxies that are not shown in Figure \ref{fig:bpt}. The orange pixels indicate regions where ionisation is due to star formation. Similarly, purple and dark purple regions indicate where ionisation is due to the presence of an AGN (Seyfert and low-ionization nuclear emission-line region (LINER), respectively). The pink regions, between the orange and purple in the colour bar, indicate ``composite'' or ``transition'' regions, where the ionisation is due to a combination of star formation and the presence of an AGN. With the exception of the noisier regions in the outskirts, which are likely not reliable, the gas in these galaxies is ionised due to star formation.}}
     \label{fig:BPT_maps_app}
\end{figure*}

\subsection{\refrep{BPT diagrams}}
\label{app_sub:bpt_diagrams}
\refrep{In this section we show the BPT diagrams corresponding to the maps shown in Figure \ref{fig:bpt} and \S \ref{app_sub:bpt_diagrams}.}

\begin{figure*}
    \begin{subfigure}{0.45\textwidth}
    \centering
        \includegraphics[width=\textwidth]{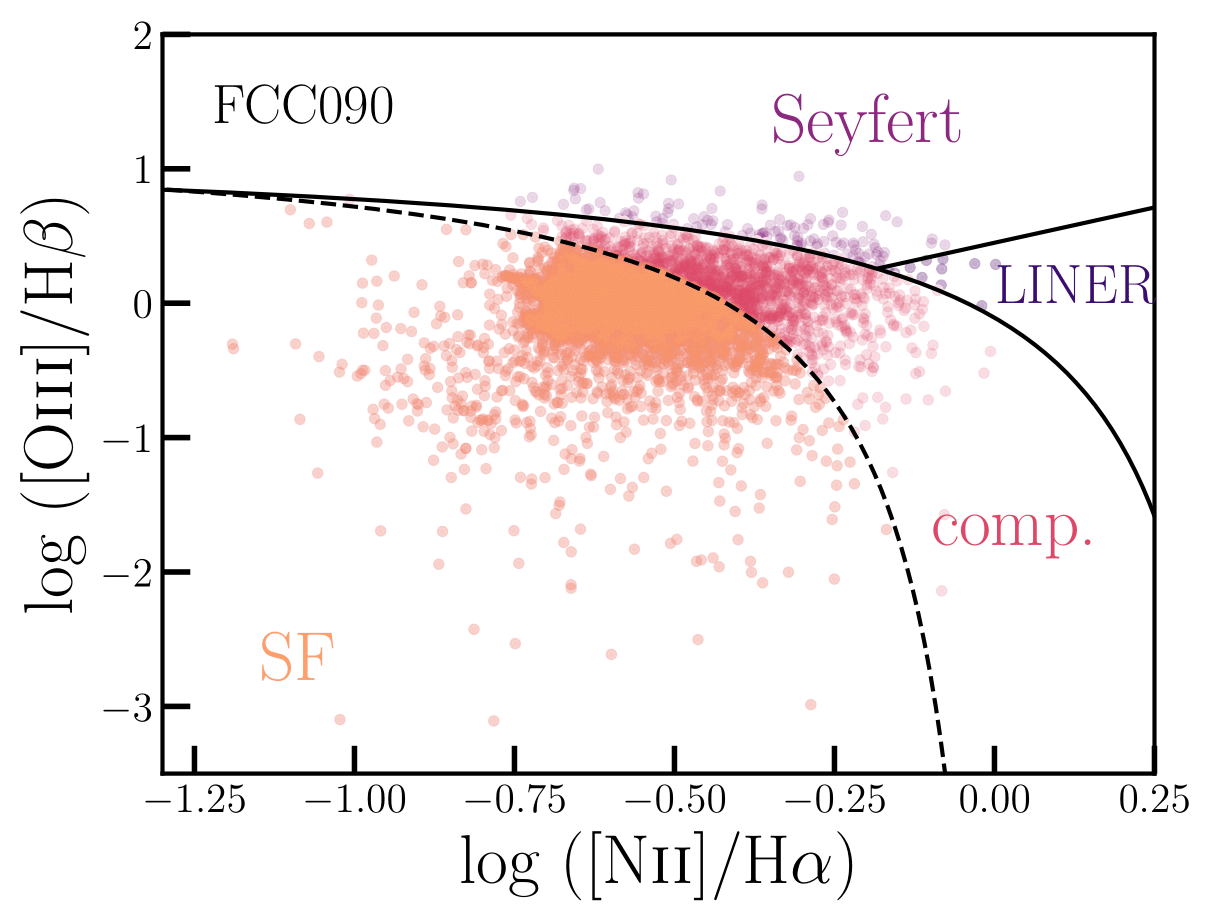}
        \caption{\refrep{BPT diagram for FCC090.}}
        \label{subfig:FCC090_bpt_diag}
    \end{subfigure}
    \begin{subfigure}{0.45\textwidth}
        \centering
        \includegraphics[width=\textwidth]{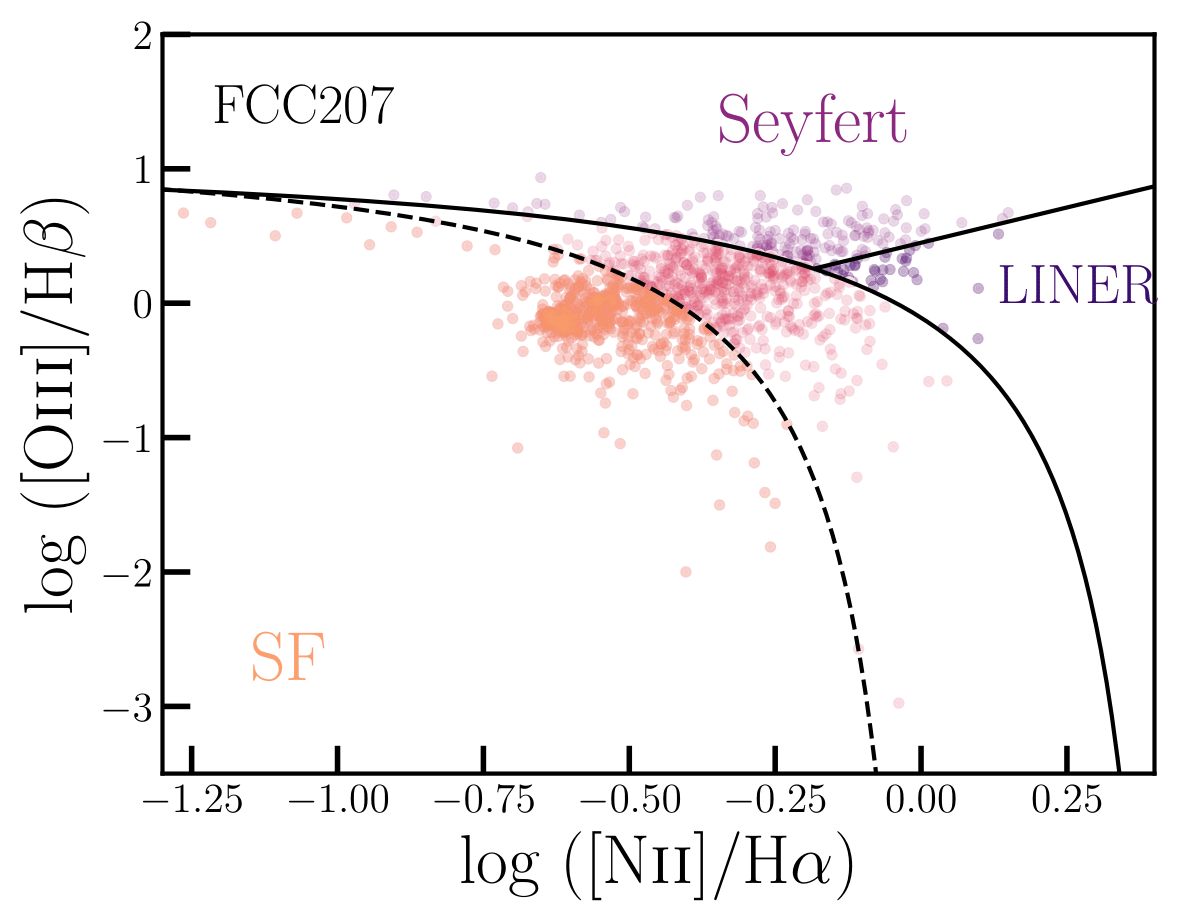}
        \caption{\refrep{BPT diagram for FCC207.}}
        \label{subfig:FCC207_bpt_diag}
    \end{subfigure}
    
    \vspace{5mm}
    
       \begin{subfigure}{0.45\textwidth}
    \centering
        \includegraphics[width=\textwidth]{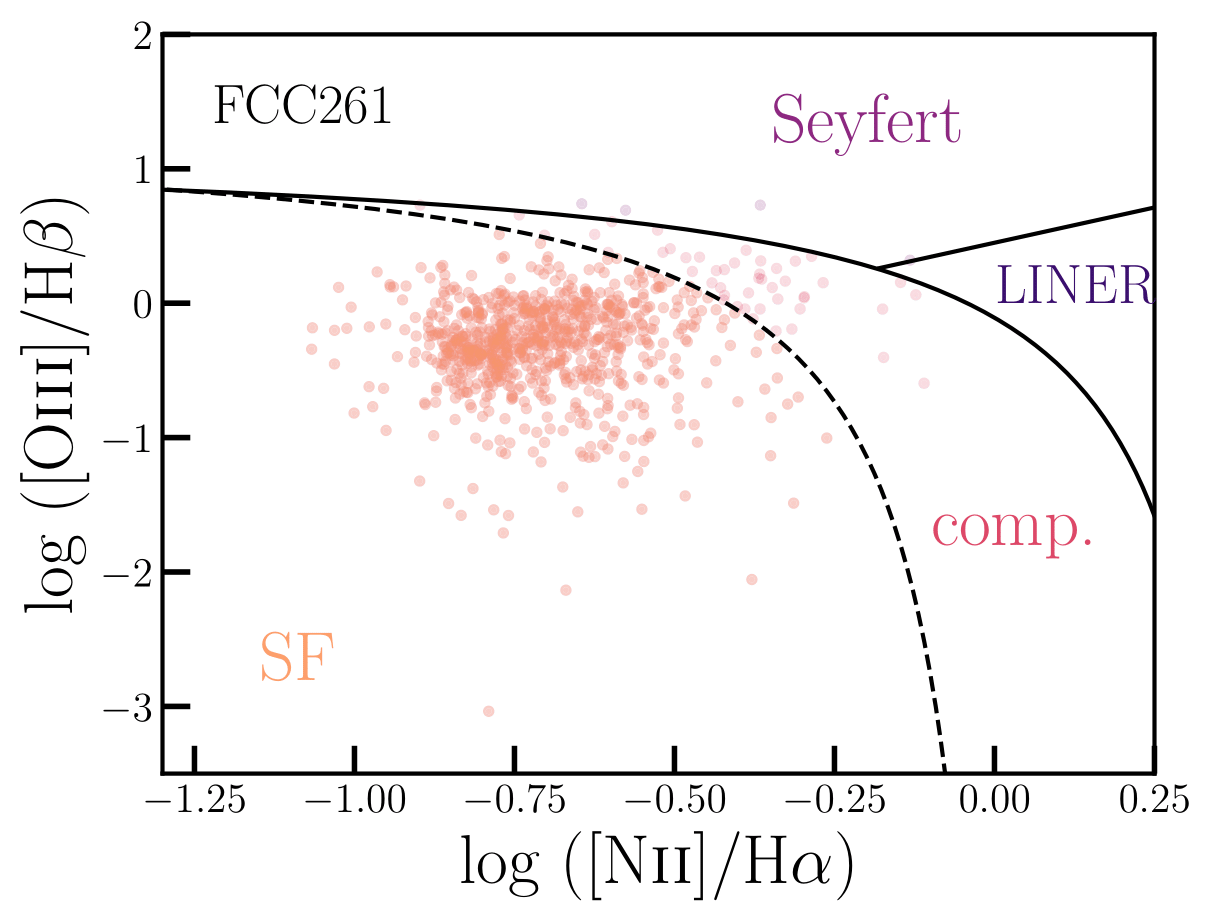}
        \caption{\refrep{BPT diagram for FCC261.}}
        \label{subfig:FCC261_bpt_diag}
    \end{subfigure}
    \begin{subfigure}{0.45\textwidth}
        \centering
        \includegraphics[width=\textwidth]{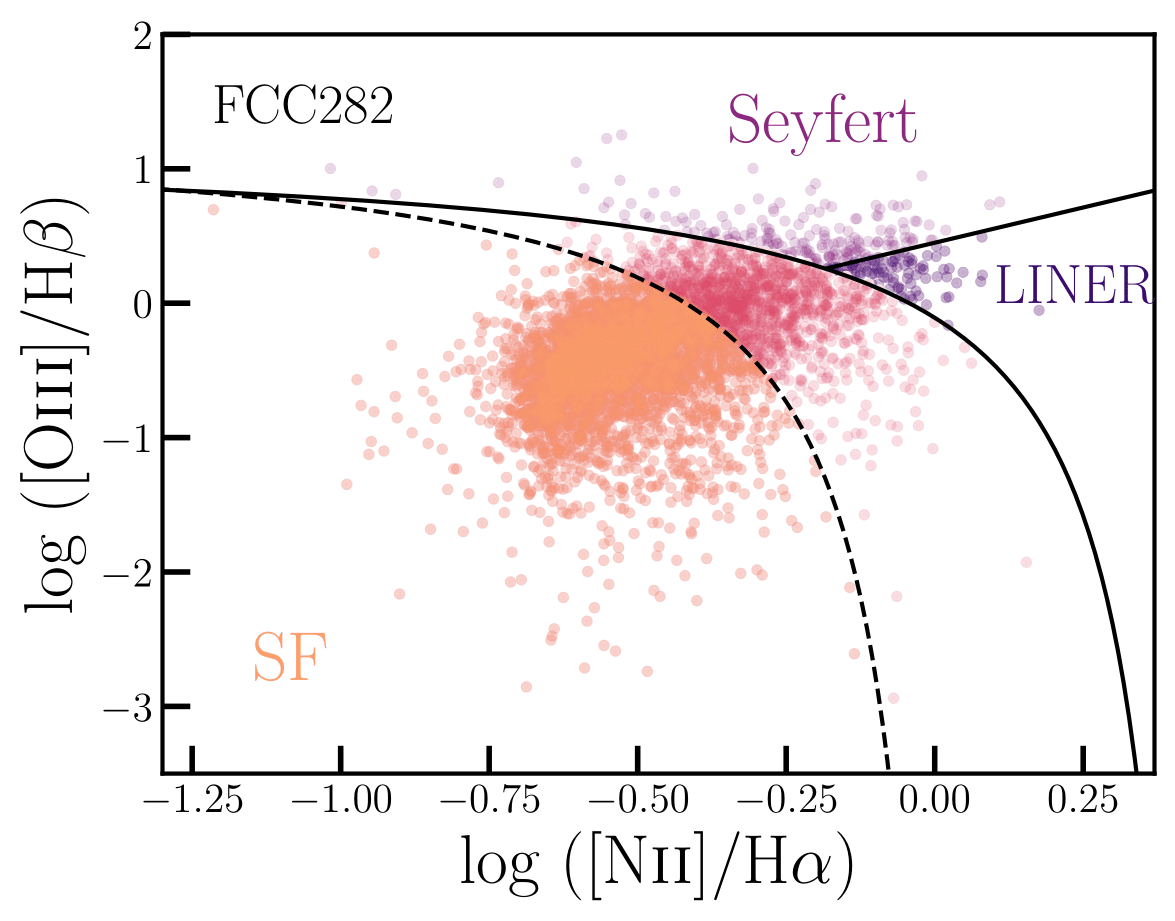}
        \caption{\refrep{BPT diagram for FCC282.}}
        \label{subfig:FCC282_bpt_diag}
    \end{subfigure}
    
        \vspace{5mm}
    
       \begin{subfigure}{0.45\textwidth}
    \centering
        \includegraphics[width=\textwidth]{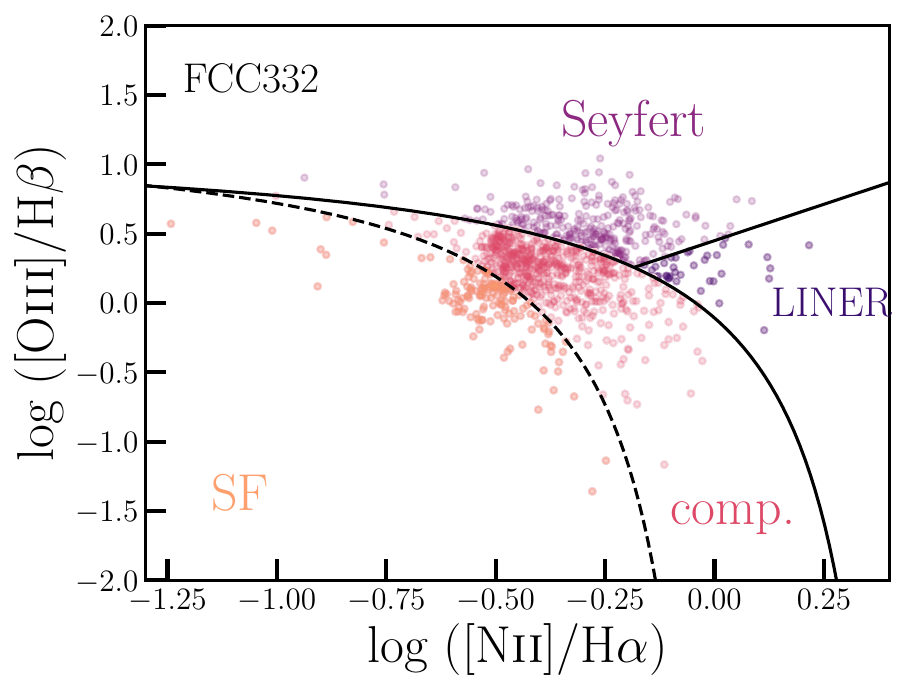}
        \caption{\refrep{BPT diagram for FCC332.}}
        \label{subfig:FCC332_bpt_diag}
    \end{subfigure}
    \begin{subfigure}{0.45\textwidth}
        \centering
        \includegraphics[width=\textwidth]{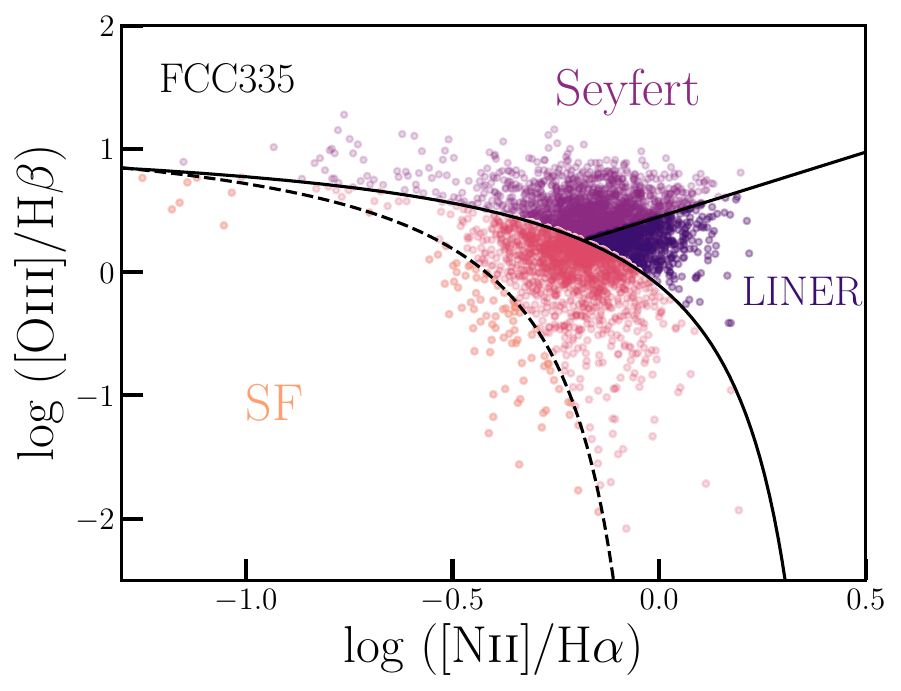}
        \caption{\refrep{BPT diagram for FCC335.}}
        \label{subfig:FCC335_bpt_diag}
    \end{subfigure}
     \caption{\refrep{BPT diagrams for the six dwarf galaxies in the sample. The boundary line from \citet{Kauffmann2003} is used to define ionisation by star formation, and spaxels above the boundary line from \citet{Kewley2001} are considered the result of AGN-like emission. Spaxels in between both lines are considered ``composite'', where AGN-like ionisation and ionisation from star formation contribute comparably. The boundary line from \citet{Schawinski2007} distinguishes between Seyfert-like and LINER-like emission. Marker colours correspond to those in Figures \ref{fig:bpt} and \ref{fig:BPT_maps_app}.}}
     \label{fig:BPT_diag_app}
\end{figure*}


\bsp	
\label{lastpage}
\end{document}